\documentclass[aps,prx,twocolumn,showpacs,superscriptaddress,floatfix,10pt,amsmath,amssymb,bibnotes,footinbib,floatfix]{revtex4-2}
\usepackage{graphicx}
\usepackage{bm,bbm,bbold}
\usepackage[urlcolor=blue]{hyperref}
\hypersetup{colorlinks=true,allcolors=blue}
\usepackage{amssymb,amsfonts} 
\usepackage{color, xcolor}
\usepackage{dsfont}

\newcommand{\StoInt}{\int_{\tau = 0}^{\tau=t}}
\newcommand{\rmd}{\mathrm{d}}
\newcommand{\rme}{\mathrm{e}}

\newcommand{\stoints}{\int_{s= 0}^{s=t}}

\newcommand{\rexp}{\mathrm{e}^{-V_{xy}/T}}
\newcommand{\Dt}{\tilde{D}_{xy}}

\newcommand{\f}{\bm}
\newcommand{\x}{{\f x}}
\newcommand{\Dx}{{\Delta x}}
\newcommand{\fDx}{{\Delta\x}}

\newcommand{\tb}{\textcolor{black}}

\begin{document}

\preprint{APS/123-QED}

\title{Stochastic Calculus  
for Pathwise Observables of Markov-Jump
  Processes:\\Unification of Diffusion and Jump Dynamics}

\author{Lars Torbj\o rn Stutzer}
\affiliation{Mathematical bioPhysics Group, Max Planck Institute for Multidisciplinary Sciences, 37077 G\"ottingen, Germany
}
\affiliation{Present address: Max Planck Institute for Dynamics and Self-organization, 37077 G\"ottingen, Germany}
\author{Cai Dieball}%
\affiliation{Mathematical bioPhysics Group, Max Planck Institute for Multidisciplinary Sciences, 37077 G\"ottingen, Germany
}
\author{Alja\v z Godec}\email{agodec@physik.uni-freiburg.de}
\affiliation{Mathematical Physics \& Stochastic Dynamics, Institute of
  Physics, University of Freiburg, 79104 Freiburg, Germany}
\affiliation{Mathematical bioPhysics Group, Max Planck Institute for Multidisciplinary Sciences, 37077 G\"ottingen, Germany
}

\date{\today}

\begin{abstract}
Path-wise observables---functionals of stochastic trajectories---are
at the heart of time-average 
statistical mechanics and are central to thermodynamic inequalities such
as uncertainty relations, speed limits, and correlation-bounds. 
They provide a means of thermodynamic
inference in the typical situation, when 
\emph{not} 
all dissipative
degrees of freedom in a system are experimentally accessible. So far, theories focusing on path-wise
observables have been developing in two major directions, diffusion
processes and Markov-jump dynamics, in a virtually disjoint
manner. Moreover, even the respective results for diffusion and jump
dynamics
were derived with a patchwork of different approaches
that are predominantly 
indirect. Stochastic calculus was recently shown
to provide a direct 
approach to path-wise observables of
diffusion processes, while a corresponding framework for jump
dynamics 
remained elusive. In our work we develop, in
an exact parallelism with continuous-space diffusion, a complete stochastic
calculus for path-wise observables of Markov-jump processes. We
formulate a ``Langevin equation'' for jump processes,
define general path-wise observables, and
establish their covariation structure, whereby we fully account for
transients and time-inhomogeneous dynamics. 
We prove
the known kinds of thermodynamic inequalities in their most general
form and
discus saturation conditions. We determine the response of path-wise observables to general (incl.\ thermal)
perturbations \textcolor{black}{ and introduce a corresponding response-function formalism.} We carry out the continuum limit to achieve
the complete unification of diffusion and jump dynamics. \textcolor{black}{In addition, we connect the framework to quantum unraveling and the Belavkin equation for open quantum systems, associating quantum and classical descriptions of thermal systems.}
Our results open new avenues 
in the direction of discrete-state analogs of generative diffusion
models and the learning of stochastic thermodynamics 
from fluctuating
trajectories. 
\end{abstract}

\maketitle


\section{Introduction}

A large variety of systems in and out of equilibrium with nominally
discrete state spaces, or 
with continuous state spaces but well-defined metastable states, can be
accurately and thermodynamically consistently described by Markov-jump
dynamics
\cite{Moro1995,Gaveau_1987,Gaveau1997,HANGGI1982207,SeifertEntropyProd,Seifert_2012,ThreeFacesI,SEIFERT2018176,SeifertStochasticTD,roldan2023martingales,LocalDetailedBalanceAcross,EspositoCoarseGraining,MasterEquationFastCoarseGraining,Schnakenberg,CurrentCharacteristicsZia,Hartich_PRX,Polettini,Aslyamov2024,dieball2024perspectivetimeirreversibilitysystems}. Intense
research over the past few decades led to quite deep understanding of
the (stochastic) thermodynamics
\cite{Gaveau1997,SeifertEntropyProd,Seifert_2012,ThreeFacesI,Seifert_2012,Maes2008,SeifertStochasticTD,Horo_Massi,ZHANG20121,Shiraishi2023,Oren,Vucelja}
and kinetics
\cite{Hiura2021,KineticTUR,Hartich_2019,Lecomte2007,Shiriashi_Saito,Dechant_Ito,HeatingFasterCooling,Oren,Uhl_2019,Artemy,Blom_X}
of Markov
dynamics under so-called local detailed balance
\cite{LocalDetailedBalanceAcross,Seifert_2012,LocalDetailedBalance,Hartich_LDB},
including fluctuation theorems
\cite{Galavotti,Chris,Crooks,Kawai,Maes_1999,SeifertEntropyProd,Kurchan},
speed limits
\cite{Kay_X,Keiji_SL,Falasco_SL,PhysRevLett.127.160601,PhysRevLett.106.250601,Kay_2,Kay_Udo,dieball2024thermodynamic},
thermodynamic uncertainty relations \tb{(TUR)} \cite{TURBiomolecular,Pal2021,Vo_2022,Neri2022,Ueda2023,ImprovingBoundsCorrelations,Hororwitz_2019,TURTimeDependentDriving,kwon2024unifiedframeworkclassicalquantum,DirectTUR,SeifertStochasticTD}, anomalous
relaxation phenomena
\cite{Lapolla,HeatingFasterCooling,AsymThermalRelaxation,Ibanez_2024},
nonequilibrium response relations
\cite{Dechant_2020,Basu_2015,Colangeli_2011,Speck,Ptaszynski2024,Aslyamov2024,HANGGI1982207,zheng2025spatialcorrelationunifiesnonequilibrium,Rotskoff2025},
and recently descriptions within the framework of information
geometry
\cite{Ruppeiner,Horo_Massi,Shiriashi_Saito,Dechant_Ito,Shiriashi_Saito,Ibanez_2024}.

From a practical perspective, state-of-the-art  experiments, in
particular single-particle tracking~\cite{Burov2011PCCP,von_Diezmann_2017} and single-molecule
spectroscopy~\cite{Gladrow2016PRL,Gnesotto2018RPP,Ritort2006JPCM,Greenleaf2007ARBBS,Moffitt2008ARB}
probe stochastic time series. Single-molecule techniques invariably track
low-, predominantly one-dimensional projections of high-dimensional
dynamics (e.g., FRET efficiencies reporting on the instantaneous distance between fluorescent
dyes~\cite{FRET,FRET2,FRET3,FRET4,SM_RNA}, molecular
extensions~\cite{Ritort2006JPCM,Felix2,Greenleaf2007ARBBS,Moffitt2008ARB,Brujic,Saleh,SM_RNA,CalmodulinStigler} or on
the instantaneous distance between plasmonic particles~\cite{Hugel,Vollmar_2024}
attached to the molecule). These projected time series are typically
analyzed by averaging along individual realizations, thus giving rise to
\emph{path-wise observables} (i.e., functionals of stochastic paths)
and leading naturally
 to the field of time-average 
statistical mechanics \cite{Gopich_2006,Gopich_2010,Eli_PRL,Rebenshtok_2008,Burov2011PCCP,hartichGodecPathProb}. 

Currently, the quantity of central interest is typically the thermodynamic
cost---the total entropy production---of nonequilibrium processes that
may be considered as the nonequilibrium ``counterpart'' of free
energy. However, one almost never has experimental access to all 
\emph{dissipative} degrees of freedom and therefore probes only 
coarse \emph{observables} (i.e., projections)  of Markov-jump 
dynamics. Diverse
approaches have meanwhile been developed for thermodynamic inference from coarse
observations
\cite{DieballCoarseGraining,SeifertStochasticTD,SeifertPartiallyAccessibleNetworks,Harunari22,Challenge,Blom2024,Snippets,Puglisi,Baiesi2024,dieball2024perspectivetimeirreversibilitysystems,Tassilo}. 
Thermodynamic inequalities, that is, diverse speed limits \cite{Kay_X,Keiji_SL,Falasco_SL,PhysRevLett.127.160601,PhysRevLett.106.250601,Kay_2,Kay_Udo,dieball2024thermodynamic} and
uncertainty relations
\cite{TURBiomolecular,Pal2021,Vo_2022,Neri2022,Ueda2023,ImprovingBoundsCorrelations,Hororwitz_2019,TURTimeDependentDriving,kwon2024unifiedframeworkclassicalquantum,DirectTUR,SeifertStochasticTD,Rahav_TUR,Plati_23,Plati_2024,Lucente_2025,Puglisi_2025},
in particular, attracted much attention, as they provide a general
means to infer a lower bound on dissipation of the entire system from
a measured observable. These \emph{thermodynamical inequalities} in
various forms (and valid under different conditions) provide important insight into how dissipation bounds
fluctuations
\cite{TURBiomolecular,Pal2021,Vo_2022,Neri2022,Ueda2023,ImprovingBoundsCorrelations,Hororwitz_2019,TURTimeDependentDriving,kwon2024unifiedframeworkclassicalquantum,DirectTUR},
transport \cite{Keiji_SL,Falasco_SL,Leighton2024,dieball2024thermodynamic}, and
correlations
\cite{ThermodynamicCorrelationInequality,Ohga2023,BoundsCorrelationTimes,DieballCorrelationBound}
and, in turn, allow for efficient thermodynamic inference based on
(generally coarse) observables. In
addition to these three major classes of thermodynamic
inequalities, alternative inference methods have been developed as
well, based on the kinetic uncertainty relation \cite{Vo_2022}, statistics of waiting times
\cite{SeifertPartiallyAccessibleNetworks,Skinner2021}, first-passage
times \cite{Neri_2022, Neri2022,Hiura2021, Pal2021}, via the so-called
variance-sum rule \cite{Ritort2024, DiTerlizzi_2024}, from the
statistics of coarsely observed trajectories \cite{Kapustin2024,
  Blom2024,Tassilo}, and others. 

However,
despite the general consensus about their importance and
applicability, there is no unified approach  to deriving
and proving the above results. For example, there are strategies involving the
Cram\'er-Rao inequality \cite{Dechant_2019, TURArbitraryInit, Vo_2022,
  ImprovingBoundsCorrelations, Shiraishi2021}, a variational approach
\cite{BoundsCorrelationTimes}, the scaled cumulant generating function
\cite{TURTimeDependentDriving}, large-deviation theory
\cite{Gingrich2016, Proesmans_2017}, and 
the master fluctuation theorem \cite{Ueda2023}, to name a few.
Because the above methods 
are generally indirect, not much is known about the sharpness and conditions
required to saturate the various inequalities. 

In the case of
diffusion processes, a direct method based on
stochastic calculus was recently introduced 
\cite{DieballCurrentVariance,DirectTUR,dieball2024thermodynamic,CrutchfieldUnderdampedTUR,DieballCorrelationBound,Rotskoff2025},
which allows for a direct and unified approach. While the
mathematical intricacies 
in correctly handling nowhere differentiable
functions seem to still pose a barrier to the proliferation of the
above ideas into the
physics literature, the underlying concept of It\^o calculus is well
established.  

In contrast, Markov-jump processes enjoy a much
greater popularity in physics, and in particular in stochastic
thermodynamics \cite{Seifert_2012,SeifertStochasticTD,SEIFERT2018176,Shiraishi2023, BookPeliti}. Here,
master-equation approaches \cite{Gaveau1996,Gaveau1997,Lecomte2007}
(and corresponding ``Quantum Hamiltonian'' analogy
\cite{Harris_2007}), Feynman-Kac ``tilting''
\cite{Barato_2015,hartichGodecPathProb}, as well as approaches relying
on path measures \cite{Maes_2008, Maes2008,Seifert_2012,ZHANG20121}
are widely adopted in the community. Nevertheless, as far as proving
thermodynamic inequalities is concerned, these methods 
are \emph{not} direct and do \emph{not} provide insight about
the sharpness and saturation conditions. A stochastic-calculus
approach in the spirit of
\cite{DieballCurrentVariance,DirectTUR,dieball2024thermodynamic,CrutchfieldUnderdampedTUR,DieballCorrelationBound,Rotskoff2025}
is therefore desired  to provide a
unification of results for diffusion and Markov-jump dynamics, and
specifically of different approaches to Markov-jump dynamics. 

The first steps in this direction, instigated in the Appendix of
\cite{DirectTUR}, have indeed already been made in the context of TURs
\cite{kwon2024unifiedframeworkclassicalquantum} and linear response \cite{zheng2025spatialcorrelationunifiesnonequilibrium}. However,
the full battery of results,
which is required for a true unification and for the
approach to be able compete with existing methods, remains to be developed. 

Here, we carry out the complete program. We develop, in exact
parallelism with the continuous diffusion counterpart, the stochastic
equations of motions --- a Langevin equation for Markov-jump processes
with a central noise-time-Correlation Lemma --- and define general
pathwise observables (i.e., densities and currents). The results are
throughout presented in direct analogy with the
continuous-space framework. We allow for time-inhomogeneous dynamics
(e.g., time-dependent driving). Equivalently to diffusions \cite{DieballCoarseGraining},
Stratonovich increments emerge that govern current-type observables
and allow for a direct proof of the complete covariations (incl.\ in transients)
of general density- and current observables that takes the form of
(generalized) Green-Kubo relations. We use the developed stochastic
calculus to prove \emph{directly} the known thermodynamic inequalities (TUR,
correlation TUR, transport bounds, bounds on correlations etc.) in their most
general form, assess their sharpness, and establish the respective
saturation conditions. Moreover, fully within the continuous-space stochastic-calculus
approach \cite{Rotskoff2025} we derive the response
of an observable to a general (incl.\ thermal) perturbations for
Markov-jump dynamics. Finally,
we carry out the continuum limit to achieve the unification of
diffusion and jump dynamics. We conclude with an outlook on open
questions and future directions.  

The manuscript is structured as follows. In Sec.~\ref{summary} we first
provide a summary of our main results. Next, in Sec.~\ref{setup} we
develop the stochastic calculus for jump processes, formalize general
pathwise observables and establish their complete covariation
structure. In Sec.~\ref{applications} we use the stochastic calculus
to prove the (quite complete) list of thermodynamic inequalities in
their most general form and apply them to thermodynamic inference on a
selection of biophysical model systems. Following this, we show in
Sec.~\ref{sec:Perturbation} how the stochastic calculus can be applied
to derive results for the response to general perturbations for
discrete-state systems, \textcolor{black}{establish the corresponding response-function formalism, and show how this relates to} the
fluctuation-dissipation theorem in equilibrium. In
Sec.~\ref{sec:ContinuousLimit} we take the continuum limit to show
the equivalence of stochastic calculi for continuous- and
discrete-state systems and in Sec.~\ref{t-inhomo} we extend the results to
time-inhomogeneous dynamics. \textcolor{black}{Subsequently, in Sec.~\ref{Quantum}, we discuss how the classical stochastic-calculus approach is connected to quantum unraveling and the (discrete) Belavkin equation and state its classical analogue. } 
We conclude in Sec.~\ref{Outlook} with an
outlook. The mathematical proofs of the noise-time correlation Lemma
and increment correlations, and other technical details are given in the Appendix~\ref{Proofs}.

\section{Summary of Main Results}\label{summary}

\begin{figure*}
    \centering
    \includegraphics[width=\linewidth]{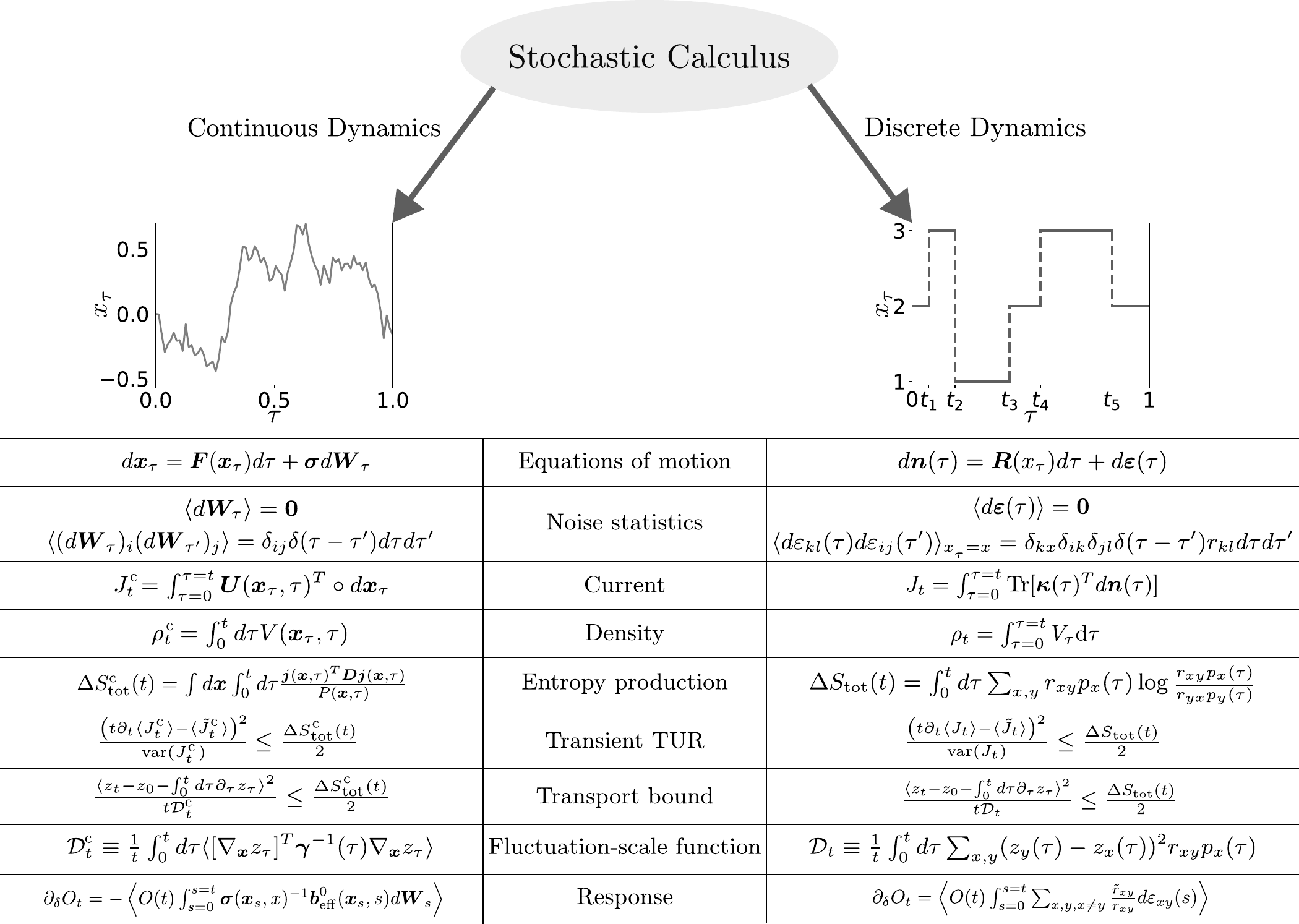}
    \caption{Comparison of key quantities in continuous and discrete
      space, namely the equations of motion, noise statistics,
      fundamental path observables, thermodynamic bounds,
      fluctuation-scale function, and response to perturbations
      (incl.\ temperature changes).}
    \label{fig:ComparisonTable}
\end{figure*}

We aim to bridge the gap between continuous-space dynamics and Markov
jump processes (MJP) using stochastic calculus. Throughout we will
develop the results for MJP in analogy with those of diffusions, all
results will therefore be shown in parallel to make the analogy
as clear and natural as possible. The central results of the stochastic calculus for
continuous-space diffusion and Markov-jump processes is highlighted in Fig.~\ref{fig:ComparisonTable}.

In the context of continuous-space dynamics we will focus on
overdamped Langevin dynamics in $d$ dimensions governed by \cite{gardiner2004handbook,Seifert_2012,DirectTUR, Dieball_2022_CorrelationsFluctuations}
\begin{align}
    \rmd\f{x}_\tau = \f F(\f{x}_\tau)\rmd\tau + \f{\sigma}\rmd\f{W}_\tau,
    \label{eq:SummaryLangevinEq}
\end{align}
for the $d$-dimensional position $\f x_\tau$ in a drift field
$\f{F}(\f{x}_\tau)$ subject to thermal noise $\f{\sigma}\rmd
\f{W}_\tau$, where $\rmd\f{W}_\tau$ are increments of the
$d$-dimensional Wiener process (i.e.,  $d$-dimensional ``white noise'')\textcolor{black}{, with
$\langle \rmd\f{W}_\tau\rangle = 0$ and $\langle
(\rmd\f{W}_\tau)_i(\rmd\f{W}_{\tau'})_j\rangle =
\delta_{ij}\delta(\tau-\tau')\rmd\tau\rmd\tau'$, and $\f \sigma$ is the noise amplitude assumed to have full rank (nonzero determinant)}. In
Sec.~\tb{\ref{t-inhomo}} we will also allow for time-inhomogeneous (temporally driven)
dynamics, while for clarity we will develop the results within
the time-homogeneous setting. 

We will prove in Sec.~\ref{setup} that trajectories \tb{$x_\tau$} of MJP follow the
matrix stochastic differential equation (i.e., a ``discrete-space Langevin equation'')
\begin{align}
    \rmd \f{n}(\tau) = \f{R}(x_\tau, \tau)\rmd \tau + \rmd \f{\varepsilon}(\tau),
    \label{eq:SummaryEoM}
\end{align}
where the displacements have elements $(\rmd
\f{n}(\tau))_{xy}=\rmd n_{xy}(\tau)$ describing jumps from state $x$
to $y$ in $[\tau, \tau+\rmd \tau]$. The matrix $(\f{R}(x_\tau,
\tau))_{xy}\rmd \tau=\textcolor{black}{\delta_{x_\tau x}}r_{xy}\rmd \tau$ encodes the
expected (average) drift in $[\tau, \tau+\rmd \tau]$
with time-independent transition rate $r_{xy}$ (following from the
corresponding master equation with generator
$(\f{L})_{xy}=-\delta_{xy}\sum_{y\ne x}r_{xy}+(1-\delta_{xy})r_{yx}$), and 
the noise is encoded
in the time-inhomogeneous shifted Poisson process $(\rmd
\f{\varepsilon}(\tau))_{xy} = \rmd \varepsilon_{xy}(\tau)$
\cite{roldan2023martingales, Rogers_Williams_2000_martingales,
  DirectTUR, kwon2024unifiedframeworkclassicalquantum}. This noise has
zero mean $\langle \rmd \varepsilon_{xy}(\tau)\rangle=0$ and the
``white noise'' property $\langle \rmd \varepsilon_{xy}(\tau)\rmd
\varepsilon_{ij}(\tau')\rangle_{x_\tau=k} =
\delta_{ix}\delta_{jy}\delta_{ik}\delta(\tau-\tau')r_{ij}\rmd \tau
\rmd \tau'$, where the expectation is conditioned on $x_\tau = k$. \textcolor{black}{The Kronecker $\delta$ show that the noise increments on different edges are independent and only diagonal elements survive, 
e.g., $\langle \rmd \varepsilon_{ij}(\tau)\rmd
\varepsilon_{ij}(\tau')\rangle_{x_\tau=k} =
\delta_{ik}\delta(\tau-\tau')r_{ij}\rmd \tau
\rmd \tau'$. Moreover, since $|\rmd \varepsilon_{ij}|\sim\sqrt{\rmd\tau}$, higher moments scale as $|\rmd \varepsilon_{ij}|^n\sim(\rmd\tau)^{n/2}$, so that all higher moments and cumulants with $n\geq 3$ vanish for small $\rmd\tau$. Note that the Langevin Eq.~\eqref{eq:SummaryEoM} has already been introduced in \cite{kwon2024unifiedframeworkclassicalquantum}.}

We prove the central time-noise correlation lemma with $\rmd \tau_i(\tau) = \textcolor{black}{\delta_{x_\tau i}}\rmd \tau$,
\begin{align}
    &\frac{\langle
    \rmd\varepsilon_{kl}(\tau)\rmd\tau_i(\tau')\rangle_{x_\tau =
      x}}{\rmd\tau\rmd\tau'}\\&
  = \delta_{xk}\textcolor{black}{\mathbb{1}}_{\tau<\tau'}\left[P(i, \tau'|l, \tau) - P(i,
    \tau'| k, \tau)\right]r_{kl} 
+ \mathcal{O}(\rmd\tau)\nonumber,
    \label{eq:SummaryNoiseTimeCorrelationLemma}
\end{align}
which gives access to nontrivial correlations between differentials in Eq.~\eqref{eq:SummaryEoM}
\begin{widetext}
\begin{equation}
    \begin{aligned}[b]
        \langle \rmd n_{xy}(\tau')\rmd\tau_i(\tau)\rangle^{x_{\tau'}=x}_{x_{\tau}=i} &= r_{xy}P(x, \tau' |i,\tau)p_i(\tau)\rmd \tau \rmd\tau',\\
    \langle \rmd n_{xy}(\tau)\rmd\tau_i(\tau')\rangle_{x_\tau=x}^{x_{\tau'}=i} 
    &= r_{xy}P(i, \tau',|y, \tau)p_x(\tau)\rmd \tau \rmd\tau',\\
    \langle \rmd n_{xy}(\tau)\rmd n_{ij}(\tau')\rangle_{x_\tau=x}^{x_{\tau'}=i} 
    &= r_{ij}r_{xy}P(i, \tau', |y, \tau)p_x(\tau)\rmd \tau \rmd\tau',
    \label{eq:SummaryIncrementCorrelation}
    \end{aligned}
\end{equation} 
\end{widetext}
where $P(x, \tau'|i,\tau)$ is the transition probability  (i.e., the
``propagator'') from state $i$ at $\tau$ \textcolor{black}{to state $x$ at $\tau'\geq \tau$} and $p_i(\tau)$ is the probability to find the system in state
$i$ at time $\tau$. Note that the 
averages are ``pinned'',
e.g., $\langle
\cdot\rangle^{x_{\tau'}=x}_{x_{\tau}=i}$ is the average conditioned on
${x_{\tau}=i}$ and ${x_{\tau'}=x}$ at times $\tau\leq \tau'$. The above
correlations were shown
in~\cite{kwon2024unifiedframeworkclassicalquantum} using a different
approach.

These correlations are required for the covariation structure of
observables, such as currents and densities. Specifically,
time-integrated currents are defined as \cite{DirectTUR}
\begin{align}
    J_t^\mathrm{c} &= \StoInt \f{U}(\f{x}_\tau, \tau)^T\circ\rmd \f{x}_\tau,\tag{\textcolor{black}{\rm Stratonovich}}
    \\
    J_t &= \StoInt \mathrm{Tr}[\f{\kappa}(\tau)^T\rmd\f{n}(\tau)],
    \label{eq:SummaryCurrent}
\end{align}
for continuous-space and MJP, respectively, where the former is a
Stratonovich integral over some vector-valued $\f{U}(\f{x}_\tau,
\tau)$ \textcolor{black}{and the $T$ denotes the transpose. Recall that the Stratonovich integral is a Riemann–Stieltjes-type integral defined as the limit of 
of functions evaluated at midpoints, e.g., $\f f(\f x_\tau)^T\circ\rmd \f x_\tau = \f f(\frac{\f x_{\tau+\rmd\tau} + \f x_\tau}{2})^T(\f x_{\tau+\rmd\tau}-\f x_\tau)$.  To give an intuition for $\f U$, consider the example $[\f U(\f x_\tau,\tau)]_i = \textcolor{black}{\mathbb{1}}_{\f x_\tau\in\{-h/2, h/2\}^d}$ the indicator valued vector being non-zero if $\f x_\tau$ is in the hypercube with side length $h$, then the current ``counts up'' the displacement of the trajectory $\{\f x_\tau\}_{0\leq\tau\leq t}$ in that hypercube. Note the superscript c throughout this manuscript refers to continuous space quantities and we write superscript (c) as a placeholder when the corresponding relation is valid in both continuous and discrete space systems}. For MJP the transitions are weighted by a time-dependent
antisymmetric transition weight $\f{\kappa}(\tau) = -\f{\kappa}(\tau)$ \textcolor{black}{which ensures a sign change in the current when reversing time \cite{Seifert_2012}}. Hence,
we identify \emph{Stratonovich integrals for MJP} as integrals over traces of
a two-point function and the jump increments.
The above currents naturally decompose 
into a stochastic  and ``usual'' integral \cite{DirectTUR}, $J^{(\mathrm{c})}_t =
J_t^\mathrm{(c), I} + J_t^\mathrm{(c), II}$, where the stochastic
part is 
\begin{align}
    J_t^\mathrm{c, I} &= \StoInt \f{U}(\f{x}_\tau, \tau)^T\f{\sigma}\rmd \f{W}_\tau\tag{\textcolor{black}{\rm It\^o}}\\
    J_t^\mathrm{I} &= \StoInt \mathrm{Tr}[\f{\kappa}(\tau)^T\rmd\f{\varepsilon}(\tau)],
    \label{eq:SummaryNoiseCurrent}
\end{align}
respectively, while the \textcolor{black}{``}usual'' part reads $J_t^\mathrm{(c), II}=\int_0^t\mathcal{U}_\tau^{(c)}\rmd\tau$ 
where $\mathcal{U}_\tau^c\equiv \f{U}(\f{x}_\tau,
\tau)^T \f{F}(\f{x}_\tau) + \nabla \cdot \left[\f{D}\f{U}(\f{x}_\tau,
  \tau)\right]$\textcolor{black}{, $\f D = \f \sigma \f \sigma^T/2$ is the diffusion matrix,} and
$\mathcal{U}_\tau\equiv\mathrm{Tr}[\f{\kappa}(\tau)^T\f{R}(x_\tau)]$.
Thus, we identify in the second line of
Eq.~\eqref{eq:SummaryNoiseCurrent} the Ito-type integral for
observables of MJP by analogy to diffusions. 

Similarly, time-integrated densities are defined as \cite{DirectTUR}
\begin{align}
    \rho_t^\mathrm{c} &= \int_0^t \rmd \tau V(\f{x}_\tau, \tau),\nonumber\\
    \rho_t &= \StoInt \textcolor{black}{V_\tau\rmd\tau},
    \label{eq:SummaryDensity}
\end{align}
for continuous and discrete dynamics, respectively, 
where $V(\f{x}_\tau,\tau)$ and \textcolor{black}{$V_\tau\equiv V(x_\tau,\tau)=\sum_k \delta_{x_\tau k}V_k(\tau)$} are scalar-valued functions.  

(Co)variances of stationary densities and currents can be expressed in terms of integral operators \cite{Dieball_2022_CorrelationsFluctuations, DieballCoarseGraining}
\begin{align}
    \hat{\mathcal{I}}^{t, \mathrm{c}}_{M, N}[\cdot] &= \int_0^t\rmd \tau \int_{\tau}^t\rmd \tau' \int \rmd \f{z}\int \rmd \f{z}'M(\f{z})^T[\cdot]N(\f{z}'),\nonumber\\
    \hat{{I}}^t_{\textcolor{black}{G}^\alpha, \textcolor{black}{G}^\beta}[\cdot] &= \int_0^t\rmd\tau\int_\tau^t\rmd\tau' \sum_{x,y}\textcolor{black}{G}_{xy}^\alpha\sum_{i,j}\textcolor{black}{G}^\beta_{ij}[\cdot],
    \label{eq:SummaryIntegrationOperator}
\end{align}
for continuous and discrete dynamics respectively. The operator
in the first line of Eq~\eqref{eq:SummaryIntegrationOperator} is
already known \cite{DieballCoarseGraining, DieballCurrentVariance}. For
$X_t^\mathrm{(c)} \in\{J_{t, \textcolor{black}{k}}^{\mathrm{(c)}}, \rho_{t, \textcolor{black}{k}}^{\mathrm{(c)}}\}$ and $Y_t^\mathrm{(c)} \in\{J_{t, \textcolor{black}{l}}^{\mathrm{(c)}},
\rho_{t, \textcolor{black}{l}}^{\mathrm{(c)}}\}$, $k,l\in\{1,2\}$, we find 
\begin{align}
    \mathrm{cov}(X_t^\mathrm{c}, Y_t^\mathrm{c}) &= \hat{\mathcal{I}}^{t, \mathrm{c}}_{X, Y}[{\Xi}_{m}^{\f{z}\f{z}'}] - \langle X_t^\mathrm{c}\rangle \langle Y_t^\mathrm{c}\rangle,\nonumber
    \\
    \mathrm{cov}(X_t, Y_t) &=\hat{{I}}^t_{\textcolor{black}{G}^X, \textcolor{black}{G}^Y}[{\Xi}^m_{ijxy}]-
    \langle X_t\rangle \langle Y_t\rangle,
\label{covariation}    
\end{align}
where $X_t^\mathrm{(c)}, Y^\mathrm{(c)}_t$ are the appropriate pathwise observables
and $\textcolor{black}{G}^X, \textcolor{black}{G}^Y$ functions that define them
(\textcolor{black}{ e.g.,} $\boldsymbol{\kappa}$ in the case of \textcolor{black}{currents for}
MJP). Moreover,
$m\in\{1,2,3\}$ indicates density-density ($m=1$), current-density ($m=2$), and current-current ($m=3$) covariances. For the density covariance ($m=1$), $X_t^\mathrm{(c)} = \rho_{t, \textcolor{black}{k}}^{\mathrm{(c)}}$ and $Y_t^\mathrm{(c)} = \rho_{t, \textcolor{black}{l}}^{\mathrm{(c)}}$, we get
\begin{align}
    {\Xi}_{1}^{\f{z}\f{z}'} &= \langle \textcolor{black}{\mathbb{1}}\rangle_{\f{x}_{\tau}=\f{z}}^{\f{x}_{\tau'}=\f{z}'} + \langle \textcolor{black}{\mathbb{1}}\rangle_{\f{x}_{\tau}=\f{z}'}^{\f{x}_{\tau'}=\f{z}},\\
    {\Xi}^1_{ijxy}&= \frac{\langle \rmd \tau_x(\tau)\rmd\tau_i(\tau')\rangle_{x_\tau=x}^{x_{\tau'}=i}}{\rmd\tau\rmd\tau'}+\frac{\langle \rmd \tau_i(\tau)\rmd\tau_x(\tau')\rangle^{x_{\tau'}=x}_{x_{\tau}=i}}{\rmd\tau\rmd\tau'},\nonumber
\end{align}
for current-density covariances ($m=2$) we recover
\begin{align}
    {\Xi}_2^{\f{z}\f{z}'} &= \frac{\langle \circ \rmd \f{x}_{\tau}\rangle_{\f{x}_{\tau}=\f{z}}^{\f{x}_{\tau'}=\f{z}'}}{\rmd \tau} + \frac{\langle \circ \rmd \f{x}_{\tau'}\rangle_{\f{x}_{\tau}=\f{z}'}^{\f{x}_{\tau'}=\f{z}}}{\rmd \tau'},\\
    {\Xi}_{ijxy}^2 &=\frac{\langle \rmd n_{xy}(\tau)\rmd\tau_i(\tau')\rangle_{x_\tau=x}^{x_{\tau'}=i}}{\rmd\tau\rmd\tau'} + \frac{\langle \rmd n_{xy}(\tau')\rmd\tau_i(\tau)\rangle^{x_{\tau'}=x}_{x_{\tau}=i}}{\rmd\tau\rmd\tau'},\nonumber
\end{align}
and for the current-current covariance ($m=3$) we find
\begin{align}
    {\Xi}_3^{\f{z}\f{z}'} &= \frac{\langle \circ \rmd \f{x}_{\tau}\circ \rmd \f{x}_{\tau'}^T\rangle_{\f{x}_{\tau}=\f{z}}^{\f{x}_{\tau'}=\f{z}'}}{\rmd \tau\rmd \tau'} + \frac{\langle \circ \rmd \f{x}_{\tau'}\circ \rmd \f{x}_{\tau}^T\rangle_{\f{x}_{\tau}=\f{z}'}^{\f{x}_{\tau'}=\f{z}}}{\rmd \tau\rmd \tau'}\\
    {\Xi}_{ijxy}^3 &=\frac{\langle \rmd n_{xy}(\tau)\rmd n_{ij}(\tau')\rangle_{x_\tau=x}^{x_{\tau'}=i}}{\rmd\tau\rmd\tau'} + \frac{\langle \rmd n_{xy}(\tau')\rmd n_{ij}(\tau)\rangle^{x_{\tau'}=x}_{x_{\tau}=i}}{\rmd\tau\rmd\tau'}.\nonumber
\end{align}
These above correlations can all be evaluated explicitly. For MJP we find, using $\int_0^t\rmd \tau\int_0^{\tau}\rmd \tau'\to\int_0^t\rmd t'( t-t')$ \cite{DieballCoarseGraining}, that
\begin{widetext}
\begin{align}
    \mathrm{cov}({\rho}_{t, \textcolor{black}{k}}, {\rho}_{t, \textcolor{black}{l}}) =& \hat{{I}}^{t}_{V^k\delta, V^l\delta}\left[P_\mathrm{s}(x,\tau;i,\tau') +P_\mathrm{s}(x,\tau';i,\tau) - p_x^\mathrm{s}p_i^\mathrm{s}\right]\nonumber
    \\
    \mathrm{cov}({J}_{t, \textcolor{black}{k}}, {\rho}_{t, \textcolor{black}{l}}) =& \hat{{I}}^{t}_{\kappa^k, V^l\delta}\left[\hat{j}_{xy}P(x, t'| i)p_{i}^\mathrm{s} + \hat{j}_{yx}^\ddag P(i, t'| y)p_y^\mathrm{s}- J^\mathrm{s}_{xy}p_{i}^\mathrm{s}\right],\nonumber\\
    \mathrm{cov}(J_{t, \textcolor{black}{k}}, {J}_{t, \textcolor{black}{l}}) =&  t\sum_{x, y}\kappa_{xy}^k \kappa_{xy}^l r_{xy}p_x^\mathrm{s} + \hat{{I}}^{t}_{\kappa^k, \kappa^l}\left[\hat{j}_{xy}\hat{j}_{ji}^\ddag P(x, t'| j)p_{j}^\mathrm{s} + \hat{j}_{ij}\hat{j}_{yx}^\ddag P(i, t'| y)p_y^\mathrm{s} - J^\mathrm{s}_{xy}J^\mathrm{s}_{ij}\right],
    \label{eq:SummaryCovariance}
\end{align}
\end{widetext}
where
$V^{\alpha/\beta}\delta$ is a shorthand notation for
$V^{\alpha/\beta}_i\delta_{ij}$, $J^\mathrm{s}_{xy}$ is the local steady-state current
between $x$ and $y$, and $\hat{j}_{xy}^\ddag$ is the
so-called dual-reverse current operator (reversed steady-state current)
\cite{Dieball_2022_CorrelationsFluctuations, DieballCoarseGraining,
  DieballCurrentVariance} which for MJP reads
$\hat{j}_{xy} = r_{xy}\to\hat{j}_{xy}^\ddag =
r_{yx}p_y^\mathrm{s}/p_x^\mathrm{s}$ \cite{Seifert_2012}. \textcolor{black}{In Eq.~\eqref{eq:SummaryCovariance}, the joint stationary probabilities for state $x$ at $\tau$ and $i$ at $\tau'$ are denoted $P_\mathrm{s}(x,\tau;i,\tau')$, while the marginal stationary distributions to be in a state $x$ is $p^\mathrm{s}_x$. Additionally, we use the fact that the propagator for time-homogeneous systems is time-translation invariant $P(i, \tau'|x, \tau) = P(i, \tau'-\tau|x)$.}
The results for continuous-space
diffusion are given in
Refs.~\cite{DieballCoarseGraining,DieballCurrentVariance}.
For stationary initial conditions and with time-independent state
function and transition weights, Eqs.~\eqref{eq:SummaryCovariance} have the
form of generalized Green-Kubo-like relations, in the sense that they relate
(co)variances of observables to time integrals of (generalized)
correlation functions of density- and current-observables (see
Refs.~\cite{Dieball_2022_CorrelationsFluctuations, DieballCoarseGraining}
for a discussion of diffusion processes). Notably,
Eqs.~\eqref{eq:SummaryCovariance} have a striking similarity to the
continuous space expressions \cite{DieballCoarseGraining,
  Dieball_2022_CorrelationsFluctuations}. Later we show how to to
generalize Eqs.~\eqref{eq:SummaryCovariance} to include time-dependent transition weights and state functions, as well as transient dynamics.

\subsection*{Thermodynamic inequalities}

Since we now have a stochastic calculus for MJP at our disposal
(incl.\ fluctuation and correlation relations), we
can employ the same strategy as in \cite{DirectTUR} and derive
thermodynamic inequalities from the Cauchy-Schwarz inequality by
appropriately choosing an auxiliary functional. 
The first family of inequalities we prove in Sec.~\ref{applications}
are generalized (incl.\ transient) correlation TURs \cite{DirectTUR, ImprovingBoundsCorrelations}
\begin{align}
    &\frac{\left(t\partial_t\langle J^\mathrm{(c)}_t\rangle -\langle \tilde{J}_t^\mathrm{(c)}\rangle - c(t)\left\{\left[t\partial_t- 1\right]\langle \rho_t^\mathrm{(c)}\rangle -\langle\tilde{\rho}_t^\mathrm{(c)}\rangle\right\}\right)^2}{\mathrm{var}\left(J_t^\mathrm{(c)} - c(t)\rho_t^\mathrm{(c)}\right)}\nonumber\\&\leq\frac{\Delta S_\mathrm{tot}^\mathrm{(c)}(t)}{2},\label{eq:SummarycTUR}
\end{align}
where $c(t):\textcolor{black}{\mathbb{R}}\to\textcolor{black}{\mathbb{R}}$ is a deterministic weight
function for the density and $\Delta S_\mathrm{tot}^\mathrm{(c)}$ is
the discrete (continuous) space entropy production \textcolor{black}{ \cite{SeifertEntropyProd}
\begin{align}
    \Delta S_\mathrm{tot}^\mathrm{c}(t) &=\int_0^t\rmd \tau\int\rmd \f x\frac{\f j(\f x,\tau)^T\f D^{-1}\f j(\f x, \tau)}{P(\f x, \tau)},\\
    \Delta S_\mathrm{tot}(t) &= \int_0^t\rmd \tau\sum_{i,j}r_{ij}p_i(\tau)\log\frac{r_{ij}p_i(\tau)}{r_{ji}p_j(\tau)}.
\end{align}
Here, $\f j(\f x, \tau) = (\f F(\f x) - \f D\nabla)P(\f x, \tau)$ is the probability current entering the Fokker Planck equation. }In analogy
with  \cite{DirectTUR} we introduced the discrete space
modified time-integrated density and current $\tilde{\rho}_t = \StoInt
\tau\partial_\tau\textcolor{black}{ V_\tau\rmd {\tau}}$ and
$\tilde{J}_t=\StoInt\tau\mathrm{Tr}\left[(\partial_\tau\f\kappa(\tau))^T\rmd\f   n(\tau)\right]$ that account for the dependence of \textcolor{black}{$V_i(\tau)$} and $\f
\kappa(\tau)$ on $\tau$. \textcolor{black}{To compare, the continuous space modified current and density read $\tilde{J}_t^\mathrm{c} = \int_{\tau=0}^{\tau=t}\tau\partial_\tau \f U(\f x_\tau, \tau)^T\circ\rmd \f x_\tau$ and $\tilde{\rho}_t^\mathrm{c} = \int_{0}^{t}\tau\partial_\tau V(\f x_\tau, \tau)\rmd \tau$, respectively. The transient correlation TUR for MJP in Eq.~\eqref{eq:SummarycTUR} has not yet been proven in this generality since previous results were restricted to steady states \cite{ImprovingBoundsCorrelations} or did not consider correlations \cite{Koyuk2019}.}

Setting $c(t)=0$ we recover the transient TUR. The CTUR (in contrast
to the TUR) allows to turn the Cauchy-Schwarz inequality into an
equality for all
times given a
suitable choice of  \textcolor{black}{$ V_\tau$} and $\f \kappa(\tau)$. This choice,
however, requires explicit and detailed information about the system
and is therefore not applicable in most realistic applications. We can
nevertheless optimize Eq.~\eqref{eq:SummarycTUR} with respect to
$c(t)$ for given \textcolor{black}{$V_\tau$} and $\f \kappa(\tau)$  to find
\begin{align}
    c^*(t) &= \frac{a(t)\mathrm{cov}({\rho}_t, {J}_t) - b(t)\mathrm{var}({J}_t)}{a(t)\mathrm{var}({\rho}_t) - b(t)\mathrm{cov}({\rho}_t, {J}_t)}\,,
    \label{eq:Summaryoptimal_c}
\end{align}
where $a(t)=t\partial_t\langle J^\mathrm{(c)}_t\rangle -\langle
\tilde{J}_t^\mathrm{(c)}\rangle$ and $b(t)=\left[t\partial_t-
  1\right]\langle \rho_t^\mathrm{(c)}\rangle
-\langle\tilde{\rho}_t^\mathrm{(c)}\rangle$. Note that this optimal
CTUR contains only operationally accessible quantities and affords an
improved thermodynamic inference compared to using currents or densities
alone. Interestingly, in contrast to the steady-state result~\cite{ImprovingBoundsCorrelations}, choosing an arbitrary $c(t)$ may even deteriorate
the inference compared to the bare current or density TURs.  

The second thermodynamic inequality we prove is the thermodynamic
bound on the transport of any scalar observable $z_{\textcolor{black}{\tau}}\equiv z(x_\tau,\tau)$
\cite{dieball2024thermodynamic}
\begin{align}
    \frac{\langle  z_{\textcolor{black}{t}} - z_{\textcolor{black}{0}} - \int_0^t\rmd \tau\partial_\tau
      z_{\textcolor{black}{\tau}}\rangle^2}{t\mathcal{D}^\mathrm{(c)}_t}\leq \frac{\Delta
      S_\mathrm{tot}^\mathrm{(c)}}{2}\,,
\label{TB_1}    
\end{align}
\textcolor{black}{
where in continuous space it
was shown \cite{BoundsCorrelationTimes,dieball2024thermodynamic} that
\begin{align}
    \mathcal{D}_t^\mathrm{c} = \frac{1}{t}\int_0^t\rmd \tau\left\langle\left[\nabla_{\f x} z_\tau\right]^T\f\gamma^{-1}(\tau)\nabla_{\f x}z_\tau\right\rangle,
\end{align}
while for MJP we find 
\begin{align}
    \mathcal{D}_t\equiv\frac{1}{t}\int_0^t\rmd \tau \sum_{x,y}[z_y(\tau) - z_x(\tau)]^2r_{xy}p_x(\tau).\label{fluctuation scale MJP}
\end{align}
Here, $\mathcal{D}^\mathrm{(c)}_t$ is the so-called
\emph{fluctuation-scale function} accounting for how much the
observation stretches microscopic coordinates. For example, note that stretching $z_\tau\mapsto az_\tau$ for some constant $a\in\mathbb R$ does not alter Eq.~\eqref{TB_1} as the stretch in the nominator is canceled by the change in $\mathcal{D}^\mathrm{(c)}_t$. In Eq.~\eqref{fluctuation scale MJP},}
\textcolor{black}{$z_x(\tau)$ is the specific value of the function $z_\tau$ if $x_\tau=x$, i.e., $z_\tau\equiv z(x_\tau,\tau)=\sum_k \delta_{x_\tau k}z_k(\tau)$.}
The fluctuation-scale function for both, continuous and discrete space
dynamics, is operationally accessible by considering increments $\rmd
z(\tau) = z(x_{\tau+\rmd\tau}, \tau+\rmd\tau) - z(x_\tau, \tau)$ along
a trajectory and recognizing (as in continuous space  \cite{BoundsCorrelationTimes,dieball2024thermodynamic}) that
\begin{align}
    \mathcal{D}_t^\mathrm{(c)} = \frac{1}{t}\int_{\tau=0}^{\tau=t}\mathrm{var}(\rmd z(\tau))\,.
\end{align}
Contrary to the CTUR, the transport bound cannot be saturated in
general. We provide a counterexample to illustrate how an attempt to
saturate Eq.~\eqref{eq:Summaryoptimal_c} in general fails.

The last family of thermodynamic bounds we prove are thermodynamic
bounds on observable correlations (or, in short, ``correlation bounds'') as a
generalization of the bounds in Ref.~\cite{BoundsCorrelationTimes}. It
involves correlations of time-independent observables $V_\tau \equiv V(x_\tau) = \sum_k \textcolor{black}{\delta_{x_\tau k}}V_k$ and $
z_\tau \equiv z(x_\tau)=\sum_k \textcolor{black}{\delta_{x_\tau k}}z_k$ and their time-average along a trajectory, e.g.,
$\overline{V}_t = t^{-1}\int_0^tV_\tau\rmd\tau$, resulting in
a lower bound of the integrated entropy production rate
$\Dot{S}_\mathrm{tot}(\tau)$ weighted by a ``pseudo variance''
$\mathrm{pvar}_\mathrm{ps}^F(z_\tau) =
\sum_{x,y}p_{xy}^\mathrm{ps}(\tau)(z_x+z_y - 2F(\tau))^{\textcolor{black}{2}}$
\cite{DieballCorrelationBound} 
\begin{widetext}
    \begin{align}
        \frac{\left[2\mathrm{cov}({V}_t, \overline{{z}}_t) - \frac{2}{t}\int_0^t\rmd \tau \mathrm{cov}(V_\tau, z_\tau)\right]^2}{t\mathrm{var}(\overline{V}_t)} - \mathcal{D}_t + \frac{2}{t}\left[\mathrm{var}(z_t) - \mathrm{var}(z_0)\right]\leq \frac{1}{t}\int_0^t\rmd\tau \Dot{S}_\mathrm{tot}(\tau)\mathrm{pvar}_\mathrm{ps}^F(z_\tau),
        \label{eq:SummaryTransientCB}
    \end{align}
\end{widetext}
where $F(\tau)$ may be any time-(in)dependent function. The tilted two-point probability\textcolor{black}{, 
\begin{align}
    p_{xy}^\mathrm{ps}(\tau) = \frac{p_x(\tau)r_{xy}Z_{xy}(\tau)^2}{\Sigma_\mathrm{ps}},\label{SummaryTwoPointProb}
\end{align} where $Z_{xy}(s)=[r_{xy}p_x(\tau)-r_{yx}p_y(\tau)]/[r_{xy}p_x(\tau)+r_{yx}p_y(\tau)]$ and $\Sigma_\mathrm{ps}=\sum_{x,y}p_x(\tau)r_{xy}Z_{xy}(\tau)^2$,}
requires
knowledge about transition rates, hence the pseudo variance is
\emph{not} operationally accessible. However, it can be controlled by choosing
bounded observables. \textcolor{black}{From the expression Eq.~\eqref{SummaryTwoPointProb}, we can see that $p_{xy}^\mathrm{ps}(\tau)$ tilts the probability weight onto dissipating transitions. Hence, the pseudo-variance projects the observable onto dissipating transitions. Further, we emphasize that the pseudo-variance is a mathematical construct, which needs to be bounded further, hence the physical interpretation is not relevant. }In NESS, Eq.~\eqref{eq:SummaryTransientCB} reads
\begin{align}
    &\frac{4\left[\mathrm{cov}(V_t, \overline{{z}}_t) - \mathrm{cov}_\mathrm{s}({V}, {z})\right]^2}{t\mathrm{var}(\overline{V}_t)} - \mathcal{D}_t \nonumber \\
    &\leq \frac{\Delta S_\mathrm{tot}(t)}{2t}\mathrm{pvar}_\mathrm{ps}^F({z})\,.
    \label{eq:SummaryStationaryCB}
\end{align}
In Sec.~\ref{applications} we provide an example to showcase how the correlation
bound can be applied to systems where both, the CTUR and transport
bound, fail to infer a non-zero entropy production.

\subsection*{Time Inhomogeneous Dynamics}

The generalization of the above results to systems exposed to
time-dependent driving with constant protocol speed $v$ (altering the
generator as $\f{L}\to\f{L}(v\tau)$) is in most cases
straightforward. Indeed, the covariation results~\eqref{covariation}, the
CTUR~\eqref{eq:SummarycTUR}, and the transport
bound~\eqref{TB_1} generalize immediately. While the results for the
covariation and the 
transport bound remain unaltered (up to the obvious change in the
expectation operation $\langle\cdot\rangle$), the CTUR becomes altered with
 the differential operator $\hat{\Lambda} \equiv t\partial_t -
 v\partial_v$ and reads
\begin{align}
    \frac{\left(\hat{\Lambda} \langle J_t\rangle - c(t)\left\{\left[\hat{\Lambda} - 1\right]\langle \rho_t\rangle \right\}\right)^2}{\mathrm{var}(J_t-c(t)\rho_t)}\leq \frac{\Delta S_\mathrm{tot}}{2},\label{eq:SummaryTimeDependentcTUR}
\end{align}
Conversely, the correlation bound does not easily generalize to include time-dependent driving.

\subsection*{Response to External Perturbations}

The response of a single-time observable $O(t)$ of a continuous-space
diffusion to a perturbation in
the noise amplitude, which effectively is a temperature perturbation \cite{Kubo_1957},
$\f{\sigma}\to\f{\sigma}+\delta\tilde{\f{\sigma}}$ \textcolor{black}{with perturbation noise amplitude $\tilde{\f \sigma}$} was recently
derived using stochastic calculus and reads \cite{Klinger2025}
\begin{align}
    \partial_\delta O_t &\equiv \lim_{\delta\to0}\frac{1}{\delta}\left(\langle O(t)\rangle_\delta - \langle O(t)\rangle\right)\label{eq:SummaryPerturbation1}\\
    &=-\left\langle O(t)\stoints\f\sigma(x_s, s)^{-1}\f{b}_\mathrm{eff}^0(x_s, s)\rmd \f{W}_s\right\rangle\label{eq:SummaryPerturbation2}
\end{align}
where $\langle \cdot\rangle$ is the average w.r.t. the unperturbed
process in Eq.~\eqref{eq:SummaryLangevinEq} [now with multiplicative noise $\f \sigma(x, \tau)$] and $\langle \cdot\rangle_\delta$ is the average w.r.t. the perturbed process
\begin{align}
    \rmd\f{x}^\delta_\tau = F(\f{x}^\delta_\tau)\rmd\tau + \left[\f{\sigma}(\f x_\tau^\delta, \tau)+\delta\tilde{\f{\sigma}}(\f x_\tau^\delta, \tau)\right]\rmd\f{W}_\tau.
    \label{eq:SummaryLangevinEqPerturbed}
\end{align}
The effective drift $[\f{b}_\mathrm{eff}^0(\f x, s)]_i =
  \frac{1}{2}\left[\f{\Sigma}(\f x, s) \f s^0(\f x, s)\right]_{i} +
  \frac{1}{2}\sum_j\partial_{x_j}\Sigma_{ij}(\f x, s)$ can be written
  in terms of the  score function $\f s^0(\f{x},t)=\nabla_{\f{x}}\log
  p(\f{x},t)$ of the instantaneous probability distribution of
  $\f{x}_\tau$ in Eq.~\eqref{eq:SummaryLangevinEq} and $\f{\Sigma} = \f{\sigma}\tilde{\f\sigma}^T + \tilde{\f\sigma}\f{\sigma}^T$ \cite{Klinger2025}.

We derive the equivalent results for MJP perturbing a general control parameter $\chi\to\chi+\delta$, where we find that
Eq.~\eqref{eq:SummaryPerturbation1} takes the form
\begin{align}
    \partial_\delta O_t =& \left\langle O(t)\stoints\sum_{x,y\neq x}\frac{\tilde{r}_{xy}}{r_{xy}}\rmd\varepsilon_{xy}(s)\right\rangle\label{eq:SummaryDiscreteResult},
\end{align}
where the unperturbed system evolves according to
Eq.~\eqref{eq:SummaryEoM} whereas the perturbed system obeys
\begin{align}
    \rmd \f{n}^\delta(\tau) = \f{R}^\delta(x^\delta_\tau)\rmd \tau + \rmd \f{\varepsilon}^\delta(\tau)
    \label{eq:SummaryPerturbedEoM},
\end{align}
with rates $r^\delta_{xy}$ perturbed as
\begin{align}
     r_{xy}^\delta &= r_{xy} + \delta \tilde{r}_{xy} + \mathcal{O}(\delta^2)\label{eq:SummaryPerturbedRates}.
\end{align}
\textcolor{black}{Note that similar results have recently been obtained in with a martingale approach \cite{faggionato2024}.}

For equilibrium systems where 
\begin{align}
    r_{xy} = D\mathrm{e}^{\left(\lambda E_x - (1-\lambda)E_y\right)/T}\label{eq:SummaryEQrates},
\end{align}
with mixing parameter $\lambda\in[0,1]$ and free energies $E_x$ of state $x$, we find that Eq.~\eqref{eq:SummaryDiscreteResult} for temperature perturbations $T\to T+\delta$ equals the equilibrium correlation function of $O$ and free energy $E$
\begin{align}
    \partial_\delta O_t =& \frac{1}{T^2}\left(\langle O E\rangle_\mathrm{eq} - \langle O(t) E(0)\rangle_\mathrm{eq} \right) \equiv \frac{1}{T^2}C_{OE}(t).\label{eq:SummaryEquilibriumCorrelation}
\end{align}

The result in Eq.~\eqref{eq:SummaryDiscreteResult} relates the
response to an observable $O$ to \emph{a
general small perturbation} of the dynamics to correlation functions
of the unperturbed dynamics (which, notably, for both MJP and
diffusions are \emph{not} necessarily
operationally accessible). In the case of equilibrium dynamics,
Eq.~\eqref{eq:SummaryDiscreteResult} reduces to the
\textit{Fluctuation Dissipation Theorem} (FDT) in
Eq.~\eqref{eq:SummaryEquilibriumCorrelation}. Therefore,
Eq.~\eqref{eq:SummaryDiscreteResult} can in some sense be seen as a
(mathematical) extension of the FDT to irreversible and transient
systems.\\
\indent We further generalize the response in
Eq.~\eqref{eq:SummaryDiscreteResult} to perturbations of
\emph{pathwise observables} of MJP (see Appendix~\ref{sec:pathObs}). Without loss of generality we set
\begin{align*}
    O_{\textcolor{black}{1}}(t) &= \stoints \mathrm{Tr}[\f{b}(s)^T\rmd\f{\varepsilon}(s)],\\
    \textcolor{black}{O_2(t)}&\textcolor{black}{=\int_0^t g_s\rmd s,}
\end{align*}
for some time-dependent hollow matrix $\f b(s)$ \textcolor{black}{and state function $g_s\equiv \sum_x \delta_{x_s x}g_x(s)$. For a perturbation strength $h(s)$, i.e., $r_{xy}\to r_{xy} + h(s)\tilde{r}_{xy}$, the result Eq.~\eqref{eq:SummaryDiscreteResult} for $O_{\rm p}(t)= O_1(t)+O_2(t)$ and its path-wise extension in Sec.~\ref{subsec:respose pathwise}
can be recast into the 
familiar response-function formalism
\begin{align}
 \Delta O_{\rm p}(t) &\equiv \langle O_{\rm p}(t)\rangle_h -\langle O_{\rm p}(t)\rangle \nonumber\\&= \int_0^t\rmd s h(s)\chi_{O_{\rm p}}(t,s) +\mathcal{O}(h^2),
\end{align} 
with response function $\chi_{O_{\rm p}}(t,s)$ 
\begin{align}
    \chi_{O_{\rm p}}(t,s) &= \chi_{O_1}(t,s) +\chi_{O_2}(t,s)\\
    &= \textcolor{black}{\mathbb{1}}_{s<t}\sum_{x,y\neq x} b_{xy}(s)\tilde{r}_{xy}p_x(s) + \int_0^t\rmd \textcolor{black}{\mathbb{1}}_{z>s}\langle g_zR(s)\rangle\nonumber\,,
\end{align}
with $R_x(s)=\sum_y\tilde{r}_{yx}p_y(s)/p_x(s)$.
}

The results in Eqs.~\eqref{eq:SummaryPerturbation2} and
~\eqref{eq:SummaryDiscreteResult} may be useful in the context of
computer simulations of high-dimensional systems, where knowing the
drift, diffusivity, and $\rmd \f{W}_s$ (and their discrete-state
counterparts) allows to evaluate perturbations without the need to
solve the high-dimensional Fokker-Planck (or master) equation of the
perturbed system. In this context,
Eq.~\eqref{eq:SummaryDiscreteResult} is expected to be useful,
e.g. when the diagonalization of the transition matrix becomes
computationally too demanding.

\textcolor{black}{\subsection*{Stochastic Calculus and Open Quantum Systems}}
\textcolor{black}{
Open discrete quantum systems coupled to a thermal bath are described by the Belavkin equation \cite{Carollo_2019}
\begin{align}
    \rmd \rho_s^\mathrm{u} = \mathcal{B}(\rho_s^\mathrm{u})\rmd s + \sum_i\left(\frac{\mathcal{J}_i(\rho_s^\mathrm{u})}{\mathrm{Tr}[\mathcal{J}_i(\rho_s^\mathrm{u})]} - \rho_s^\mathrm{u}\right)\rmd \tilde{n}_{is},\label{SummaryBelavkin}
\end{align}
where the pure state $\rho_s^\mathrm{u}$ is a rank-1 matrix and $\mathcal{J}_i(\rho_s^\mathrm{u}) = \hat{J}_i\rho_s^\mathrm{u}\hat{J}_i^\dagger$ with jump operator $\hat{J}_i$ for the $i$th quantum jump. The deterministic drift in Eq.~\eqref{SummaryBelavkin} has the form
\begin{align}
    \mathcal{B}(\rho_s^\mathrm{u}) = -i\hat{H}_\mathrm{eff}\rho_s^\mathrm{u} + \rho_s^\mathrm{u}\hat{H}_\mathrm{eff}^\dagger - \rho_s^\mathrm{u}\mathrm{Tr}(-i\hat{H}_\mathrm{eff}\rho_s^\mathrm{u} + \rho_s^\mathrm{u}\hat{H}_\mathrm{eff}^\dagger),
\end{align}
where the the effective Hamiltonian $\hat{H}_\mathrm{eff} = \hat{H} - \frac{i}{2}\sum_j\hat{J}_j^\dagger\hat{J}_j$ includes the Hamiltonian $\hat{H}$ and $\rmd \tilde{n}_{is}$ is a stochastic differential which takes values 1 if the $i$th transition occurs in $[s,s+\rmd s]$ and 0 otherwise. The statistical properties of the noise are $\langle \rmd \tilde{n}_{is}\rangle = \mathrm{Tr}(\mathcal{J}_i(\rho_s))$ and $\rmd \tilde{n}_{is}\rmd \tilde{n}_{js} = \delta_{ij}\rmd \tilde{n}_{is}$.
\\
\indent We show that the classical analogue to Eq.~\eqref{SummaryBelavkin} reads
\begin{align}
    \rmd \delta_{x_s x} =& \sum_{y\neq x}\left(r_{yx}\delta_{x_s y}\rmd s - \langle \rmd n_{yx}\rangle -r_{xy}\delta_{x_s x}\rmd s + \langle \rmd n_{xy}\rangle \right)\nonumber\\
    &+\sum_{y\neq x}(\rmd \varepsilon_{yx} + \langle \rmd n_{yx}\rangle - \rmd \varepsilon_{x y}-\langle \rmd n_{xy}\rangle).\label{SummaryClassicalBelavkin}
\end{align}
Taking the average over all noise histories in Eqs.~\eqref{SummaryBelavkin} and~\eqref{SummaryClassicalBelavkin} recovers the Lindblad and master equation, respectively. Hence, Eqs.~\eqref{SummaryBelavkin} and~\eqref{SummaryClassicalBelavkin} are the stochastic evolution of states in the system. These, as well as other functionals of stochastic increments, we can study as we know the underlying equations of motion, e.g., Eq.~\eqref{eq:SummaryEoM}.
}

In the following section we provide detailed definitions and derivations of the above
results and present applications. Technical steps and Lemmas are
deferred to the Appendix.



\section{Theory}\label{setup}

\subsection{Stochastic Equations of Motion}

To facilitate the comparison between discrete and continuous state
spaces we briefly summarize the basics of overdamped (for simplicity
additive-noise) diffusions. 

Let $(\f{x}_\tau)_{0\leq \tau\leq t}$ be a trajectory of $\f{x}_\tau\in\textcolor{black}{\mathbb{R}}^d$ evolving according to overdamped Langevin dynamics \cite{gardiner2004handbook,Seifert_2012, Klinger2025}
\begin{align}
    \rmd\f{x}_\tau = \f F(\f{x}_\tau)\rmd\tau + \f{\sigma}\rmd\f{W}_\tau,
    \label{eq:LangevinEq}
\end{align}
where $\rmd\f{W}_\tau$ is the $d$-dimensional Wiener process with
$\langle \rmd\f{W}_\tau\rangle = 0$ and $\langle
(\rmd\f{W}_\tau)_i(\rmd\f{W}_{\tau'})_j\rangle =
\delta_{ij}\delta(\tau-\tau')\rmd\tau\rmd\tau'$, $\f F$ is a drift field, and $\f{\sigma}$ is the noise
amplitude yielding the diffusion matrix $\f{D} =
\f{\sigma}\f{\sigma}^T/2$. While some bounds for continuous trajectories
presented in the next section also apply to underdamped trajectories,
we here limit the discussion to overdamped systems. This restriction
allows for consistent comparisons with jump dynamics, since these descriptions are valid for
negligible and marginalized momenta \cite{gardiner2004handbook,
  Moro1995}, respectively.  The evolution of the probability density
$P(\f{x}, \tau)$ of $\f{x}_\tau$ is governed by 
the
Fokker-Planck equation \cite{Risken1996}
$\partial_\tau P(\f{x}, \tau) = -{\nabla} \cdot \f{j}(\f{x}, \tau)$,
written as a continuity equation with probability current
$\f{j}(\f{x},\tau) = \left[\f{F}(\f{x}) - \f{D}\nabla\right]P(\f{x},
\tau)$.

\begin{figure}
    \centering
     \includegraphics[width=.48\textwidth]{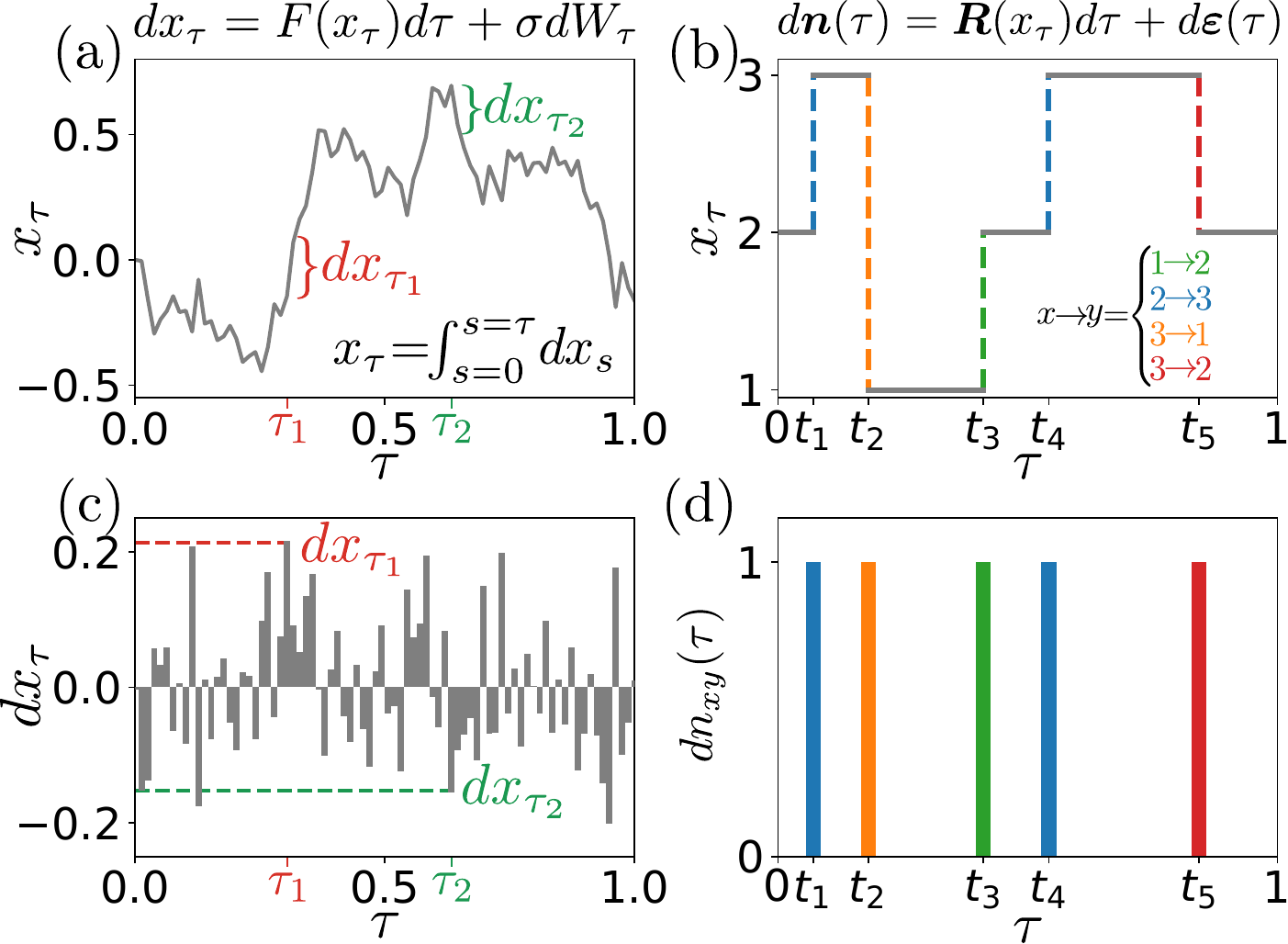}
    \caption{Schematic comparison of discrete and continuous trajectories. In (a), a continuous trajectory is shown. (b) shows a discrete trajectory in a three-state MJP. The transitions $2\to 3$ (blue), $3\to 1$ (orange), $1\to 2$ (green), and $3\to 2$ (red) are marked with the respective colors.  The displacement and jump increments $\rmd x_\tau$ and $\rmd n_{xy}(\tau)$  are presented in (c) and (d) for the continuous and discrete trajectories. The colors in (d) correspond to the different transitions in (b).}
    \label{fig:TrajectoryComparison}
\end{figure}

Consider now a discrete state space $\mathcal{N}\subseteq
\textcolor{black}{\mathbb{N}}$ which, without loss of generality, we can enumerate as
$\mathcal{N}=\{1, 2, \dots, N\}$. Focusing for the time being on
time-homogeneous systems (i.e., time-independent transition rates
$r_{ij}$ between states $i$ and $j$) the state probability vector
$p_i(\tau)$ evolves according to the master equation \cite{Schnakenberg, SEIFERT2018176}
\begin{align}
    \frac{\rmd}{\rmd \tau} p_i(\tau) = \sum_{j\neq i}\left[r_{ji}p_j(\tau) - r_{ij}p_i(\tau)\right],
    \label{eq:MasterEquation}
\end{align}
where the summation is only over $j$. In terms in the incoming and outgoing fluxes $J_{xy}(\tau) =
r_{xy}p_x(\tau)$ we have as $ \frac{\rmd}{\rmd \tau} p_i(\tau) \tb{=}\sum_{j\ne
  i}[J_{ij}(\tau)-J_{ji}(\tau)]$, and in vector notation
$\frac{\rmd}{\rmd \tau}\f p(\tau)=\f{L} \f p(\tau)$ with generator $\f{L}$
having entries $L_{ij} = \delta_{ij}r_{ii} + (1-\delta_{ij})r_{ji}$
and exit rate $r_{ii} = -\sum_{j\neq i} r_{ij}$. The transition
probability (or ``propagator'') is therefore given by $P(i, \tau'-\tau|j)=P(i, \tau'|j, \tau) = \left(\exp{(\tau'-\tau)\f{L}}\right)_{ij}$.

A path is fully determined by 
the sequence of states $(x_i)$ and transition times $t_i$,
$(x_\tau)_{0\leq\tau\leq t} = (x_0, t_0=0; x_1, t_1;\dots)$. The \textcolor{black}{normalized}
probability of such a trajectory is then \cite{Seifert_2012,LocalDetailedBalance,hartichGodecPathProb,HartichOptimalInferenceStrategies,Seifert_2025}
\begin{align}
    \textcolor{black}{\mathbb{P}}\left[(x_\tau)_{0\leq\tau\leq t}\right] = p_{x_0}(0) \rme^{\sum_k \tau_k(t) r_{kk}}\prod_{i,j}(r_{ij})^{n_{ij}(t)},
    \label{eq:MJP_PathMeasure}
\end{align}
where $p_{x_0}(0)$ is the probability of the initial state $x_0$ and
$n_{ij}(t)$ and $\tau_k(t)$ are the number of transitions
$i\to j$ and the total time spent in
state $k$ along the
trajectory of length $t$, respectively, and can be formulated as \cite{DirectTUR}
\begin{align}
    n_{ij}(t) &= \int_{\tau=0}^{\tau=t}\rmd n_{ij}(\tau),\nonumber\\
    \tau_{i}(t) &= \int_{\tau=0}^{\tau=t}\rmd\tau_{i}(\tau).
\end{align}
Suppose $x_\tau = i\neq j$, then $\rmd
n_{ij}(\tau)\sim\mathrm{Poi}(r_{ij}\rmd\tau)$ is Poisson distributed
and describes the number of transitions $i$ to $j$ that occur in the
time interval $[\tau, \tau+\rmd\tau]$. In particular
\cite{roldan2023martingales, Rogers_Williams_2000_martingales, DirectTUR, kwon2024unifiedframeworkclassicalquantum},
\begin{align}
    \rmd n_{ij}(\tau)\xrightarrow{\rmd\tau\to0}
    \begin{cases}
    1, & \text{$x_\tau=i$ and transition $i\to j$}\\
    & \text{occurs in $[\tau, \tau+\rmd\tau]$,}\\
    0, & \text{else}.
    \end{cases}
\end{align}
Note that $n_{ij}(\tau)$ is \emph{not} continuous (and therefore
\emph{not} differentiable). 
Similarly, the differential describing the time spent in a state is \cite{roldan2023martingales, Rogers_Williams_2000_martingales, DirectTUR, kwon2024unifiedframeworkclassicalquantum}
\begin{align}
    \rmd\tau_i(\tau) = 
    \begin{cases}
        \rmd\tau, & x_\tau = i,\\
        0, & \text{else}.
    \end{cases}
    \label{eq:dtau}
\end{align}
For any state $i$ along the trajectory, the probability to transition
to another state $j$ in the 
$\rmd\tau$ is given by $r_{ij}\rmd\tau$. Therefore, it is natural to define a noise \cite{DirectTUR, kwon2024unifiedframeworkclassicalquantum, zheng2025spatialcorrelationunifiesnonequilibrium, roldan2023martingales}
\begin{align}
    \rmd\varepsilon_{ij}(\tau) \equiv \rmd n_{ij}(\tau) - r_{ij}\rmd\tau_i(\tau),
    \label{eq:DiscreteNoise}
\end{align}
by subtracting the expected number of jumps, $r_{ij}\rmd\tau_i(\tau)$,
from the actual jumps $\rmd n_{ij}(\tau)$. This noise has also been
studied in the context of martingales \cite{roldan2023martingales}.

Rearranging Eq.~\eqref{eq:DiscreteNoise} yields the consistent stochastic equation of motion for $x\neq y$ for MJP
\begin{align}
    \underbrace{\rmd n_{xy}(\tau)}_{\text{``displacement"}} = \underbrace{r_{xy}\textcolor{black}{\delta_{x_\tau x}}}_{\text{``drift"}}\rmd\tau + \underbrace{\rmd\varepsilon_{xy}(\tau)}_{\text{``noise"}}.
    \label{eq:eom_MJP}
\end{align}
Defining the matrix
$(\f{R}(x_\tau))_{ij}\equiv(1-\delta_{ij})r_{ij}\textcolor{black}{\delta_{x_\tau i}}$,
we can write Eq.~\eqref{eq:eom_MJP} as a stochastic matrix equation
\begin{align}
    \rmd\f{n}(\tau) = \f{R}(x_\tau)\rmd\tau + \rmd\f{\varepsilon}(\tau).
    \label{eq:Matrix_eom_MJP}
\end{align}
The differential matrices have entries $(\rmd \f{n}(\tau))_{ij} = \rmd
n_{ij}(\tau)$ and $(\rmd \f{\varepsilon}(\tau))_{ij} = \rmd
\varepsilon_{ij}(\tau)$, where
\begin{align}
    \!\!\!\langle \rmd\varepsilon_{kl}(\tau)\rmd\varepsilon_{ij}(\tau')\rangle_{x_\tau=x} &= \delta_{kx}\delta_{ik}\delta_{jl}\delta(\tau-\tau')r_{kl}\rmd\tau\rmd\tau'\label{depsilonCorr},
\end{align}
in complete analogy with the Wiener process $\rmd  W_\tau \propto \sqrt{\rmd \tau}$.
Equation~\eqref{eq:Matrix_eom_MJP} is a
matrix evolution equation, while the continuous evolution,
Eq.~\eqref{eq:LangevinEq}, is vectorial. However, it has the same
structure as the Langevin equation, as it consist of a
``displacement'', ``deterministic drift'', and ``noise'', see
Eq.~\eqref{eq:eom_MJP}. We highlight the comparison in
Tab.~\ref{tab:ComparisonEOMterms}. It should be stressed that the
diagonal terms in Eq.~\eqref{eq:Matrix_eom_MJP} are to be neglected,
as they do \emph{not} have any physically relevant meaning.  

\begin{table}
    \centering
    \caption{\textcolor{black}{Overview of quantities entering the stochastic equations of motion in continuous and discrete space.}}
    \begin{tabular}{c c c}
        \hline
        Meaning & Continuous & Discrete\\
        \hline
         Displacement & $\rmd\f{x}_\tau$ & $\rmd\f{n}(\tau)$ \\
         Deterministic Drift & $\f{F}(\f{x}_\tau)\rmd\tau$ & $\f{R}(x_\tau)\rmd\tau$\\
         Noise & $\f{\sigma}\rmd\f{W}_\tau$ & $\rmd\f{\varepsilon}(\tau)$\\
         \hline
    \end{tabular}
    \label{tab:ComparisonEOMterms}
\end{table}

Note that $\rmd \f{\varepsilon}(\tau)$ is \emph{not} Gaussian noise,
but rather a shifted (due to the $\f{R}(x_\tau)\rmd \tau$ term)
time-inhomogeneous Poisson process. Explicitly, because for $x_\tau=x$
and $y\neq x$ the differential $\rmd
n_{xy}(\tau)\sim\mathrm{Poi}(r_{xy}\rmd \tau)$ is Poisson, the values
$\rmd \varepsilon_{xy}(\tau)$ can take for any $x_\tau$ are $\rmd
\varepsilon_{xy}(\tau)\in\{0,-r_{xy}\rmd \tau, 1-r_{xy}\rmd \tau,
2-r_{xy}\rmd \tau, 3-r_{xy}\rmd \tau, \dots\}$, where $\rmd
\varepsilon_{xy}(\tau)=0$ happens if $x_\tau\neq x$. In case
$x_\tau=x$, one can identify $P(\rmd \varepsilon_{xy}(\tau) =
k-r_{xy}\rmd \tau|x_\tau=x) = \mathrm{poi}_{r_{xy}\rmd\tau}(k)$ for
any $k\in\textcolor{black}{\mathbb{N}}_{\geq 0}$\textcolor{black}{, where $\mathrm{poi}_\lambda(k) = \lambda^k\mathrm{e^{-\lambda}/k!}$ is the Poisson distribution function}. In the limit of small $\rmd \tau$\textcolor{black}{,} to
linear order in $\rmd \tau$\textcolor{black}{,} the noise therefore takes values  $\rmd
\varepsilon_{xy}(\tau)\in\{0,-r_{xy}\rmd \tau,  1-r_{xy}\rmd \tau\}$
with probabilities $1-p_x(\tau), \left(1-r_{xy}\rmd\tau\right)p_x(\tau),
 r_{xy}\rmd \tau p_x(\tau)$, respectively. The probability
for any $k\geq 2$ are $\mathcal{O}(\rmd \tau^k)$ and can therefore be
neglected. 

Additionally, the scaling of the noise $\rmd
\f{\varepsilon}(\tau)\sim\sqrt{\rmd\tau}$ [see Eq.~\eqref{depsilonCorr}] implies that
$|\rmd\f{\varepsilon}(\tau)/\rmd\tau|\sim1/\sqrt{\rmd\tau}$ and,
therefore, $|\rmd\f{n}(\tau)/\rmd\tau|\sim1/\sqrt{\rmd\tau}$. The limit
$\rmd\tau\to0$ does \emph{not} exist. This is similar
to the well-known (and often ignored) issue with writing the
Langevin equation in the ``physicist's" way using $\Dot{\f x}_\tau$. 

Further similarities in the equations of motion become obvious, e.g.,
in Fig.~\ref{fig:TrajectoryComparison}. Any trajectory is fully
characterized by $\rmd \f{n}(\tau)$ or $\rmd \f{x}_\tau$ (see
Fig.~\ref{fig:TrajectoryComparison}). Moreover, as shown in
Sec.~\ref{sec:EoM_to_Ensemble}, the master equation follows from Eq.~\eqref{eq:Matrix_eom_MJP} in the same manner as
the Fokker-Planck equation follows from Eq.~\eqref{eq:LangevinEq}.

We further require the statistical properties of the above stochastic
differentials. Conditioning the expectation on being in a specific
state, e.g. $x_\tau = x$, we recover for $\tau\leq \tau'$
\begin{align}
    \langle \rmd n_{ij}(\tau)\rangle_{x_\tau=x} &= \delta_{ix}r_{ij}\rmd\tau\nonumber,\\
    \langle \rmd\tau_{i}(\tau)\rangle_{x_\tau=x} &= \delta_{ix}\rmd\tau\nonumber,\\
    \langle \rmd\varepsilon_{ij}(\tau)\rangle_{x_\tau=x} &= 0\nonumber,\\
    \langle \rmd\tau_{i}(\tau)\rmd\tau_{j}(\tau')\rangle_{x_\tau=x} &= \delta_{ix}P(j, \tau'|i, \tau)\rmd\tau'\rmd\tau.\label{eq:ListExpectations}
\end{align}
We are also required to evaluate cross-correlations between the noise
and the time spent in a state, which follows from the noise-time correlation lemma
\begin{align}
    &\frac{\langle \rmd\varepsilon_{kl}(\tau) \rmd\tau_i(\tau')\rangle_{x_\tau = x}}{\rmd\tau\rmd\tau'}\label{eq:NoiseTimeCorrelationLemma} \\&= \delta_{xk}\textcolor{black}{\mathbb{1}}_{\tau<\tau'}\left[P(i, \tau'|l, \tau) - P(i, \tau'| k, \tau)\right]r_{kl} + \mathcal{O}(\rmd\tau)\nonumber.
\end{align}
The above lemma has been shown before
\cite{kwon2024unifiedframeworkclassicalquantum}; in
App.~\ref{sec:NoiseTimeProof} we provide a new alternative proof. Note
that, due to the Markovianty of the system, we do \emph{not} need to restrict
on $\tau\leq \tau'$ in evaluating the expectation, as $\tau>\tau'$ vanishes. 

The above correlators allow, for example, to evaluate the covariance
between the number of jumps \textcolor{black}{at} different space-time points along the
trajectory, i.e., for $\tau\leq\tau'$, \cite{kwon2024unifiedframeworkclassicalquantum}
\begin{align}
        \langle& \rmd n_{kl}(\tau)\rmd n_{ij}(\tau^\prime)\rangle_{x_\tau=x}=\rmd\tau\rmd\tau^\prime \delta_{kx}r_{kl}\\
        &\times\left[
        \delta(\tau - \tau^\prime)\delta_{ik}\delta_{jl} + \textcolor{black}{\mathbb{1}}_{\tau=\tau'}\delta_{ik}r_{ij}+ \textcolor{black}{\mathbb{1}}_{\tau < \tau^\prime}r_{ij}P(i, \tau^\prime|l, \tau)\right].\nonumber
\end{align}
The middle term can be neglected when integrating over $\tau$ and
$\tau'$ as it has measure zero. The first term accounts for jumps
occurring at $\tau=\tau'$, while the last term accounts for the jumps
$k\to l$ at time $\tau$ that result in $i\to j$ at some later time
$\tau'> \tau$.  

We may rewrite the path measure Eq.~\eqref{eq:MJP_PathMeasure} with
the stochastic differentials discussed above as
\begin{widetext}
\begin{align}
    \textcolor{black}{\mathbb{P}}[(x_\tau)_{0\leq \tau\leq t}] = p_{x_0}(0)\exp\left[\stoints\sum_{x} r_{xx}\rmd \tau_x(s) + \stoints\sum_{x,y\neq x}\log r_{xy}\rmd n_{xy}(s)\right],\label{eq:StocDiffPathMeasure}
\end{align}
\end{widetext}
where the second sum is taken over both $x$ and $y\ne x$.
The advantage of writing the path measure as in
Eq.~\eqref{eq:StocDiffPathMeasure} instead of
Eq.~\eqref{eq:MJP_PathMeasure} is that it allows for possible time dependencies of the rates.

\subsection{Pathwise Observables: Currents and Densities}

We now introduce pathwise observables, i.e., time-integrated densities
and currents that were recently analyzed extensively in the context of
diffusions \cite{DirectTUR, Dieball_2022_CorrelationsFluctuations, DieballCoarseGraining, DieballCurrentVariance}. As before, we start with the  
continuous space setting, where a time-integrated current is defined as \cite{DirectTUR}
\begin{align}
    J_t^\mathrm{c} = \StoInt \f{U}(\f{x}_\tau, \tau)^T\circ\rmd \f{x}_\tau,
    \label{eq:continuousCurrent}
\end{align}
as a Stratonovich integral over some (square integrable) vector-valued window function
$\f{U}(\f{x}_\tau, \tau)$, and a time-integrated density is defined as \cite{DirectTUR}
\begin{align}
    \rho_t^\mathrm{c} = \int_0^t \rmd \tau V(\f{x}_\tau, \tau),
    \label{eq:continuousDensity}
\end{align}
where $V(\f{x}_\tau, \tau)$ is some  (square integrable) scalar-valued function. 

In the context of MJP the time-integrated currents are weigh the transitions
occurring along a trajectory with a fully
anti-symmetric, possibly time-dependent weight
$\f{\kappa}(\tau)$, i.e., $\kappa_{ij}(\tau) =
-\kappa_{ji}(\tau)$. The current defined this way is manifestly
anti-symmetric such that the current changes sign upon time
reversal. Note that $\f{\kappa}(\tau)$ describes experimentally accessible
transitions. We therefore define a (scalar) time-integrated current as \cite{DirectTUR}
\begin{align}
    J_t &\equiv \StoInt \mathrm{Tr}[\f{\kappa}(\tau)^T\rmd\f{n}(\tau)].
    \label{eq:DiscreteCurrent}
\end{align}
A comparison of Eqs.~\eqref{eq:continuousCurrent} and
\eqref{eq:DiscreteCurrent} highlights the fact that integrating over
time and tracing over $\rmd \f{n}(\tau)$ is the MJP correspondence to
the Stratonovich integral~\eqref{eq:continuousCurrent}. The stochastic integral Eq.~\eqref{eq:DiscreteCurrent} is time antisymmetric in analogy to the current defined for diffusion processes in the form of a Stratonovich integral Eq.~\eqref{eq:continuousCurrent}. The current in
Eq.~\eqref{eq:DiscreteCurrent} naturally decomposes (as for diffusions \cite{DirectTUR})
into a ``usual time integral'' $J_t^\mathrm{II}$ and the stochastic
integral, $J_t=J_t^\mathrm{I}+J_t^\mathrm{II}$ where
\begin{subequations}
    \begin{align}
    J_t^\mathrm{I} &= \StoInt \mathrm{Tr}[\f{\kappa}(\tau)^T\rmd\f{\varepsilon}(\tau)],\label{eq:DissipativeDiscreteCurrent}\\
    J_t^\mathrm{II} &=\int_0^t \mathrm{Tr}[\f{\kappa}(\tau)^T\f{R}(x_\tau)]\rmd\tau\label{eq:TimeDiscreteCurrent},
\end{align}
\end{subequations}
i.e., integrals with traces over $\rmd \f{\varepsilon}(\tau)$ are not necessarily time antisymmetric and therefore may be
identified (per analogy) as It\^o integrals due to the correspondence
to the stochastic part of Eq.~\eqref{eq:continuousCurrent}
\cite{DirectTUR}. This analogy (i) allows us to
easier distinguish between the integrals over noise or jump increments
and (ii) is fully consistent with observables of continuous-space
diffusion. By means of the
latter we will later demonstrate how continuous
space It\^o integrals emerge from the corresponding integrals of  MJP
with traces over $\rmd \f{\varepsilon}(\tau)$ in the continuum
limit. This will later provide a unified description of MJP and overdamped
diffusions within the framework of stochastic calculus.

Similarly, a density assigns each state $i$ a weight $V_i(\tau)$ and then integrates these along the trajectory. These weights can be combined into the state function \textcolor{black}{ $V_\tau\equiv V(x_\tau,\tau) = \sum_k \delta_{x_\tau k}V_k(\tau)$}. With \textcolor{black}{${V}_\tau$} we define the density \cite{DirectTUR}
\begin{align}
    \rho_t = \StoInt \textcolor{black}{{V}_\tau \rmd{\tau}}.
    \label{eq:DiscreteDensity}
\end{align}

\begin{figure}
    \centering
        \includegraphics[width=\linewidth]{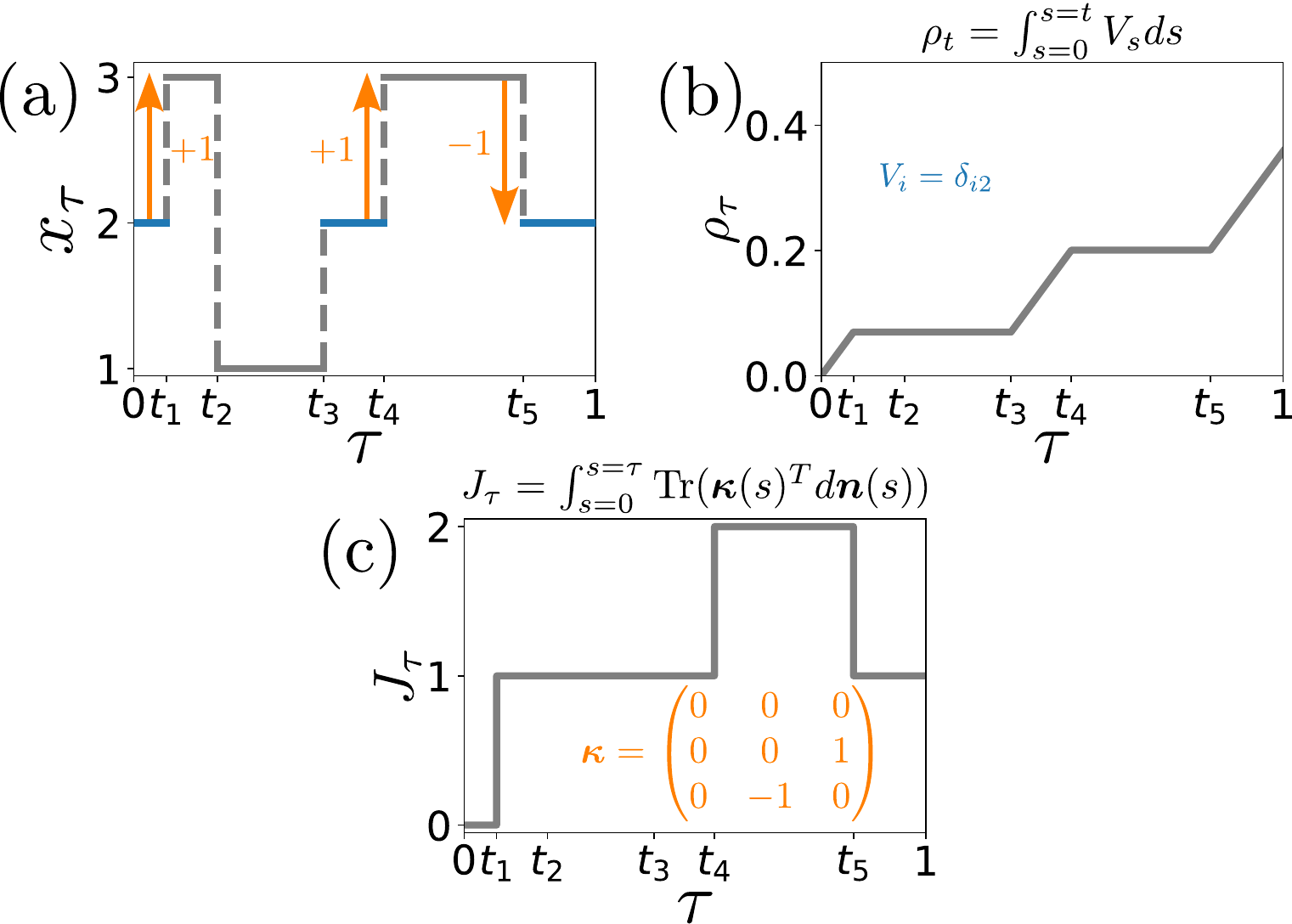}
    \caption{Example trajectory and examples of an arising current and density. The part of the trajectory in (a) which contributes to the density $\rho_\tau$ with state function, $\textcolor{black}{{V}_i(\tau) = \delta_{i2}}$, is marked in blue. Similarly, the transitions contributing to the current $J_\tau$ with transition weights, $\kappa_{ij} = \delta_{i2}\delta_{j3} - \delta_{i3}\delta_{j2}$, are marked by orange arrows with numbers representing the weight of the transition. The resulting density and current can be seen in (b) and (c), respectively.}
    \label{fig:CurrentDensityExample}
\end{figure}

An exemplary trajectory with corresponding time-integrated density and
current is shown in Fig.~\ref{fig:CurrentDensityExample} to visualize Eqs.~\eqref{eq:DiscreteCurrent} and \eqref{eq:DiscreteDensity}.

\subsection{From Stratonovich Increments to Observable Correlations}

In the case of diffusion processes, correlations of pathwise observables were shown to provide deep information about
the passage of ``pinned'' trajectories through the respective window
in terms of generalized Green-Kubo like relations
\cite{DieballCoarseGraining}. They are furthermore relevant in the
context of thermodynamic inequalities, e.g., TURs \cite{DirectTUR,
  CrutchfieldUnderdampedTUR} and thermodynamic correlation bounds
(CB) \cite{BoundsCorrelationTimes,
  DieballCorrelationBound}.  Following
Refs.~\cite{DieballCoarseGraining, DieballCurrentVariance,
  Dieball_2022_CorrelationsFluctuations} we now show how the
steady-state correlations of stochastic differentials relate to
covariances of pathwise observables of MJP. In particular, we
emphasize how these compare to the corresponding continuous-space results. 

In order to express  (co)variances of continuous and discrete
currents, Eq.~\eqref{eq:continuousCurrent} and
Eq.~\eqref{eq:DiscreteCurrent}, and densities,
Eq.~\eqref{eq:continuousDensity} and Eq.~\eqref{eq:DiscreteDensity},
in a compact \emph{unifying} form \cite{DieballCoarseGraining, DieballCurrentVariance, Dieball_2022_CorrelationsFluctuations} 
we define the integration operators
\begin{subequations}
    \begin{align}
        \hat{\mathcal{I}}^{t, \mathrm{c}}_{M, N}[\cdot] &= \int_0^t\rmd \tau \int_\tau^t\rmd \tau' \int \rmd \f{z}\int \rmd \f{z}'M(\f{z})^T[\cdot]N(\f{z}'),\label{eq:ContinuousIntegrationOperator}\\
        \hat{{I}}^t_{\textcolor{black}{G}^\alpha, \textcolor{black}{G}^\beta}[\cdot] &= \int_0^t\rmd\tau\int_\tau^t\rmd\tau' \sum_{x,y}\textcolor{black}{G}_{xy}^\alpha\sum_{i,j}\textcolor{black}{G}^\beta_{ij}[\cdot],
    \label{eq:DiscreteIntegrationOperator}
    \end{align}
\end{subequations}
where $[\cdot]$ in the first line denotes functions which depend on
$\f{z}$, $\f{z}'$, $\tau$, and $\tau'$. Similarly, the functions entering
$\hat{{I}}^t_{\textcolor{black}{G}^\alpha, \textcolor{black}{G}^\beta}[\cdot]$ depend on $x$, $y$, $i$, $j$,
$\tau$, and $\tau'$. The functions $M(\f{z})$ and $N(\f{z})$ may be
scalar- or vector-valued functions, while $\textcolor{black}{G}^\alpha$ and $\textcolor{black}{G}^\beta$ are
two-point quantities, e.g.,
$\f{\kappa}$. Equation~\eqref{eq:DiscreteIntegrationOperator}
corresponds to Eq.~\eqref{eq:ContinuousIntegrationOperator} where each
continuous spatial integral has been changed to a discrete double
sum. While one may expect the integrals to be substituted by a single
sum, the double sum is indeed required to include non-local quantities, e.g.,
currents.   

We will now consider two observables $X_t^\mathrm{(c)}
\in\{J_{t, \textcolor{black}{k}}^{\mathrm{(c)}}, \rho_{t, \textcolor{black}{k}}^{\mathrm{(c)}}\}$ and
$Y_t^\mathrm{(c)} \in\{J_{t, \textcolor{black}{l}}^{\mathrm{(c)}}, \rho_{t, \textcolor{black}{l}}^{\mathrm{(c)}}\}$,
where the superscript \textcolor{black}{(c)} is used to distinguish discrete \textcolor{black}{(no superscript)} from the
continuous \textcolor{black}{(superscript $\mathrm{c}$)} setting and
$k,l\in\{1,2\}$ is the index for currents $J_{t, \textcolor{black}{1}}^{\mathrm{(c)}},
J_{t, \textcolor{black}{2}}^{\mathrm{(c)}}$ and densities $\rho_{t, \textcolor{black}{1}}^{\mathrm{(c)}},
\rho_{t, \textcolor{black}{2}}^{\mathrm{(c)}}$. These currents and densities may have
different current-/density-defining functions, which we take into
consideration here.  

As shown in Refs.~\cite{DieballCoarseGraining, DieballCurrentVariance, Dieball_2022_CorrelationsFluctuations}, the continuous (co)variances can be written as compactly in terms of Eq.~\eqref{eq:ContinuousIntegrationOperator} together with fundamental increment correlations. We show here that the same structure appears for the discrete case. In fact, the covariances can be written as
\begin{align}
    \mathrm{cov}(X_t^\mathrm{c}, Y_t^\mathrm{c}) &= \hat{\mathcal{I}}^{t, \mathrm{c}}_{X, Y}[{\Xi}_{m}^{\f{z}\f{z}'}] - \langle X_t^\mathrm{c}\rangle \langle Y_t^\mathrm{c}\rangle,\nonumber
    \\
    \mathrm{cov}(X_t, Y_t) &=\hat{{I}}^t_{U^X, U^Y}[{\Xi}^m_{ijxy}]- \langle X_t\rangle \langle Y_t\rangle,
\end{align}
where \textcolor{black}{the $\Xi$ matrices are defined through increment correlations, see Eqs.~\eqref{eq:Xi1},~\eqref{eq:Xi2}, and~\eqref{eq:Xi3}. The index $m$ we use to distinguish between different observable correlations: }$m=1$ for $X_t^\mathrm{(c)} = \rho_{t, \textcolor{black}{k}}^{\mathrm{(c)}}$ and $Y_t^\mathrm{(c)} = \rho_{t, \textcolor{black}{l}}^{\mathrm{(c)}}$, w.l.o.g. $m=2$ for $X_t^\mathrm{(c)} = J_{t, \textcolor{black}{k}}^{\mathrm{(c)}}$ and $Y_t^\mathrm{(c)} = \rho_{t, \textcolor{black}{l}}^{\mathrm{(c)}}$, and $m=3$ for $X_t^\mathrm{(c)} = J_{t, \textcolor{black}{k}}^{\mathrm{(c)}}$ and $Y_t^\mathrm{(c)} = J_{t, \textcolor{black}{l}}^{\mathrm{(c)}}$. For density-density covariance ($m=1$), the increment correlations read \cite{DieballCoarseGraining}
\begin{align}
    {\Xi}_{1}^{\f{z}\f{z}'} &= \langle \textcolor{black}{\mathbb{1}}\rangle_{\f{x}_{\tau}=\f{z}}^{\f{x}_{\tau'}=\f{z}'} + \langle \textcolor{black}{\mathbb{1}}\rangle_{\f{x}_{\tau}=\f{z}'}^{\f{x}_{\tau'}=\f{z}},\label{eq:Xi1}\\
    {\Xi}^1_{ijxy}&= \frac{\langle \rmd \tau_x(\tau)\rmd\tau_i(\tau')\rangle_{x_\tau=x}^{x_{\tau'}=i}}{\rmd\tau\rmd\tau'}+\frac{\langle \rmd \tau_x(\tau')\rmd\tau_i(\tau)\rangle^{x_{\tau'}=x}_{x_{\tau}=i}}{\rmd\tau\rmd\tau'}.\nonumber
\end{align}
Here, $\langle\cdot\rangle_{\f{x}_{\tau}=\f{z}'}^{\f{x}_{\tau'}=\f{z}}$ is expectation conditioned on $\f{x}_{\tau}=\f z$ and $\f{x}_{\tau'}=\f z'$ with $\tau\leq \tau'$ and analogously for the discrete case. Evaluating these correlations yields the stationary covariances \cite{DieballCoarseGraining, DieballCurrentVariance}
\begin{widetext}
\begin{align}
    \mathrm{cov}({\rho}_{t, \textcolor{black}{k}}^{\mathrm{c}}, {\rho}_{t, \textcolor{black}{l}}^{\mathrm{c}}) =& \hat{\mathcal{I}}^{t, \mathrm{c}}_{V^k, V^l}\left[P_\mathrm{s}(\f{z},\tau;\f{z}',\tau') + P_\mathrm{s}(\f{z}',\tau;\f{z},\tau') - p_\mathrm{s}(\f{z})p_\mathrm{s}(\f z')\right],\nonumber\\
    \mathrm{cov}({\rho}_{t, \textcolor{black}{k}}, {\rho}_{t, \textcolor{black}{l}}) =& \hat{{I}}^{t}_{V^k\delta, V^l\delta}\left[P_\mathrm{s}(x,\tau;i,\tau') + P_\mathrm{s}(x,\tau';i,\tau) - p_x^\mathrm{s}p_i^\mathrm{s}\right].\label{eq:DensityCov}
\end{align}
\end{widetext}
The two-point function $V^k\delta$ stands for the (effectively
one-point) quantity $V^k_i\delta_{ij}$ \footnote{Note that the $\delta$ is introduced for convenience to allow for compact notation using the integration operator $\hat{{I}}^t_{U^\alpha, U^\beta}[\cdot]$.} and $P_\mathrm{s}(\cdot;\cdot)$
denotes the joint probability for stationary systems, i.e., $P_\mathrm{s}(x,\tau;i,\tau')=P(i,\tau'|x,\tau)p_x^\mathrm{s}$. 

\begin{figure}[ht!!]
    \centering
    \includegraphics[width=\linewidth]{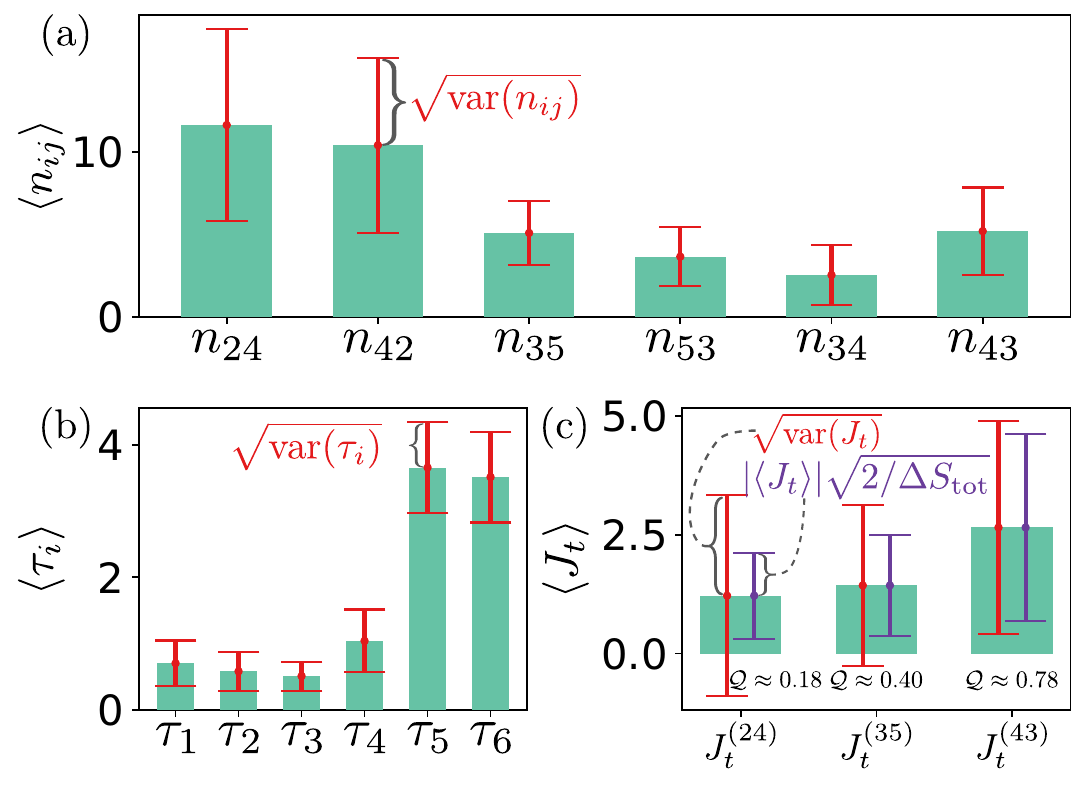}
    \caption{Statistics of transitions and waiting times in the stationary SAT model. The average number of transition $\langle n_{ij}\rangle$ are shown in $(a)$ for a selection of edges. In (b), the average time spent $\langle \tau_i\rangle$ in each state is presented. The steady-state currents resulting from the transitions in (a) are presented in (c). The red errorbars are the respective standard deviations $\sqrt{\mathrm{var}(J_t)}$. In (c), the lower bound $|\langle J_t\rangle|\sqrt{2/\Delta S_\mathrm{tot}}\leq\sqrt{\mathrm{var}(J_t)}$ resulting from the steady-state TUR is added as purple errorbars. We use $t=10$ in every panel. }
    \label{fig:Histogram}
\end{figure}

Similarly, for current-density covariances ($m=2$), we need the following increment correlations \cite{DieballCoarseGraining}
\begin{align}
    {\Xi}_2^{\f{z}\f{z}'} &= \frac{\langle \circ \rmd \f{x}_{\tau}\rangle_{\f{x}_{\tau}=\f{z}}^{\f{x}_{\tau'}=\f{z}'}}{\rmd \tau} + \frac{\langle \circ \rmd \f{x}_{\tau'}\rangle_{\f{x}_{\tau}=\f{z}'}^{\f{x}_{\tau'}=\f{z}}}{\rmd \tau'},\label{eq:Xi2}\\
    {\Xi}_{ijxy}^2 &=\frac{\langle \rmd n_{xy}(\tau)\rmd\tau_i(\tau')\rangle_{x_\tau=x}^{x_{\tau'}=i}}{\rmd\tau\rmd\tau'} + \frac{\langle \rmd n_{xy}(\tau')\rmd\tau_i(\tau)\rangle^{x_{\tau'}=x}_{x_{\tau}=i}}{\rmd\tau\rmd\tau'}\nonumber,
\end{align}
which include correlations of a
Stratonovich increment $\circ\rmd \f{x}_\tau$ and $\rmd n_{xy}(\tau)$
with a future event. The evaluation of such correlations 
is possible by means of the
generalized time-reversal symmetry, i.e., dual-reversal symmetry
\cite{Hatano_2001,Dieball_2022_CorrelationsFluctuations, DieballCoarseGraining,
  DieballCurrentVariance}. In both cases, the dual system (denoted by
$^\ddag$) has the same steady state probability whereas the irreversible
drift $\f{j}_\mathrm{s}(\f{x})$ has the opposite sign
$\f{j}^\ddag_\mathrm{s}(\f{x}) = -\f{j}_\mathrm{s}(\f{x})$. It is useful to define
current and dual current operators; in the continuous-space setting,
these are \cite{Dieball_2022_CorrelationsFluctuations, DieballCoarseGraining,
  DieballCurrentVariance}
$\hat{\f{j}}_{\f{x}}=p_s(\f{x})^{-1}\f{j}_\mathrm{s}(\f{x}) -
p_\mathrm{s}(\f{x}) \f D \nabla_{\f{x}}\textcolor{black}{\left[\textcolor{black}{p_\mathrm{s}(\f x)^{-1}}\right]}$ and
$\hat{\f{j}}_{\f{x}}^\ddag=p_s(\f{x})^{-1}\f{j}_\mathrm{s}(\f{x}) +
p_\mathrm{s}(\f{x}) \f D \nabla_{\f{x}}\textcolor{black}{\left[\textcolor{black}{p_\mathrm{s}(\f x)^{-1}}\right]}$. The
corresponding operators for discrete-state systems are much simpler
and read
$\hat{j}_{xy} = r_{xy}$ and $\hat{j}_{xy}^\ddag =
r_{yx}p_y^\mathrm{s}/p_x^\mathrm{s}$, respectively. The latter is the dual current
operator reversing the steady-state currents
\cite{Seifert_2012} . When these current operators act on
$p_x^\mathrm{s}$, they result in the probability flux and dual
probability flux on edge $x\to y$, i.e., $\hat{j}_{xy}p^\mathrm{s}_x =
r_{xy}p_x^\mathrm{s} = J_{xy}^\mathrm{s}$ and
$\hat{j}^\ddag_{xy}p^\mathrm{s}_x = r_{yx}p_y^\mathrm{s} =
J_{yx}^\mathrm{s}$. By means of the flux operators, it can be shown that  
\begin{align}
    \langle \rmd n_{xy}(\tau')\rmd\tau_i(\tau)\rangle^{x_{\tau'}=x}_{x_{\tau}=i} &= \hat{j}_{xy}P(x, \tau'-\tau|i)p^\mathrm{s}_i\rmd \tau \rmd\tau'\nonumber\\ 
    &= r_{xy}P(x, \tau'-\tau|i)p^\mathrm{s}_i\rmd \tau \rmd\tau',
    \label{eq:IncrementCorrelation1}
\end{align}
which is the current on edge $x\to y$ at time $\tau'>\tau$ in addition
to $x_{\tau}=i$. With an approach analogous to Ref.~\cite{DieballCoarseGraining}, we use the dual current operator to express the correlation of an initial transition with future events. To be precise, we find that (see App.~\ref{sec:JumpTimeProof})
\begin{align}
    \langle \rmd n_{xy}(\tau)\rmd\tau_i(\tau')\rangle_{x_\tau=x}^{x_{\tau'}=i} &= \hat{j}^\ddag_{yx}P(i, \tau'-\tau|y)p^\mathrm{s}_y\rmd \tau \rmd\tau'\nonumber\\ 
    &= r_{xy}P(i, \tau'-\tau|y)p^\mathrm{s}_x\rmd \tau \rmd\tau',
    \label{eq:IncrementCorrelation2}
\end{align}
as well as (see App.~\ref{sec:JumpJumpProof})
\begin{align}
    \langle \rmd n_{xy}(\tau)\rmd n_{ij}(\tau')\rangle_{x_\tau=x}^{x_{\tau'}=i} &= \hat{j}_{ij}\hat{j}^\ddag_{yx}P(i, \tau'-\tau|y)p^\mathrm{s}_y\rmd \tau \rmd\tau'\nonumber\\ 
    &= r_{ij}r_{xy}P(i, \tau'-\tau|y)p^\mathrm{s}_x\rmd \tau \rmd\tau'.
    \label{eq:IncrementCorrelation3}
\end{align}
Equations~\eqref{eq:IncrementCorrelation2} and
\eqref{eq:IncrementCorrelation3} enable the calculation of
(co)variances of $n_{ij}$ and thus provide valuable insight about,
e.g., how noisy are transitions in a given system (for an explicit
example see Figs.~\ref{fig:Histogram}a and \ref{fig:Histogram}b on the statistics of
transitions and dwell-time, respectively, in a model of secondary active transport (SAT) in
Sec.~\ref{applications}). Moreover, we can rewrite the current-density
covariances compactly and in a unified form as \cite{DieballCoarseGraining, DieballCurrentVariance}
\begin{widetext}
\begin{align}
    \mathrm{cov}({J}_{t, \textcolor{black}{k}}^{\mathrm{c}}, {\rho}_{t, \textcolor{black}{l}}^{\mathrm{c}})&= \hat{\mathcal{I}}^{t, \mathrm{c}}_{\f{U}^k, V^l}\left[\hat{\f{j}}_{\f{z}'}P(\f{z}', t'|\f{z})p_\mathrm{s}(\f z) + \hat{\f{j}}^\ddag_{\f{z}}P(\f{z}', t'|\f{z})p_\mathrm{s}(\f z) -  \f{j}_\mathrm{s}(\f{z})p_\mathrm{s}(\f z')\right],\nonumber \\
    \mathrm{cov}({J}_{t, \textcolor{black}{k}}, {\rho}_{t, \textcolor{black}{l}}) &= \hat{{I}}^{t}_{\kappa^k, V^l\delta}\left[\hat{j}_{xy}P(x, t'| i)p_{i}^\mathrm{s} + \hat{j}_{yx}^\ddag P(i, t'| y)p_y^\mathrm{s}- J^\mathrm{s}_{xy}p_{i}^\mathrm{s}\right].\label{eq:DensCurCov}
\end{align}
The only time dependence in the integrands in
Eq.~\eqref{eq:DensCurCov} is the time-translation invariant
propagators $P(i, t| y)$, that is, the integrands depend only on the
time difference $t$. Hence, the  time integrals in
Eqs.~\eqref{eq:ContinuousIntegrationOperator} and
\eqref{eq:DiscreteIntegrationOperator} may be simplified to
$\int_0^t\rmd \tau\int_0^{\tau}\rmd \tau'\to\int_0^t\rmd t'( t-t')$ \cite{DieballCoarseGraining}. 
\end{widetext}

It remains to consider current-current covariances, which require the correlations of Stratonovich increments [in case of MJP, Stratonovich differentials $\rmd n_{xy}(\tau)$ are defined by analogy; see Eq.~\eqref{eq:DiscreteCurrent}] at different times \cite{DieballCoarseGraining}
\begin{align}
    {\Xi}_3^{\f{z}\f{z}'} &= \frac{\langle \circ \rmd \f{x}_{\tau}\circ \rmd \f{x}_{\tau'}^T\rangle_{\f{x}_{\tau}=\f{z}}^{\f{x}_{\tau'}=\f{z}'}}{\rmd \tau\rmd \tau'} + \frac{\langle \circ \rmd \f{x}_{\tau'}\circ \rmd \f{x}_{\tau}^T\rangle_{\f{x}_{\tau}=\f{z}'}^{\f{x}_{\tau'}=\f{z}}}{\rmd \tau\rmd \tau'}\label{eq:Xi3},\\
    {\Xi}_{ijxy}^3 &=\frac{\langle \rmd n_{xy}(\tau)\rmd n_{ij}(\tau')\rangle_{x_\tau=x}^{x_{\tau'}=i}}{\rmd\tau\rmd\tau'} + \frac{\langle \rmd n_{xy}(\tau')\rmd n_{ij}(\tau)\rangle^{x_{\tau'}=x}_{x_{\tau}=i}}{\rmd\tau\rmd\tau'}.\nonumber
\end{align}
Using again the generalized time-reversal symmetry, we can evaluate these correlations to obtain 
\begin{widetext}
\begin{align}
    \mathrm{cov}({J}_{t, \textcolor{black}{k}}^{\mathrm{c}}, J_{t, \textcolor{black}{l}}^{\mathrm{c}})=& 2t\int\rmd \f{z} (\f{U}^k(\f z))^T \f{D} \f{U}^l(\f z)p_\mathrm{s}(\f z) \nonumber\\&+ 
    \hat{\mathcal{I}}^{t, \mathrm{c}}_{\f{U}^k, \f{U}^l}\left[\hat{\f{j}}_{\f{z}}\left(\hat{\f{j}}^\ddag_{\f{z}'}\right)^T P(\f{z}, t'|\f{z}')p_\mathrm{s}(\f z') + \hat{\f{j}}^\ddag_{\f{z}}\hat{\f{j}}_{\f{z}'}^TP(\f{z}', t'|\f{z})p_\mathrm{s}(\f z) -  \f{j}_\mathrm{s}(\f{z})\f{j}_\mathrm{s}(\f{z}')^T\right]\nonumber\\
    \mathrm{cov}(J_{t, \textcolor{black}{k}}, {J}_{t, \textcolor{black}{l}}) =&  t\sum_{x, y}\kappa_{xy}^k \kappa_{xy}^l r_{xy}p_x^\mathrm{s} + \hat{{I}}^{t}_{\kappa^k, \kappa^l}\left[\hat{j}_{xy}\hat{j}_{ji}^\ddag P(x, t'| j)p_{j}^\mathrm{s} + \hat{j}_{ij}\hat{j}_{yx}^\ddag P(i, t'| y)p_y^\mathrm{s} - J^\mathrm{s}_{xy}J^\mathrm{s}_{ij}\right],\label{eq:CurrentCov}
\end{align}
\end{widetext}
where the analogy between continuous and discrete state spaces is
evident. Note that one can derive the results for covariances in
different, less direct, ways, e.g., using Feynman-Kac and (spectral)
perturbation theory (see \cite{hartichGodecPathProb,Dieball_2023} and references therein).

Equations~\eqref{eq:DensityCov}, \eqref{eq:DensCurCov}, and
\eqref{eq:CurrentCov} have the form of generalized Green-Kubo
relations, connecting (co)variances of pathwise observables to
generalized  (auto)correlation functions, which in the
continuous-space setting have been discussed in
Ref.~\cite{DieballCoarseGraining}. Here we highlight the similarities and differences between discrete space and continuous space.
\begin{figure}[ht!!]
    \centering
     \includegraphics[width=\linewidth]{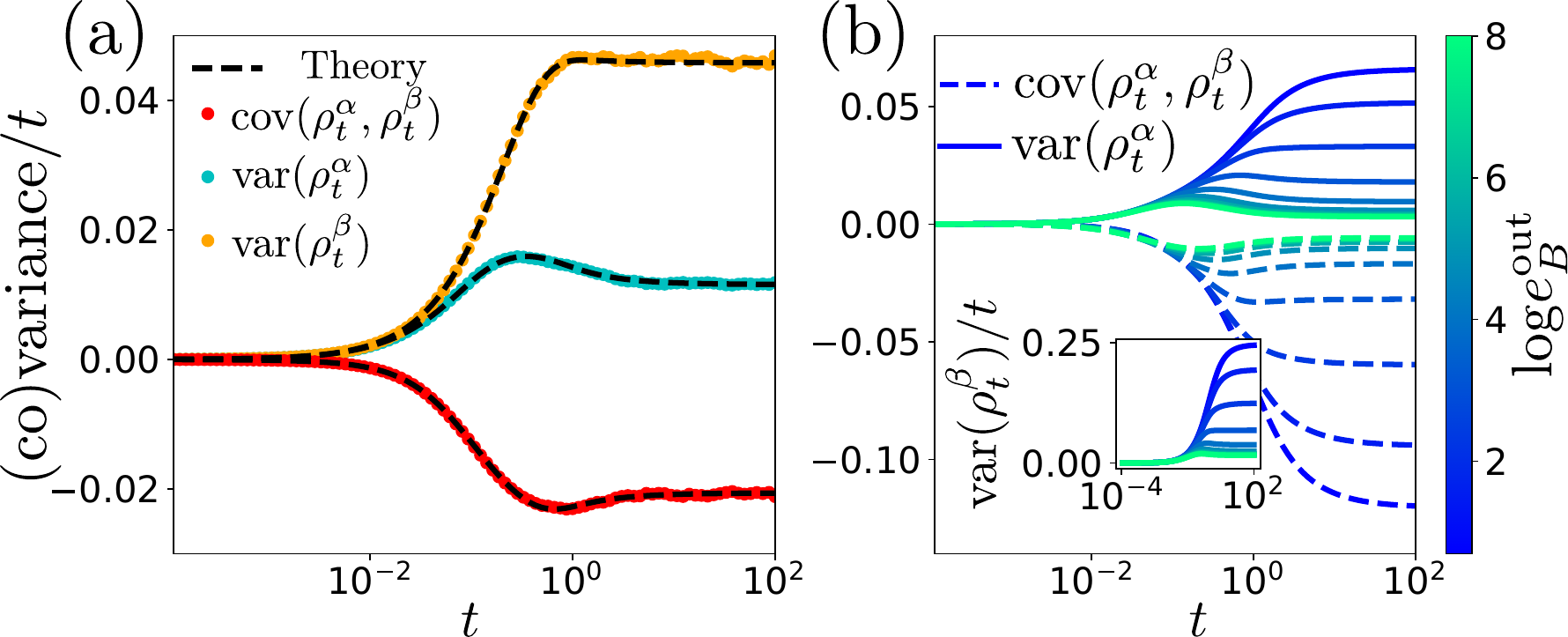}
    \caption[Density variance and covariance in SAT]{Two densities, ${\rho}_t^\alpha$ and ${\rho}_t^\beta$, are used to visualize density (co)variances in the SAT model, see Sec.~\ref{SAT}. These correspond to the state functions $V_i^\alpha = \delta_{i1} + \delta_{i2}$ and $V_i^\beta = \delta_{i5} + \delta_{i6}$ measuring if molecules of type $A$ and $B$ are in the channel, respectively. In (a), the (scaled) variances $\mathrm{var}({\rho}^{\alpha/\beta}_t)/t$ are shown together with the (scaled) covariance between them $\mathrm{cov}({\rho}^{\alpha}_t,{\rho}^{\beta}_t)/t$ from numerical simulations (colored) and analytical solutions Eq.~\eqref{eq:DensityCov} (dashed lines). The analytical variance of ${\rho}_t^\alpha$ (solid lines) and covariance (black dashed lines), both modulated by $1/t$, are shown in (b) for various values of $e^\mathrm{out}_B$. The inset shows $\mathrm{var}({\rho}_t^\beta)/t$. The numerics in (a) are evaluated using $N=10^4$ trajectories sampled using the celebrated Gillespie algorithm \cite{CelebratedGillespie, CelebratedGillespie2}. The initial condition is $p_i=(\delta_{i1}+\delta_{i3}+\delta_{i5})/3$ and the values of the parameters can be found in Tab. \ref{tab:ParameterSAT}. In (b), $e^\mathrm{out}_B = 40$ is used.}
    \label{fig:SAT}
\end{figure}

The mathematical structure of the results is (up to the analogy) the
same, which should not come as a surprise, at least (but not only) in
the sense  that the MJP process can (under appropriate conditions on
the drift 
$\f F$ in Eq.~\eqref{eq:SummaryLangevinEq}, i.e., that the reversible contribution stems from
a generalized potential with deep minima separated by high barriers
\cite{Moro1995,Gaveau_1987,Gaveau1997,LocalDetailedBalanceAcross,Hartich_PRX})
be seen as the long-time coarse-grained limit of the continuous
system.   

The continuous space correlations in
Eqs.~\eqref{eq:Xi1},~\eqref{eq:Xi2}, and~\eqref{eq:Xi3} are valid for
all $t\geq 0$ and similarly for systems with nominally discrete
degrees of freedom (such as spin systems). Conversely, if the MJP is
meant as an approximate, long-time effective 
description of a system with continuous degrees of freedom, 
Eqs.~\eqref{eq:Xi1},~\eqref{eq:Xi2}, and~\eqref{eq:Xi3} described
covariances of pathwise observables at sufficiently long times.   

A non-zero $\mathrm{cov}(J_{t}^\mathrm{c},\rho_{t}^\mathrm{c})$
necessarily indicates that detailed balance is broken \cite{DieballCoarseGraining}. The same observation can be made in the discrete case, as the equilibrium dual operator $\hat{j}_{xy}^{\ddag, \mathrm{eq}} = \hat{j}_{xy}^\mathrm{eq}$ together with $\f{\kappa}=\f{\kappa}(\tau) = -\f{\kappa}^T$ yields $\mathrm{cov}_\mathrm{eq}(J_t,\rho_t) = 0$.

To provide preliminary intuition, density (co)variances (scaled by $t^{-1}$) are shown in Fig.~\ref{fig:SAT} for the
secondary active transport (SAT) model, which
will be discussed in more detail in Sec.~\ref{applications}. In addition, in Fig.~\ref{fig:Histogram}c, we show the mean current $J_t$ arising in the stationary SAT model with the magnitude of the typical fluctuations using the standard deviation, $\sqrt{\mathrm{var}(J_t)}$, and how these can be bounded from below using the stationary thermodynamic uncertainty relation (TUR).

\subsection{Transient Covariances}\label{sec:transCovariances}

The generalizations of Eqs.~\eqref{eq:DensityCov},~\eqref{eq:DensCurCov}, and \eqref{eq:CurrentCov} to transient dynamics has been performed in Ref.~\cite{Dieball_2022_CorrelationsFluctuations} for continuous space dynamics. We give a brief overview of the generalization for MJP in this section. The first difference arises in the integration operator Eq.~\eqref{eq:DiscreteIntegrationOperator}
\begin{align}
    \hat{{I}}^t_{\textcolor{black}{G}^\alpha, \textcolor{black}{G}^\beta}[\cdot] &= \int_0^t\rmd\tau\int_0^t\rmd\tau' \sum_{x,y}\textcolor{black}{G}_{xy}^\alpha(\tau)\sum_{i,j}\textcolor{black}{G}^\beta_{ij}(\tau')[\cdot],
    \label{eq:generalDiscreteIntegrationOperator}
\end{align}
where $U^\alpha$ and $U^\beta$ are explicitly time-dependent. Note that also the integration limits differ, as the explicit time-dependence of the $U$s and probabilities requires the full range of both temporal integrals. It is at this point easier to consider the covariances not in terms of probability flux operators, but rather express them in terms of transition rates. The discrete covariances in Eqs.~\eqref{eq:DensityCov},~\eqref{eq:DensCurCov}, and \eqref{eq:CurrentCov} generalize to
\begin{widetext}
\begin{align}
    \mathrm{cov}({\rho}_{t, \textcolor{black}{k}}, {\rho}_{t, \textcolor{black}{l}}) =& \hat{{I}}^{t}_{V^k\delta, V^l\delta}\left[P(x,\tau;i,\tau') - p_x(\tau)p_i(\tau')\right]\nonumber,
    \\
    \mathrm{cov}({J}_{t, \textcolor{black}{k}}, {\rho}_{t, \textcolor{black}{l}}) =& \hat{{I}}^{t}_{\kappa^k, V^l\delta}\left[\textcolor{black}{\mathbb{1}}_{\tau>\tau'}r_{xy}P(x, \tau| i, \tau')p_{i}(\tau') + \textcolor{black}{\mathbb{1}}_{\tau<\tau'}r_{yx} P(i, \tau'| y, \tau)p_y(\tau)- r_{xy}p_x(\tau)p_i(\tau')\right]\nonumber,
    \\
    \mathrm{cov}(J_{t, \textcolor{black}{k}}, {J}_{t, \textcolor{black}{l}}) =&  \int_0^t\rmd\tau\sum_{x, y}\kappa_{xy}^k(\tau) \kappa_{xy}^l(\tau) r_{xy}p_x(\tau)\nonumber\\&+ \hat{{I}}^{t}_{\kappa^k, \kappa^l}\left[\textcolor{black}{\mathbb{1}}_{\tau>\tau'}r_{xy}r_{ij} P(x, \tau| j, \tau')p_{i}(\tau') + \textcolor{black}{\mathbb{1}}_{\tau<\tau'}r_{ij}r_{xy} P(i, \tau'| y, \tau)p_y(\tau) - r_{xy}p_x(\tau)r_{ij}p_i(\tau')\right]\label{eq:TransientCovariances},
\end{align}
\end{widetext}
For a generalization to time-dependent rates see Sec.~\ref{t-inhomo}, and for time-inhomogeneous diffusion see Ref.~\cite{Dieball_2022_CorrelationsFluctuations}.

\section{Application of Stochastic Calculus to Thermodynamic Bounds}\label{applications}   

\subsection{Model systems}\label{models}

\begin{figure*}[ht]
    \centering
        \includegraphics[width=\linewidth]{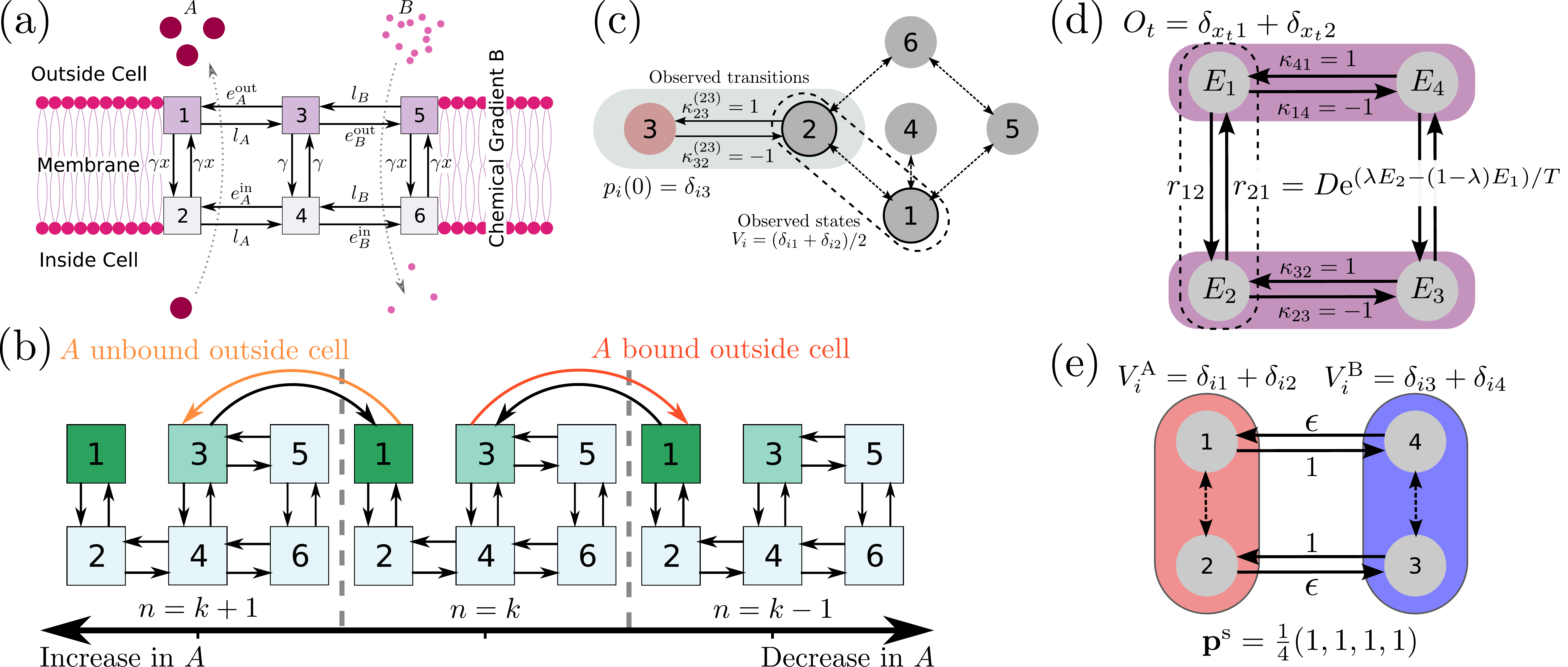}
    \caption{Overview of models used to visualize the results: (a)
      Markov model for secondary active transport (SAT);~the states
      $(1,3,5)$ correspond to the protein funnel being open toward the
      cell exterior while states $(2,4,6)$ describe the funnel
      being open toward the cell interior \textcolor{black}{(see Tab.~\ref{tab:SATstates})}. The transition rate for the
      conversion of the opening from the interior to the exterior
      (and vice versa) is $\gamma$, which in the presence of ligands
      $A$ or $B$  is scaled by a dimensionless factor $x$. $A$ and $B$
      leave the funnel with rate $l_{A/B}$ and enter it with rate
      $\textcolor{black}{e}^{\mathrm{out}/\mathrm{in}}_{A/B}$ outside/inside the
      cell. Of particular interest in the (extended) SAT model is the
      impact of $\textcolor{black}{e}^\mathrm{out}_B$, i.e., the concentration of
      the helper molecule.~(b)~Schematic of how the SAT model in (a) may
      be extended to a ``train'' of SAT models to capture the number of
      $A$ molecules, $n$, transported through the cell membrane.~(c)~Schematic of the Markov model for
      calmodulin folding. We focus on the inference of dissipation by
      observing a current with weights
      $\kappa_{ij}^{(23)}=\delta_{i2}\delta_{j3}-\delta_{j2}\delta_{i3}$
      and/or density with weight $V_i=(\delta_{i1} +
      \delta_{i2})/2$.~(d)~Four-state ring model with equilibrium transition
      rates parameterized by  free
      energies $E_i$ for state $i$, local diffusion coefficient $D$,
      temperature $T$, and  mixing parameter $\lambda$ [see
      Eq.~\eqref{eq:EQrates}], and upon constant driving,
      respectively.
      We are interested in the response to temperature perturbations
      and investigate the effect on a single-time observable $\textcolor{black}{A}_t =
      \delta_{x_t 1} + \delta_{x_t 2}$ and a current with transition weights
      $\kappa_{ij} =
      \delta_{i4}\delta_{j1}-\delta_{i1}\delta_{j4}+\delta_{i3}\delta_{j2}-\delta_{i2}\delta_{j3}$. (e)~4-state
      ring model with constant driving
      having rates 1 in the clockwise and $\epsilon$ in
      countercockwise direction; the stationary distribution is
      independent of $\epsilon$. The model is used for a comparison of
      the quality of thermodynamic bounds in situations where only the
      mesostates A and B, corresponding to states $(1,2)$ and $(3,4)$,
      respectively, are observed without the ability to distinguish
      the respective transition pathways between them.}
    \label{fig:Models}
\end{figure*}

While discussing various applications of the developed
stochastic calculus for MJP we will throughout \textcolor{black}{allude} to biophysically
relevant model systems shown in Fig.~\ref{fig:Models}. 
Thereby we aim mostly at illustrating how to apply
the theoretical results rather than to discuss the respective
biological implications. \textcolor{black}{The three models we introduce in the following sections are the secondary active transport across cell membranes (and an extension of this model), the calmodulin folding dynamics, and a four-state toy model in various settings. }


\subsubsection{Secondary Active Transport}\label{SAT}

\begin{table}[]
    \textcolor{black}{
    \centering
    \caption{Overview of states in SAT model.}
    \begin{tabular}{c l}
    \hline
    State & Meaning \\
    \hline
     1    &  Funnel outside with $A$ bound\\
     2    &  Funnel inside with $A$ bound\\
     3    &  Funnel outside with nothing bound\\
     4    &  Funnel inside with nothing bound\\
     5    &  Funnel outside with $B$ bound\\
     6    &  Funnel inside with $B$ bound\\
     \hline
    \end{tabular}
    \label{tab:SATstates}
    }
\end{table}

We first consider a Markov model of secondary active transport (SAT)
of molecule $A$  through a protein funnel facilitated by molecule
$B$ \cite{Berlaga2021, Berlaga2022, Berlaga2022_2}. The model, shown
in Fig.~\ref{fig:Models}a \textcolor{black}{and listed in Tab.~\ref{tab:SATstates}}, consists of six states; three [i.e.,\ $(1, 3, 5)$]  corresponding to the funnel
being open toward the exterior of the cell and other three
[i.e.,\ $(2, 4, 6)$] open toward the interior of cell. \textcolor{black}{Hence, the state space is $\mathcal{N}=\{1,2,3,4,5,6\}$.}
In the absence
of $A$ or $B$ bound, the funnel changes conformation with rate $\gamma$. Conversely, the crossing
via $A$ or $B$-filled states occurs with $\gamma x$ for a dimensionless
speedup factor $x$. $A$ and $B$ leave or enter
the funnel inside and outside the cell with rates $l_{A/B}$ and  ${\rm
  e}^{\mathrm{out/in}}_{A/B}$, respectively. \textcolor{black}{The adopted parametrization is listed in Tab.~\ref{tab:ParameterSAT} in Appendix~\ref{AppSimPara}.}

We further consider a periodic extension of coupled SAT models to describe the \textcolor{black}{total}
transport of molecules $A$ through the membrane
(Fig.~\ref{fig:Models}b). We leave the rates in
the system unchanged, whereas the state space is extended to
$\mathcal{N}\times\textcolor{black}{\mathbb{Z}}$, with the additional degree of freedom
accounting for the total number of $A$ molecules outside the cell. 

\textcolor{black}{The transport can be split into two categories: symporters and antiporters. These refer to to molecule $A$ being transported parallel or anti-parallel to molecule $B$, respectively. We are interested in the
fraction of the total steady-state entropy production that we can
estimate via the thermodynamic bounds for various
values of transition rates $\textcolor{black}{e}^\mathrm{out}_B$. These transition rates
are assumed proportional to the concentration of $B$ outside the cell
\cite{Berlaga2021}, allowing us to ``control'' the value of $\textcolor{black}{e}^\mathrm{out}_B$, leaving it as free parameter we can tune.}

\subsubsection{Calmodulin Folding Dynamics}\label{calmodulin}

Being a $\mathrm{Ca}^{2+}$ receptor,
calmodulin plays an important role in physiological signaling processes
\cite{CalmodulinIntroductionFrits}. The Markov model for the
calmodulin folding pathway \cite{CalmodulinStigler} is commonly used
as a paradigmatic example 
in, e.g., the study of first passage times \cite{FirstPassageRick} and
stochastic thermodynamics \cite{TURTimeDependentDriving, degunther_2024}. The
respective states represent different conformations of the protein and
the transitions are changes of conformation
\cite{CalmodulinStigler}. Here we adopt the transition rates from
Ref.~\cite{FirstPassageRick} summarized in
Tab.~\ref{tab:Calmodulin_Rates} \textcolor{black}{in Appendix~\ref{AppSimPara}} while the
topology of the folding network is shown in
Fig.~\ref{fig:Models}c. The calmodulin example will be used to gauge
the quality of thermodynamic inference via the derived thermodynamic
inequalities.  

\subsubsection{Four-State Ring Toy Model}\label{4state}

Here we consider, as a minimal example, a four-state model with states
$\{1, 2, 3, 4\}$ on a ring to visualize results. We will choose two
different parameterizations of the model, visualized in Figs.~\ref{fig:Models}d and \ref{fig:Models}e.

In the equilibrium (detailed-balance) setting, we assign to each state
a free energy $E_i$, $i\in\{1,2,3,4\}$, such that the stationary
probability is $p_i^\mathrm{eq}\propto\rme^{-E_i/T}$ (note that we
explicitly include the temperature here). For any $\lambda\in[0,1]$, the following transition rates obey detailed balance \cite{Maes2008}
\begin{align}
    r_{xy}=D\rme^{(\lambda E_x - (1-\lambda)E_y)/T}.
    \label{eq:EQrates}
\end{align}
We use the free energies {listed in Tab.~\ref{tab:FreeEnergiesRing} \textcolor{black}{in Appendix~\ref{AppSimPara}}.}

In the constant driving setting, we set the clockwise rates to unity
whereas the counterclockwise rates are controlled by the parameter
$\epsilon$. This way the steady-state distribution
$p_i^\mathrm{s}=p=1/4$ is independent of $\epsilon$.

\subsection{Thermodynamic inequalities and inference}\label{bounds}

To highlight the power of the developed stochastic calculus for
observables of MJP  we now apply it to \emph{directly} prove a
broad selection of thermodynamic inequalities.  
Throughout we draw comparisons to continuous-space results and
highlight some fundamental differences between the two. We immediately
gauge the quality of the bounds by means of model examples and discuss
the respective saturation conditions. 
The section thus entails a brief discussion of the crucial steps in
the proofs of the thermodynamic bounds for diffusions. 

\subsubsection{Entropy and Pseudo-Entropy Production}
The central object of interest for thermodynamic inference is the mean
total entropy production, which consist of the entropy change in the
system as well as the surrounding medium. For continuous systems, the
total entropy production (EP) is \cite{Seifert_2012, SeifertEntropyProd,Seifert_2025}
\begin{align}
    \Delta S_\mathrm{tot}^\mathrm{c}(t) = \int_0^t\rmd\tau\int\rmd\f{x}\frac{\f{j}(\f{x},\tau)^T\f{D}(x)\f{j}(\f{x}, \tau)}{P(\f{x}, \tau)}\,.
    \label{eq:ContinuousEP}
\end{align}
In the direct proofs in Refs.~\cite{DirectTUR, dieball2024thermodynamic}, the stochastic integral
\begin{align}
    A_t^\mathrm{c} \equiv \StoInt \frac{\f{j}(\f{x}_\tau, \tau)\textcolor{black}{^T}}{P(\f{x}_\tau, \tau)}\left[2\f{D}(\f{x}_\tau)\right]^{-1}\f{\sigma}(\f{x}_\tau)\rmd\f{W}_\tau,
\end{align}
is a central object, chosen to immediately yield $\langle
(A_t^\mathrm{c})^2\rangle = \Delta S_\mathrm{tot}^\mathrm{c}(t)/2$ which
allows for crucial simplifications in the proofs.

Conversely, in discrete-state systems the total entropy production reads \cite{ThreeFacesI,SeifertEntropyProd,Seifert_2025}
\begin{align}
    \Delta S_\mathrm{tot}(t) = \frac{1}{2}\int_0^t\rmd\tau \sum_{i,j} (r_{ij}p_i(\tau)- r_{ji}p_j(\tau)\log\frac{r_{ij}p_i(\tau)}{r_{ji}p_j(\tau)}.
    \label{eq:EP}
\end{align}
As a key quantity in the various proofs of thermodynamic bounds we introduce the auxiliary integral
\begin{align}
    A_t &= \StoInt \mathrm{Tr}[\textcolor{black}{\f Z(\tau)}^T\rmd\f{\varepsilon}(\tau)],
    \label{eq:AuxInt}
\end{align}
where
\begin{align}
    (\textcolor{black}{\f Z(\tau)})_{ij} = \frac{r_{ij}p_i(\tau) - r_{ji}p_j(\tau)}{r_{ij}p_i(\tau) + r_{ji}p_j(\tau)}.\label{ZMatrix}
\end{align}
Using Eq.~\eqref{eq:ListExpectations} the mean is $\langle A_t\rangle
= 0$ and the variance can be identified as a ``pseudo entropy
production'' \cite{Shiraishi2021, Keiji_SL} 
\begin{align}
    \langle A_t^2\rangle = \frac{1}{2}\int_0^t\rmd\tau \sum_{i,j}\frac{\left[r_{ij}p_i(\tau) - r_{ji}p_j(\tau)\right]^2}{r_{ij}p_i(\tau) + r_{ji}p_j(\tau)}.
    \label{eq:PseudoEntopyPoduction}
\end{align}
Applying the logarithmic inequality $(a-b)^2/(a+b)\leq (a+b)
\log(a/b)/2$ $\forall a,b>0$, the total entropy production
Eq.~\eqref{eq:EP} can be bounded from below using
Eq.~\eqref{eq:PseudoEntopyPoduction}, i.e., $\langle A_t^2\rangle\leq
\Delta S_\mathrm{tot}(t)/2$. The logarithmic inequality is saturated
only for $a=b$ where the right hand side vanishes, and therefore (for
discrete state spaces)  $\langle A_t^2\rangle=
\Delta S_\mathrm{tot}(t)/2$ only at equilibrium where $\Delta
S_\mathrm{tot}(t)=0$. 
 
Note that the inability to saturate $\langle A_t^2\rangle\leq \Delta
S_\mathrm{tot}(t)/2$ outside of equilibrium
is the first fundamental difference between discrete and continuous
systems. The former bounds the pseudo-entropy production (which in
turn bounds the entropy production) and the
latter directly the total entropy production. 
As a consequence, when discussing the
saturation of bounds in discrete space we refer to lower bounds
on the pseudo-entropy production, not the total entropy
production. However, we will show in Sec.~\ref{sec:ContinuousLimit},
if the continuum limit exist the discrete
pseudo-entropy production in Eqs.~\eqref{eq:PseudoEntopyPoduction} and
\eqref{eq:EP} both converge to continuous entropy production in
Eq.~\eqref{eq:ContinuousEP}, which may be saturated far from
equilibrium \cite{DirectTUR}.

\subsubsection{General TUR and Correlation TUR}\label{cTUR}
In the direct proof of the TUR in Refs.~\cite{DirectTUR,
  kwon2024unifiedframeworkclassicalquantum}, the time-integrated
current is split as $J_t^\mathrm{c} = J_t^\mathrm{c,I}+J_t^\mathrm{c, II}$, with
\begin{align}
    J_t^\mathrm{c, I} &= \StoInt \f{U}(\f{x}_\tau, \tau)^T\f{\sigma}\rmd \f{W}_\tau\nonumber,\\
    J_t^\mathrm{c, II} &= \int_0^t {U}(\f{x}_\tau, \tau)\rmd \tau,
\end{align}
where ${U}(\f{x}_\tau, \tau) = \f{U}(\f{x}_\tau, \tau)^T \f{F}(\f{x}_\tau) + \nabla \cdot \left[\f{D}\f{U}(\f{x}_\tau, \tau)\right]$. 

The proof of the transient TUR for MJP, which can be found, e.g., in
Ref.~\cite{kwon2024unifiedframeworkclassicalquantum}, is analogous to
the continuous case \cite{DirectTUR}. We will reiterate the proof
here, focusing on the use of stochastic calculus, as well as including
densities to obtain the correlation TUR (CTUR).

We start from Eq.~\eqref{eq:Matrix_eom_MJP} and split the current in
Eq.~\eqref{eq:DiscreteCurrent} as
\begin{align}
    J_t  &= J_t^\mathrm{I} + J_t^\mathrm{II},
\end{align}
where
\begin{align}
    J_t^\mathrm{I} &= \StoInt \mathrm{Tr}[\f{\kappa}(\tau)^T\rmd\f{\varepsilon}(\tau)],\nonumber\\
    J_t^\mathrm{II} &=\int_0^t \mathrm{Tr}[\f{\kappa}(\tau)^T\f{R}(x_\tau)]\rmd\tau,
\end{align}
which are a stochastic and a ``usual'' integral
respectively. Moreover, we have $\langle J_t^\mathrm{I}\rangle=0$ and
\begin{align}
    \langle J_t^\mathrm{II}\rangle = \langle J_t\rangle = \int_0^t\rmd\tau \sum_{x,y}\kappa_{xy}(\tau)r_{xy}p_x(\tau).
    \label{eq:MeanCurrent}
\end{align}
To prove the TUR we use the Cauchy-Schwarz inequality
\begin{align}
  \langle A_t(J_t - \langle J_t \rangle)\rangle^2\equiv   \langle A_t\Delta J_t\rangle^2\leq \mathrm{var}(J_t)\langle A_t^2\rangle.    \label{eq:CSI}
\end{align}
One can easily show that $\langle A_t J_t^\mathrm{I}\rangle = \langle
J_t\rangle$. Evaluating 
\begin{align}
    \langle& A_t J_t^\mathrm{II}\rangle \nonumber\\
    =&\int_0^t\rmd\tau^\prime \int_0^t \rmd\tau \sum_{i,j}\kappa_{ij}(\tau^\prime)r_{ij}\textcolor{black}{\mathbb{1}}_{\tau < \tau^\prime} \label{eq:AJ2correlator}\\&\times\sum_{x,y}\left[P(i, \tau^\prime|y, \tau) - P(i, \tau^\prime|x, \tau)\right]p_x(\tau) r_{xy}Z_{xy}(\tau)\nonumber,
\end{align}
requires Eq.~\eqref{eq:NoiseTimeCorrelationLemma} and two integrations
by part (for details see App.~\ref{sec:Current2AcorrelationExplicit}),
which yield
\begin{align}
    \langle A_t J_t^\mathrm{II}\rangle = (t\partial_t - 1)\langle J_t\rangle - \langle \tilde{J}_t\rangle,
    \label{eq:AJ2correlatorFinal}
\end{align}
where we introduced the modified current
\begin{align}
    \tilde{J}_t &\equiv \StoInt \tau\mathrm{Tr}[\left(\partial_\tau\f{\kappa}(\tau)^T\right)\rmd\f{n}(\tau)],
    \label{eq:ModifiedCurrent}
\end{align}
that accounts for the change of $\f{\kappa}(\tau)$ in time $\tau$. We
thus proved the transient TUR 
\begin{align}
    \frac{\left[t\partial_t\langle J_t\rangle  - \langle \tilde{J}_t\rangle \right]^2}{\mathrm{var}(J_t)}\leq \langle A_t^2\rangle \leq\frac{\Delta S_\mathrm{tot}(t)}{2}.
    \label{eq:TransientTUR}
\end{align}
As in the continuous case \cite{DirectTUR}, the modified current is necessary
for the validity of Eq.~\eqref{eq:TransientTUR}.

With slight modifications Eq.~\eqref{eq:TransientTUR} also holds for
time-dependent driving as proven in
Ref.~\cite{TURTimeDependentDriving} using scaled cumulant generating
  functions. In Ref.~\cite{kwon2024unifiedframeworkclassicalquantum},
the same TUR was proven using a stochastic calculus
approach. Note that Eq.~\eqref{eq:TransientTUR} is a special case of
the results in Refs.~\cite{TURTimeDependentDriving,
  kwon2024unifiedframeworkclassicalquantum} for systems with constant
rates and $\f{\kappa}(v\tau)$ with $v=1$ as the protocol velocity. In
other words, the modified current in Eq.~\eqref{eq:ModifiedCurrent}
corresponds to $v\partial_v J_t|_{v=1}$ where the generator is
constant in time.

As shown in Ref.~\cite{DirectTUR}, one needs to include a density to
saturate Eq.~\eqref{eq:TransientTUR}. By means of a calculation that
is analogous to the evaluation of density-current correlations in Sec.~\ref{sec:transCovariances} we here further show that
\begin{align}
    \langle A_t\rho_t\rangle &= (t\partial_t - 1)\langle \rho_t\rangle - \langle \tilde{\rho}_t\rangle,\nonumber\\
    \tilde{\rho}_t &\equiv \StoInt \tau\partial_\tau\textcolor{black}{{V}_\tau \rmd{\tau}}.
\end{align}
By shifting the current $\Delta J_t\to \Delta J_t - c(t)\Delta \rho_t$
with some $c:\textcolor{black}{\mathbb{R}}\to\textcolor{black}{\mathbb{R}}$ in Eq.~\eqref{eq:CSI} already yields the transient correlation TUR (CTUR)
\begin{align}
    \frac{\left[t\partial_t\langle J_t\rangle  - \langle \tilde{J}_t\rangle - c(t)\left((t\partial_t-1)\langle \rho_t\rangle - \langle \tilde{\rho}_t\rangle\right)\right]^2}{\mathrm{var}(J_t - c(t)\rho_t)}\leq \frac{\Delta S_\mathrm{tot}(t)}{2}.
    \label{eq:TransientCTUR}
\end{align}
This route to proving thermodynamic bounds may thus legitimately
called be referred to as
\emph{direct} and allows to discuss saturation in a straightforward
manner, as we show next. 

The l.h.s. of Eq.~\eqref{eq:TransientCTUR} corresponds to the
estimated total entropy production $\Sigma_\mathrm{est}(t)$. To
effectively visualize how much of the total entropy produce is
recovered in the estimate we use the quality factor
\begin{align}
    \mathcal{Q} = \frac{\Sigma_\mathrm{est}(t)}{\Delta S_\mathrm{tot}(t)}.
\end{align}
For the TUR, CTUR, and TB, the quality factor is bounded by $0\leq
\mathcal{Q}\leq 1$. We will later show that the correlation bound the quality factor only is trivially bounded from above $ \mathcal{Q}^\mathrm{CB}\leq 1$.

\subsubsection{Optimization of TUR and CTUR}\label{optim}

There are two ways to optimize Eq.~\eqref{eq:TransientCTUR}, either by
saturating Eq.~\eqref{eq:CSI} or by choosing the optimal $c(t)$ for
fixed $J_t$ and $\rho_t$. The former requires $\Delta J_t - c(t)\Delta\rho_t\propto
A_t$. Since $c(t)\neq 0$, we see that the saturation is achieved when
$V_i(\tau) = \hat{\f{e}}_i^T \f{\kappa}(\tau)\f{L}\hat{\f{e}}_i/c(t) =
\sum_j \kappa_{ij}(\tau)r_{ij}/c(t)$ and $\f{\kappa} = c' \f{Z}$ for
any $c'\in \textcolor{black}{\mathbb{R}}$. Here, $\hat{\f{e}}_i$ is the $i$th unit
vector. However, these choices do \emph{not} saturate
Eq.~\eqref{eq:TransientCTUR} but only the estimate for
the pseudo-entropy production [see first inequality in
Eq.~\eqref{eq:TransientTUR}]. In other words, the ``saturated'' CTUR
quality factor reads 
\begin{align}
    \mathcal{Q}^\mathrm{cTUR}_{\mathrm{sat}} = \frac{2\langle A_t^2 \rangle }{\Delta S_\mathrm{tot}(t)}\leq 1.
    \label{eq:SaturatedTURQ}
\end{align}
We again highlight, that $\mathcal{Q}^\mathrm{CTUR}_{\mathrm{sat}}=1$
in equilibrium or (as we show below) in the continuum limit (if
said limit exists). Moreover, the optimization requires the rates
and transient probability distribution to be known,  which makes
the saturated bound superfluous. Namely, knowing the rates and the transient probability distribution allows the EP to be determined exactly.

The saturation approach shown here is analogous to
the approach for continuous space dynamics \cite{DirectTUR} as it
involves choosing $V(\f{x}_\tau, \tau)$ such that
$c(t)\rho_t^\mathrm{c}=J_t^\mathrm{c, II}$. Then, the vector-valued
function $\f{U}(\f{x}_\tau, \tau)$ needs to be chosen proportional to the integrand
in $A_t^\mathrm{c}$.

\begin{figure}
    \centering
      \includegraphics[width=.8\linewidth]{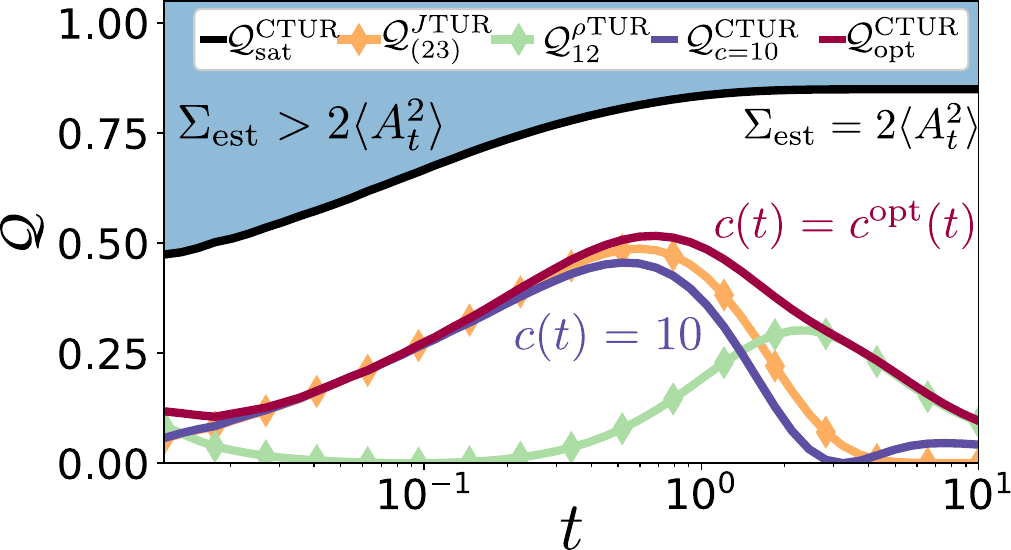}
    \caption{Saturation of CTUR in the transient calmodulin folding dynamics (see Fig.~\ref{fig:Models}c). We present the various CTUR quality factors $\mathcal{Q}$ arising from various combinations of density, $\rho_{\textcolor{black}{\tau}}$, and current, $J_t$, as a function of time. The black line denotes the saturation of the TUR, i.e., when $\f{\kappa}(\tau) = \textcolor{black}{\f Z(\tau)}$ and $c(t)\rho_t = J_t^\mathrm{II}$, corresponding to the estimated entropy production being the pseudo-entropy production $\Sigma_\mathrm{est}=2\langle A_t^2\rangle$. Above the black line is the region $\Sigma_\mathrm{est}>2\langle A_t^2\rangle$ which is not accessible using the (C)TUR. The remaining quality factors shown use the current ($\mathcal{Q}_{(23)}^{J\mathrm{TUR}}$), density ($\mathcal{Q}_{12}^{\rho\mathrm{TUR}}$), or a combination of these ($\mathcal{Q}_{c=10}^{\mathrm{CTUR}}$ and $\mathcal{Q}_{\mathrm{opt}}^{\mathrm{CTUR}}$) in orange, green, purple, and red, respectively. For the CTURs, the proportionality $c(t) = 10$ (purple) and $c(t) = c^{\textcolor{black}{\mathrm{opt}}}(t)$ (red) are chosen, the latter optimizing the quality factor for given $J_t$ and $\rho_t$. The rates can be found in Tab.~\ref{tab:Calmodulin_Rates}.}
    \label{fig:TURsaturation}
\end{figure}

\begin{figure}
    \centering
      \includegraphics[width=.7\linewidth]{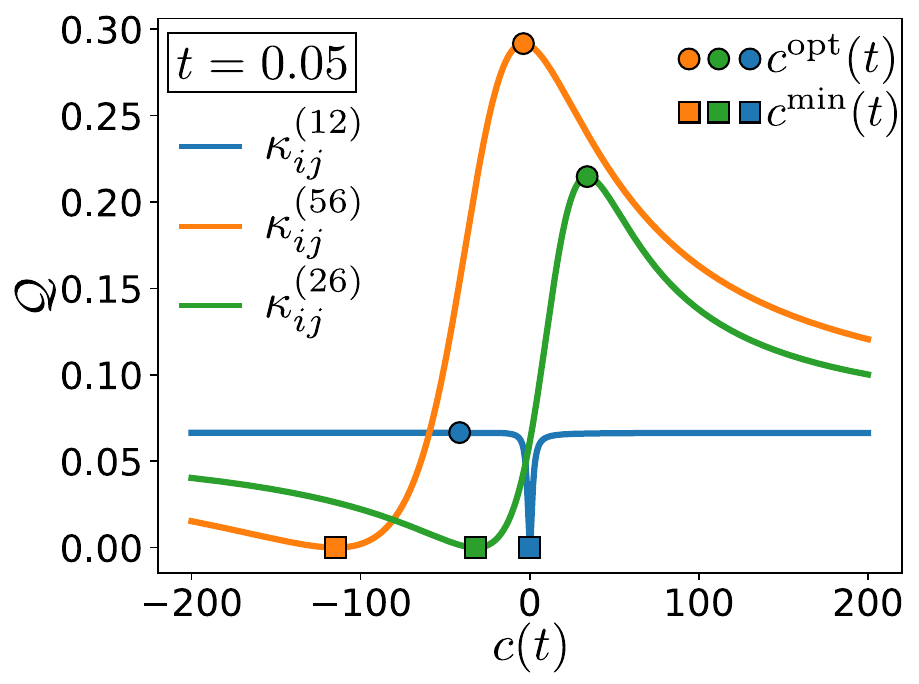}
    \caption{Quality factor as a function of density weighting function $c(t)$ in the calmodulin system at a fixed $t=0.05$. The color correspond to different transition weights $\kappa_{ij}^{(kl)}=\delta_{ik}\delta_{jl}-\delta_{il}\delta_{jk}$ between two states, i.e., $\kappa_{ij}^{(12)}$ (blue), $\kappa_{ij}^{(56)}$ (orange), and $\kappa_{ij}^{(26)}$ (green). The state function entering the density is $V_i(\tau) = \delta_{i3}$ and the initial condition is $p_i(0) = \delta_{i6}$. The circles are the quality factors for most optimal $c(t) = c^{\textcolor{black}{\mathrm{opt}}}(t)$, while the squares denote the least optimal $c(t) = c^{\textcolor{black}{\mathrm{min}}}(t)$.}
    \label{fig:optimal_c}
\end{figure}

Hence, it seems to be more practical to optimize the bound w.r.t. $c(t)$ for a given $J_t$ and $\rho_t$. The l.h.s.\ of Eq.~\eqref{eq:TransientCTUR}, which we denote $h(t)$, has two extrema in $c(t)$
\begin{align}
    c^{\textcolor{black}{\mathrm{min}}}(t) &= \frac{a(t)}{b(t)}\,,\nonumber\\
    c^{\textcolor{black}{\mathrm{opt}}}(t) &= \frac{a(t)\mathrm{cov}({\rho}_t, {J}_t) - b(t)\mathrm{var}({J}_t)}{a(t)\mathrm{var}({\rho}_t) - b(t)\mathrm{cov}({\rho}_t, {J}_t)}\,,
    \label{eq:optimal_c}
\end{align}
where  $a(t)\equiv t\partial_t\langle {J}_t\rangle -
\langle{\tilde{J}}_t\rangle$ and $b(t) \equiv (t\partial_t-1)\langle
       {\rho}_t\rangle  - \langle{\tilde{\rho}}_t\rangle$. All
       quantities entering Eq.~\eqref{eq:optimal_c} are accessible
       from $J_t$ and $\rho_t$, making the optimization practically feasible. Note that $c^{\textcolor{black}{\mathrm{min}}} (t)$ and $c^{\textcolor{black}{\mathrm{opt}}}(t)$ correspond to the minimum and maximum of $h(t)$, respectively, as 
\begin{align}
    \frac{\partial^2}{\partial c^2}h(c, t)|_{c(t) = c^{\textcolor{black}{\mathrm{min}}}(t)} &= \frac{4b(t)^2}{\mathrm{var}\left({J}_t - \frac{a(t)}{b(t)}\rho_t\right)}\geq 0\,,\\
    \frac{\partial^2}{\partial c^2}h(c, t)|_{c(t) = c^{\textcolor{black}{\mathrm{opt}}}(t)} &= \frac{-4b(t)^2\mathrm{var}\left({J}_t - \frac{a(t)}{b(t)}\rho_t\right)}{\left(\mathrm{var}\left({J}_t - c^{\textcolor{black}{\mathrm{opt}}}(t)\rho_t\right)\right)^2}\leq 0\,.\nonumber
\end{align}
Moreover, one can immediately see that using $c^{\textcolor{black}{\mathrm{min}}}(t)$ one
recovers the second law of thermodynamics, as $t\partial_t\langle
J_t\rangle  - \langle \tilde{J}_t\rangle -
c^{\textcolor{black}{\mathrm{min}}}(t)\left[(t\partial_t-1)\langle \rho_t\rangle - \langle
\tilde{\rho}_t\rangle\right]=0$. 
The optimal quality factor therefore reads
\begin{align}
    &\mathcal{Q}^\mathrm{CTUR}_\mathrm{opt} \label{eq:optimalTURQ}\\&= \frac{2\left[a(t)^2 \mathrm{var}(\rho_t) - 2a(t)b(t)\mathrm{cov}(J_t, \rho_t) + b(t)^2\mathrm{var}(J_t)\right]}{\left[\mathrm{var}(J_t)\mathrm{var}(\rho_t) - \mathrm{cov}(J_t,\rho_t)^2\right]\Delta S_\mathrm{tot}(t)}.\nonumber
\end{align}
The saturated quality factor $\mathcal{Q}_\mathrm{sat}^\mathrm{CTUR}$
in  Eq.~\eqref{eq:SaturatedTURQ} is shown in
Fig.~\ref{fig:TURsaturation} together with the optimal
$\mathcal{Q}^\mathrm{CTUR}_\mathrm{opt}$ and the quality factors of
the current TUR $\mathcal{Q}^{J\mathrm{TUR}}_{(23)}$, density TUR
$\mathcal{Q}^{\rho\mathrm{TUR}}_{12}$, and a CTUR
$\mathcal{Q}^{\mathrm{CTUR}}_{c=10}$ with $c(t)=10$ for the calmodulin
system (see Fig.~\ref{fig:Models}c). We consider
the transition weights $\kappa_{ij}^{(23)} = \delta_{i2}\delta_{j3} -
\delta_{i3}\delta_{j2}$ and state function $V_i=(\delta_{i1} +
\delta_{i2})/2$. Only the saturated quality factor remains finite for
$t\to\infty$. As expected, the optimal
$\mathcal{Q}^\mathrm{CTUR}_\mathrm{opt}$ gives the best inferred
entropy production when $\f{\kappa}(\tau)$ and \textcolor{black}{${V}_\tau$} do \emph{not}
saturate the TUR, whereas
$\mathcal{Q}^{\mathrm{CTUR}}_{c=10}\leq\mathcal{Q}^{J\mathrm{TUR}}_{(23)}$
for $t\leq3$. Hence, simply using an arbitrary $c(t)$ in the CTUR may
in fact lead to a
worse estimate for the entropy production compared to the current
TUR. The non-trivial behavior of $c(t)$ is also highlighted in
Fig.~\ref{fig:optimal_c}. 
The expressions of $c^{\textcolor{black}{\mathrm{opt}}}(t)$ and $c^{\textcolor{black}{\mathrm{min}}}(t)$
Eqs.~\eqref{eq:optimal_c} can be both positive and
negative. Furthermore, they are in no particular relation to another,
i.e., one cannot easily say if and when $c^{\textcolor{black}{\mathrm{opt}}}(t)<c^{\textcolor{black}{\mathrm{min}}}(t)$ and vice
versa. A more detailed discussion is given in App.~\ref{sec:optimal_c_cTUR}.

\subsection{Thermodynamic Transport Bound}

The thermodynamic bound on the transport [or short
``transport bound'' (TB)] of a scalar observable $z_\tau =
z(\f{x}_\tau,\tau)$   in continuous space was derived in
\cite{dieball2024thermodynamic, Leighton2022} and reads
\begin{align}
    \frac{\langle z_t - z_0 - \int_0^t\rmd\tau \partial_\tau z_\tau\rangle^2}{t\mathcal{D}_t^\mathrm{c}}\leq \frac{\Delta S_\mathrm{tot}^\mathrm{c}(t)}{2},
    \label{eq:continuousTB}
\end{align}
where $\mathcal{D}_t^\mathrm{c}$ is the continuous-space fluctuation
scale function
\begin{align}
    \mathcal{D}_t^\mathrm{c} \equiv\frac{1}{t}\int_0^t\rmd \tau \langle [\nabla_{\f{x}}z_\tau]^T \f{\gamma}^{-1}(\tau)\nabla_{\f{x}}z_\tau\rangle,
    \label{eq:continuousD}
\end{align}
that accounts for how much $z(\f{x}_\tau,\tau)$
``stretches'' microscopic coordinates $\f{x}_\tau$ \cite{dieball2024thermodynamic}.
Here, $\f{\gamma}$ is a positive definite and symmetric friction
matrix that is allowed to depend on time and the averages $\langle
\cdot \rangle$ in Eqs.~\eqref{eq:continuousTB} and
\eqref{eq:continuousD} are taken over transient trajectories. In
Refs.~\cite{dieball2024thermodynamic, BoundsCorrelationTimes} it was
further shown how $\mathcal{D}_t^\mathrm{c}$ relates to the short-time
fluctuations $\rmd z_\tau = z_{\tau+\rmd \tau} - z_\tau$, i.e., 
\begin{align}
    \mathcal{D}_t^\mathrm{c} = \int_{0}^t\rmd\tau \frac{\mathrm{var}(\rmd z_\tau)}{2\rmd \tau}.
\end{align} 
To prove the discrete-space transport bound we use Eq.~\eqref{eq:AuxInt} and the stochastic integral
\begin{align}
    B_t = \StoInt \mathrm{Tr}[\f{H}(\tau)\rmd\f{\varepsilon}(\tau)],\label{Bfunctional}
\end{align}
where $[\f{H}(\tau)]_{ij} = z_i(\tau) - z_j(\tau)$ is the difference
of some observable state function
$z_\tau\equiv z(x_\tau,\tau)=\sum_{k}\textcolor{black}{\delta_{x_\tau k}}z_k(\tau)$. The
mean of $B_t$ vanishes while the second moment reads
\begin{align}
    \langle B^2_t\rangle = \int_0^t\rmd\tau \sum_{i,j}\left[z_i(\tau)-z_j(\tau)\right]^2r_{ij}p_i(\tau),
\end{align}
which we can identify with the short-time fluctuation function
$\mathcal{D}_t = \langle B^2_t\rangle/t$. Namely, consider  the
infinitesimal difference $\rmd
z(\tau) = z_{\tau + \rmd \tau} - z_\tau$ along the trajectory. Then for any
$n\in\textcolor{black}{\mathbb{N}}_{>0}$ we then get $\langle [\rmd z(\tau)]^n \rangle =
\sum_{x,y} (z_y-z_x)^n r_{xy}p_x(\tau)\rmd \tau + \mathcal{O}(\rmd
\tau ^2)$. Thus, $\mathrm{var}[\rmd z(\tau)] = \sum_{x,y} (z_y-z_x)^2
r_{xy}p_x(\tau)\rmd \tau + \mathcal{O}(\rmd \tau ^2)$ and 
\begin{align}
    \mathcal{D}_t = \frac{1}{t}\int_{0}^t \mathrm{var}[\rmd z(\tau)]\,.
\end{align}
After performing an integration by parts, the cross correlation
between $A_t$ [given in Eq.~\eqref{eq:AuxInt}] and $B_t$ is
\begin{align}
    \langle A_tB_t\rangle &= \sum_i \left[z_i(t)p_i(t) - z_i(0)p_i(0) - \int_0^t\rmd\tau \partial_\tau z_i(\tau) p_i(\tau)\right]\nonumber\\
    &=\langle z_t -z_0 - \int_0^t\rmd\tau \partial_\tau z_\tau\rangle.
\end{align}
The average of $z_t$ is given by $\langle
z_\tau\rangle =\sum_k\sum_l\delta_{kl}z_k(\tau)p_l(t)$. By combining the various
averages entering the Cauchy-Schwarz inequality $\langle
A_tB_t\rangle\le\langle A_t^2\rangle\langle B_t^2\rangle$  we obtain the
transport bound for MJP
\begin{align}
    \frac{\langle z_t - z_0 - \int_0^t\rmd\tau \partial_\tau z_\tau\rangle^2}{t\mathcal{D}_t}\leq \frac{\Delta S_\mathrm{tot}(t)}{2}.
    \label{eq:TB}
\end{align}
Upon identifying the dynamical activity $\int_0^t\rmd\tau
\sum_{x,y\ne x}r_{xy}p_x(\tau)$
\cite{ThermodynamicCorrelationInequality, Ohga2023}, the
fluctuation-scale function $\mathcal{D}_t $ can be seen as a
$[z_y(\tau) - z_x(\tau)]^2$-weighted dynamical activity, which
justifies referring to $\mathcal{D}_t$ as the diffusion constant of
$z_\tau$ on a graph.

\subsubsection{Saturation of the Transport Bound}

While the continuous TB in Eq.~\eqref{eq:continuousTB} can be
saturated, however not uniquely, with $\nabla_{\f{x}}z_\tau = c\f{\gamma}\f{j}(\f{x}, \tau) /
P(\f{x}, \tau)$ (see Ref.~\cite{dieball2024thermodynamic} for
details), there is \emph{no} general way to saturate the discrete
TB. We provide a simple counterexample to show why this is the
case. Consider some $a,b,c>0$ and let
\begin{align}
    \f{L} = 
    \begin{pmatrix}
        -1-a & b & 1\\
        1 & -1-b & c\\
        a & 1 & -1-c
    \end{pmatrix},
\end{align}
be the generator for a fully connected three state MJP with states
$\{1,2,3\}$. Saturation of Eq.~\eqref{eq:TB} entails choosing
$z_\tau$ such that $\f{H}(\tau)^T\propto \textcolor{black}{\f Z(\tau)}$ to saturate
the Cauchy-Schwarz inequality. Explicitly, for all $\tau$
\begin{align}
    z_2(\tau) - z_1(\tau) &\overset{!}{=} Z_{12}(\tau) =  \frac{p_1(\tau) - bp_2(\tau)}{p_1(\tau) + bp_2(\tau)}\,,\nonumber\\
    z_3(\tau) - z_1(\tau) &\overset{!}{=} Z_{13}(\tau) =  \frac{ap_1(\tau) - p_3(\tau)}{ap_1(\tau) + p_3(\tau)}\,,\nonumber\\
    z_3(\tau) - z_2(\tau) &\overset{!}{=} Z_{23}(\tau) =  \frac{p_2(\tau) - cp_3(\tau)}{p_2(\tau) + cp_3(\tau)}\,.
    \label{eq:SaturationTBCounter}
\end{align}
A solution of Eqs.~\eqref{eq:SaturationTBCounter} involves $Z_{23}(\tau) = Z_{13}(\tau) - Z_{12}(\tau)$, i.e.,
\begin{align}
    \frac{p_2(\tau) - cp_3(\tau)}{p_2(\tau) + cp_3(\tau)} = \frac{2p_1(\tau)(abp_2(\tau) - p_3(\tau))}{(ap_1(\tau) + p_3(\tau))(p_1(\tau) + bp_2(\tau))}.
    \label{eq:SaturationTBCondition}
\end{align}
Even for $abc=1$, which corresponds to a system relaxing towards
equilibrium, we can always choose an initial condition $\f{p}(\tau=0)$
such that the equality in Eq.~\eqref{eq:SaturationTBCondition} is
not achieved. For example, let $a=b=c=2$ and $p_2(0) = p_3(0)=1/2$. The
l.h.s. of Eq.~\eqref{eq:SaturationTBCondition} yields $-1/3$ while the
r.h.s. is $0$, hence the equality is not achieved and the set of equations
Eqs.~\eqref{eq:SaturationTBCounter} has no solution. Thus, there is no
state function $z_\tau$ which saturates the TB.

\subsubsection{Unification of (C)TUR and TB}\label{unification}
It turns out that the TB is a special case of the CTUR. Starting from the transient CTUR \eqref{eq:TransientCTUR}, we now choose $\rho_t$ such that $J_t - c(t)\rho_t = J^\mathrm{I}_t$, i.e., $V_i(\tau) =
\sum_j \kappa_{ij}(\tau)r_{ij}/c(t)$. Then,
\begin{align}
    t\partial_t \langle \rho_t\rangle &= t\sum_i V_i(t)p_i(t)\nonumber\\
    &=t \sum_{i,j\neq i}\frac{\kappa_{ij}(\tau)r_{ij}}{c(t)}p_i(t)\nonumber\\
    &=\frac{1}{c(t)}t\partial_t\langle J_t\rangle,
\end{align}
and similarly
\begin{align}
    \langle\tilde{\rho}_t\rangle &= \int_0^t\rmd \tau\sum_i \tau\partial_\tau V_i(\tau) p_i(\tau)\nonumber\\
    &=  \int_0^t\rmd \tau\sum_{i,j\neq i} \frac{\tau\partial_\tau\kappa_{ij}(\tau)r_{ij}}{c(t)} p_i(\tau)\nonumber\\
    &= \frac{1}{c(t)}\int_0^t\rmd \tau\sum_{i,j\neq i} \tau\partial_\tau\kappa_{ij}(\tau)r_{ij} p_i(\tau)\nonumber\\
    &=\frac{1}{c(t)}\langle \tilde{J}_t\rangle.
\end{align}
An analogous calculation reveals $\langle \rho_t\rangle=\langle J_t\rangle/c(t)$ and hence using this choice of $\rho_t$ in Eq.~\eqref{eq:TransientCTUR} yields the following form of the transient CTUR 
\begin{align}
    \frac{\langle J_t\rangle ^2}{\mathrm{var}(J_t - c(t)\rho_t)} = \frac{\langle J_t\rangle ^2}{\mathrm{var}(J_t^\mathrm{I})}\leq \frac{\Delta S_\mathrm{tot}(t)}{2}\,.
    \label{eq:gTB}
\end{align}
We call Eq.~\eqref{eq:gTB} the generalized TB (gTB) because the choice
$\kappa_{ij}(\tau) = z_j(\tau) - z_i(\tau)$ yields the transport
bound. Explicitly, the mean current is
\begin{align}
    \langle J_t\rangle &=\int_0^t\rmd s \sum_{i,j\neq i}[z_j(s) - z_i(s)]r_{ij}p_i(s)\nonumber\\
    &=\int_0^t\rmd s \sum_{i}z_i(s)\underbrace{\sum_{j\neq i}[r_{ji}p_j(s)-r_{ij}p_i(s)]}_{=\partial_sp_i(s)}\label{gTB1}\\
    &=\sum_i\left(z_i(t)p_i(t) - z_i(0)p_i(0) - \int_0^t\rmd s [\partial_s z_i(s)] p_i(s)\right),\nonumber
\end{align}
and the variance of the dissipative current reads
\begin{align}
    \mathrm{var}(J_t^\mathrm{I}) &= \langle (J_t^\mathrm{I})^2\rangle - \underbrace{\langle J_t^\mathrm{I}\rangle^2}_{=0}\nonumber\\
    &=\int_0^t\rmd s \sum_{i,j\neq i}[z_j(s) - z_i(s)]^2r_{ij}p_i(s)\nonumber\\
    &= t\mathcal{D}_t.\label{gTB2}
\end{align}
With the specific choice $\kappa_{ij}(\tau) = z_j(\tau) - z_i(\tau)$, i.e., using Eqs.~\eqref{gTB1} and \eqref{gTB2}, Eq.~\eqref{eq:gTB} becomes the TB
\begin{align}
    \frac{\langle z_t - z_0 - \int_0^t\rmd\tau \partial_\tau z_\tau\rangle^2}{t\mathcal{D}_t}\leq \frac{\Delta S_\mathrm{tot}(t)}{2}.
\end{align}
Equation~\eqref{eq:gTB} therefore naturally relates the TB to the CTUR as
a special case. Indeed, the gTB is efficient, i.e. yields a non-zero
bound in both, transient and stationary systems, while the TB only
yields nontrivial bounds in transient systems. A similar connection can
be made in continuous space, where a recent TUR for underdamped
Langevin dynamics \cite{CrutchfieldUnderdampedTUR} can be seen as a
generalization of the continuous TB
\cite{dieball2024thermodynamic}. It should be noted that, while the
TB in general cannot be saturated, the gTB can be saturated with the choice of weight $\f{\kappa}(\tau) \propto \textcolor{black}{\f Z(\tau)}$. Moreover,
in some cases the transient TB of one system may be identical to the
stationary gTB of another system. The following example, an extension
of the SAT model (see Fig.~\ref{fig:Models}b), shows how the TB may give rise to the gTB in a special case.


\begin{figure*}
    \centering
        \includegraphics[width=.8\linewidth]{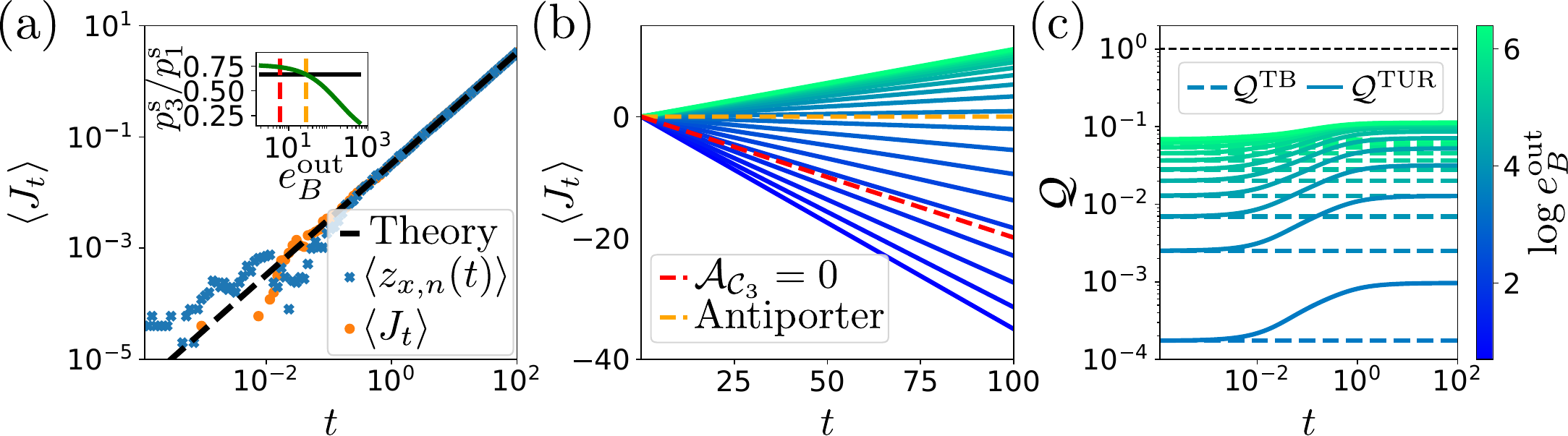}
    \caption{Comparison of TUR and transport bound with the periodic extension
      of the SAT model in Fig.~\ref{fig:Models}b. The mean of ${J}_t$ (orange) and $z_{x,n}$ (blue) from $N=50000$ trajectories sampled with the Gillespie algorithm is shown in (a) as a function of $t$, compared to the analytic $\langle {J}_t\rangle$ [see Eq.~\eqref{eq:MeanCurrent}], for $e^\mathrm{out}_B\approx 40$. The inset shows the fraction $p_3^\mathrm{s}/p_1^\mathrm{s}$ for different values of $e^\mathrm{out}_B$.  The dashed orange  and red  lines indicate the values of $e^\mathrm{out}_B$ where the antiporter regime begins ($e^\mathrm{out}_B\approx 27$) and where $\mathcal{A}_{\mathcal{C}_3}=0$ ($e^\mathrm{out}_B = 6$), respectively, while the black line is the upper limit $p_3^\mathrm{s}/p_1^\mathrm{s}$ can have so that the system is an antiporter. The mean current is shown in (b) for various values of $e^\mathrm{out}_B$. The steady-state quality factors for TUR (solid) and TB (dashed) are shown in (c) as a function of $t$ for various values of $e^\mathrm{out}_B$ in the antiporter regime. The parameters are listed in Tab. \ref{tab:ParameterSAT}.}
    \label{fig:SATperiodic}
\end{figure*}


To make reasonable comparisons \textcolor{black}{in the SAT model}, we either have to consider symporter
or antiporter systems. We are going to focus on the latter. We first vary $\textcolor{black}{e}^\mathrm{out}_B$  to assess
which values of transition rates give rise to antiporter behaviour
before adressing thermodynamic inferrence.  

We start with the cycle affinities of the
model. The six-state SAT model consist of three cycles $\mathcal{C}_1
= \{1\to2,2\to4,4\to3,3\to1\}$, $\mathcal{C}_2 =
\{3\to4,4\to6,6\to5,5\to3\}$, and
$\mathcal{C}_3=\{1\to2,2\to4,4\to6,6\to5,5\to3,3\to1\}$. The
corresponding cycle affinities are $\mathcal{A}_{\mathcal{C}_1} =
\log(e^\mathrm{out}_A/e^\mathrm{in}_A)$, $\mathcal{A}_{\mathcal{C}_2}
= \log(e^\mathrm{in}_B/e^\mathrm{out}_B)$, and
$\mathcal{A}_{\mathcal{C}_3} = \mathcal{A}_{\mathcal{C}_1} +
\mathcal{A}_{\mathcal{C}_2} =
\log(e^\mathrm{out}_Ae^\mathrm{in}_B/e^\mathrm{in}_Ae^\mathrm{out}_B)$. Since
the transport of molecule $A$ with help of $B$ is of interest,
one could naively assume that the transport relates to
$\mathcal{A}_{\mathcal{C}_3}$. However, $\mathcal{A}_{\mathcal{C}_3}$
does \emph{not} quantify the direction of transport, as
$\mathcal{A}_{\mathcal{C}_3}$ does not give any information on 
the transport of type $A$ alone. One can also convince oneself that
$\mathcal{A}_{\mathcal{C}_1}$ is insufficient as well, due to the
cycle affinity being independent of $e^{\mathrm{out}}_B$.  

However, the current (entering the stationary TUR) in the 
six-state model gives the insight we require. The relevant transition weight is
$\kappa_{ij}^{\mathrm{SAT}}=\delta_{i1}\delta_{j3}-\delta_{i3}\delta_{j1}$,
effectively counting the total number of molecules $A$ being
transported towards the exterior of the cell. The mean current is $\langle
J_t^\mathrm{SAT}\rangle_\mathrm{s} =
\sum_{x,y}\kappa_{xy}^\mathrm{SAT}r_{xy}p_x^\mathrm{s}$. If $\langle
J_t^\mathrm{SAT}\rangle_\mathrm{s}>0$, the average total number of
molecules $A$ transported towards the outside of the cell is positive,
corresponding to an 
average transport of $A$ from the inside to the outside of the
cell. Thus, the condition for an antiporter behavior is
$l_A/e^\mathrm{out}_A>p_3^\mathrm{s}/ p_1^\mathrm{s}$, which depends
on $\textcolor{black}{e}^\mathrm{out}_B$ via the steady-state distribution. With the parameters we use, see Tab.~\ref{tab:ParameterSAT},
we need $p_3^\mathrm{s}/p_1^\mathrm{s}<2/3$. The value $2/3$ is marked
as a black line in the inset of
Fig.~\ref{fig:SATperiodic}a, where the
green line corresponds to the value of $p_3^\mathrm{s}/p_1^\mathrm{s}$
as a function of $\textcolor{black}{e}^\mathrm{out}_B$. The dashed red line marks the
value of  $\textcolor{black}{e}^\mathrm{out}_B$ s.t. $\mathcal{A}_{\mathcal{C}_3}=0$,
while the dashed orange line is where
$p_3^\mathrm{s}/p_1^\mathrm{s}=2/3$. The same can be readily seen
in Fig.~\ref{fig:SATperiodic}b, where the
mean current is presented for various values of
$\textcolor{black}{e}^\mathrm{out}_B$. Clearly, the condition
$\mathcal{A}_{\mathcal{C}_3}=0$ is not sufficient for a positive current.  

Additionally, we can consider the transient extended model and apply
the TB. For the extended model, consider $p_{x,n}(\tau)$ the
probability of being in state $x$ in system $n$ at time $\tau$, where
$n\in\textcolor{black}{\mathbb{Z}}$ corresponds to the number of $A$ molecules
transported through the membrane. When marginalizing over $\textcolor{black}{\mathbb{Z}}$
we recover the steady-state probability $p_x^\mathrm{s} =
\sum_{n\in\textcolor{black}{\mathbb{Z}}}p_{x,n}(\tau)$. As a state function we choose
the number of $A$ molecules $z_\tau = n$, such that 
\begin{align}
    \langle z_\tau\rangle =
    \sum_{n\in\textcolor{black}{\mathbb{Z}}}n\sum_{x=1}^6p_{x,n}(\tau) =
    \sum_{n\in\textcolor{black}{\mathbb{Z}}}n\tilde{p}_{n}(\tau),
    \label{eq:meanZperiodicSAT}
\end{align}
where $\tilde{p}_n(\tau)$ is the probability distribution marginalized
over $\mathcal{N}$. Without loss of generality, we use $\langle
z_0\rangle=0$. Equation~\eqref{eq:meanZperiodicSAT} therefore
describes the average number of net-transitions $\langle n_{13} -
n_{31}\rangle$ between states 1 and 3 in the original system until
time $\tau$, as $z_\tau$ the net number of transitions. Hence,
one can immediately recognize that $\langle z_\tau\rangle =
\langle J_t^\mathrm{SAT}\rangle_\mathrm{s}$, which can be readily seen
in Fig.~\ref{fig:SATperiodic}a.  

The fluctuation-scale function is recognized to be
\begin{widetext}
    \begin{align}
        \mathcal{D}_t^\mathrm{ext} = \frac{1}{t}\sum_{n\in\textcolor{black}{\mathbb{Z}}} \sum_{x, y\in\mathcal{N}}\int_0^t\rmd\tau&\left[ (z_{y,n} - z_{x,n})^2\delta_{x\to y, n}+(z_{y,n+1} - z_{x,n})^2\delta_{x\to y, n \to n+1}\right.\nonumber\\
        &+\left.(z_{y,n-1} - z_{x,n})^2\delta_{x\to y, n \to n-1}\right]r_{xy}p_{x,n}(\tau)\,.\label{eq:extendedFSF}
    \end{align}
\end{widetext}
Note that $\delta_{x\to y, n \to n+1}$ is one if the transition $x\to y$ is observed while the system goes from $n$ to $n+1$ and zero otherwise. We can simplify Eq.~\eqref{eq:extendedFSF} to find
\begin{align}
    \mathcal{D}_t^\mathrm{ext} = \frac{1}{t}\sum_{n\in\textcolor{black}{\mathbb{Z}}}\int_0^t\rmd\tau\left[r_{13}p_{1,n}(\tau) + r_{31}p_{3,n}(\tau)\right],
\end{align}
because $z_{x,n}-z_{y,n} = 0$ and $z_{x,n}-z_{y,n\pm 1} = \pm
1$. Marginalizing over $n$ finally gives 
\begin{align}
    \mathcal{D}_t^\mathrm{ext} &=\left[r_{13}p_{1}^\mathrm{s} + r_{31}p_{3}^\mathrm{s}\right]
    = \sum_{x,y\in\mathcal{N}}(\kappa_{xy}^{\mathrm{SAT}})^2r_{xy}p_x^\mathrm{s}.\label{eq:finalextendedFSF}
\end{align}
Upon comparing Eq.~\eqref{eq:finalextendedFSF} to
Eq.~\eqref{eq:CurrentCov} it becomes apparent that
$\mathcal{D}_t^\mathrm{ext} = \lim_{t\to
  0}\mathrm{var}_\mathrm{s}(J_t^\mathrm{SAT})/t$. Furthermore, using
the TB on the transient extended SAT model yields the stationary gTB
of the original six-state SAT model, as the TB reduces to
Eq.~\eqref{eq:gTB}.  

The quality factors of the TUR $\mathcal{Q}^\mathrm{TUR}$ and TB $\mathcal{Q}^\mathrm{TB}$ are shown
in Fig.~\ref{fig:SATperiodic}c (solid and dashed line, respectively) for
$\textcolor{black}{e}^\mathrm{out}_B$ in the antiporter regime. Because
$t\mathcal{D}_t^\mathrm{ext}
>\mathrm{var}_\mathrm{s}(J_t^\mathrm{SAT})$ for all $t>0$, the TUR
infers a higher fraction of the entropy production, i.e., $\mathcal{Q}^\mathrm{TUR}\geq
\mathcal{Q}^\mathrm{TB}$. Additionally, one may show that taking
$\textcolor{black}{e}^\mathrm{out}_B\to\infty$ leads to the quality factors converging to
a finite value \cite{EspositoCoarseGraining,GROMACS_Thesis_Norway}. Hence, there is no improvement in the
inference of dissipation at higher concentrations of $B$
outside the cell beyond a certain point.

\subsection{Thermodynamic Correlation Bound}\label{sec:CB}
Thermodynamic bounds on correlation times were studied in the
context of Langevin dynamics in
\cite{BoundsCorrelationTimes} in stationary systems at large
times. Recently such  thermodynamic correlation inequalities were
extended to finite times and transients
\cite{DieballCorrelationBound}. Following the latter
stochastic-calculus approach, we here develop analogous results for time-homogeneous MJP, in
the general transient setting and for arbitrary times. That such
results are possible was already anticipated in
\cite{BoundsCorrelationTimes}.

The proof of the correlation bound starts from the auxiliary stochastic integral
\begin{align}
    {B}_t &= \frac{1}{\sqrt{t}}\StoInt \mathrm{Tr}\left[\left(\f{H}(\tau)  - \tilde{\f{Z}}(\tau)^T\right)\rmd\f{\varepsilon}(\tau)\right],
\end{align}
where $\tilde{Z}_{xy}(\tau) = Z_{xy}(\tau) \left[z_x(\tau) + z_y(\tau)
  - 2F(\tau)\right]$ for some time-dependent function $F(\tau)$\textcolor{black}{, $Z_{xy}$ is given by Eq.~\eqref{ZMatrix}, and $\f H$ is the same matrix also entering Eq.~\eqref{Bfunctional}}. The
second key ingredient is the shifted and rescaled density observable
\begin{align}
    {C}_t &= \frac{1}{\sqrt{t}}\StoInt\textcolor{black}{\left[V_\tau - \langle {V}_\tau\rangle\right]\rmd{\tau}},
\end{align}
where $\langle V_\tau\rangle = \sum_i V_i p_i(\tau)$. While the TUR and TB allow the state functions
$z_\tau$ and \textcolor{black}{${V}_\tau$} to be explicitly time-dependent, the
correlation bound requires these to be \emph{strictly
time-independent}. Furthermore, we write, e.g., $\langle
z_\tau \rangle = \sum_i z_i p_i(\tau)$ to explicitly highlight
that the average depends on time through $p_i(\tau)$. Additionally, we
need to introduce the re-weighted two-point probability 
\begin{align}
    p_{xy}^\mathrm{ps}(\tau) = \frac{p_x(\tau) r_{xy} Z_{xy}(\tau) ^2}{\Sigma_\mathrm{ps}}\,,
    \label{eq:PseudoProbability}
\end{align}
with the renormalization constant $\Sigma_\mathrm{ps} = \sum_{x,y}
p_x(\tau) r_{xy} Z_{xy}(\tau) ^2$ which is the pseudo-entropy production rate (pseudo-EPR). 
Expectation values w.r.t. $p_{xy}^\mathrm{ps}(\tau)$ will be denoted
by $\langle \cdot \rangle_\mathrm{ps}$. We will also introduce  the
pseudo variance $\mathrm{pvar}_\mathrm{ps}^{F}(z_\tau) =
\sum_{x,y} (z_x + z_y - 2F(\tau))^2p_{xy}^\mathrm{ps}(\tau)$, which is
the ``true'' variance only for $F(\tau) = \langle
z_\tau\rangle_\mathrm{ps}/2 = \sum_{x,y}(z_x +
z_y)p_{xy}^\mathrm{ps}(\tau)/2$ and some other positive time-dependent
quantity otherwise.  

We immediately see that
\begin{align}
    \langle C_t^2\rangle = \frac{1}{t}\int_0^t\rmd \tau \int_0^t\rmd \tau' \mathrm{cov}\left({V}_\tau, {V}_{\tau'}\right)\,,
    \label{eq:CorrelationCovariance}
\end{align}
where $V_\tau\equiv V(x_\tau)=\sum_k\delta_{k,x_\tau}V_k$, and introducing the time-averaged density
\begin{align}
    \overline{V}_t &\equiv \frac{1}{t}\StoInt V_\tau\rmd \tau\,,\nonumber\\
    \overline{z}_t &\equiv \frac{1}{t}\StoInt z_\tau \rmd \tau,
\end{align}
Eq.~\eqref{eq:CorrelationCovariance} can be identified as
\begin{align}
    t\mathrm{var}(\overline{V}_t) = \langle C_t^2\rangle.
    \label{eq:tvarDefinition}
\end{align}
With the shorthand notations \textcolor{black}{$V_x^\Delta(\tau) = V_x - \langle V_{\tau}\rangle$} and $z^F_\tau = z_\tau - \langle F(\tau)\rangle$, the cross term is computed to be
\begin{widetext}
\begin{align}
    t\langle B_t C_t \rangle &= \int_{0}^{t}\rmd\tau'\int_{0}^{t}\rmd\tau \sum_{x,y,i}\textcolor{black}{\mathbb{1}}_{\tau < \tau'}V^\Delta_i(\tau')\left(H_{yx} - Z_{xy}(\tau)\left[z_x^F(\tau) + z_y^F(\tau)\right]r_{xy}p_x(\tau)\right)\left[P(i, \tau'|y, \tau) - P(i, \tau'|x, \tau)\right],
    \label{eq:CrossCorrelationBC}
\end{align}   
where the sum over $x,y$ can be simplified because of symmetries (we suppress the time indices here) to get
\begin{align}
    &\sum_{x,y}H_{yx}r_{xy}p_x[ P(i|y)-P(i|x)]
    =-\sum_{x,y}z_x^F(r_{xy}p_x + r_{yx}p_y)[ P(i|y)-P(i|x)],
\end{align}
and 
\begin{align}
    &\sum_{x,y} Z_{xy}(z_x^F + z_y^F)r_{xy}p_x\left[P(i|y) - P(i|x)\right]
    = \sum_{x,y} z_x^F(r_{xy}p_x- r_{yx}p_y)\left[P(i|y) - P(i|x)\right].
\end{align}
\end{widetext}
Using time-homogeneity of the semi-group $\f{P}(\tau'|\tau)=
\f{P}(\tau'-\tau)\equiv{\rm e}^{\f{L}(\tau'-\tau)}$ which obeys the master
equation Eq.~\eqref{eq:MasterEquation}, i.e. $\partial_\tau
\f{P}(\tau) = \f{L}\f{P}(\tau) = \f{P}(\tau)\f{L}$, and
$\partial_\tau\f{P}(\tau'-\tau) = -\partial_{\tau'}\f{P}(\tau'-\tau)$,
the integrand in Eq.~\eqref{eq:CrossCorrelationBC} simplifies to
$2\sum_{x,i}\textcolor{black}{\mathbb{1}}_{\tau <
  \tau'}V^\Delta_i(\tau')z^F_x(\tau)p_x(\tau)\partial_{\tau'}P(i,
\tau'|x, \tau)$. Since 
\begin{align}
    &\sum_{x,i}\left[\partial_{\tau'}V^\Delta_i(\tau')\right]z^F_x(\tau)p_x(\tau)P(i, \tau'|x, \tau)\nonumber\\
    &=-\partial_{\tau'}\langle {V}_{\tau'}\rangle \underbrace{\sum_{x}z^F_x(\tau)p_x(\tau)\underbrace{\sum_{i}P(i, \tau'|x, \tau)}_{=1}}_{=\langle z_\tau\rangle - F(\tau)},
\end{align}
the particular choice $F(\tau) = \langle z_\tau\rangle$ yields
\begin{align}
    t\langle B_t C_t \rangle &= 2\int_{0}^{t}\rmd\tau'\int_{0}^{t}\rmd\tau \textcolor{black}{\mathbb{1}}_{\tau < \tau'}\partial_{\tau'}\mathrm{cov}({V}_{\tau'}, z_\tau).
    \label{eq:CrossCorrelationBCSimplified}
\end{align}    
An integration by parts in $\tau'$ with
$\partial_{\tau'}\textcolor{black}{\mathbb{1}}_{\tau < \tau'} = \delta(\tau'-\tau)$
further yields
\begin{align}
    \langle B_t C_t \rangle &=2\mathrm{cov}\left[{V}_t, \overline{{z}}_t\right] - \frac{2}{t}\int_0^t\rmd \tau \mathrm{cov}\left[{V}_\tau, z_\tau\right].
    \label{eq:CrossCorrelationBCFinal}
\end{align}
Lastly, the second moment of $B_t$ reads
\begin{align}
    \langle B_t^2\rangle =& \mathcal{D}_t + \frac{1}{t}\int_0^t\rmd\tau \Sigma^\mathrm{ps}(\tau)\mathrm{pvar}_\mathrm{ps}^F\left(z_\tau\right) \nonumber\\&- \frac{2}{t}\int_0^t\rmd\tau\sum_{x,y}\left(\left[z_y-F(\tau)\right]^2 - \left[z_x-F(\tau)\right]^2\right)\nonumber\\&\times Z_{xy}(\tau)r_{xy}p_x(\tau)\,.
    \label{eq:ComplicatedSecondMoment}
\end{align}
The integrand of the last term in Eq.~\eqref{eq:ComplicatedSecondMoment} reduces to
\begin{align}
    &\sum_{x,y}\left(\left[z_y-F(\tau)\right]^2 - \left[z_x-F(\tau)\right]^2\right)Z_{xy}(\tau)r_{xy}p_x(\tau)\nonumber \\
    &= \sum_{y} \left[z_y-F(\tau)\right]^2\partial_\tau p_y(\tau)\,.
\end{align}
Choosing $F(\tau) = \langle z_\tau\rangle$ again simplifies
Eq.~\eqref{eq:ComplicatedSecondMoment} to
\begin{align}
    \langle B_t^2\rangle =& \mathcal{D}_t + \frac{1}{t}\int_0^t\rmd\tau \Sigma^\mathrm{ps}(\tau)\mathrm{pvar}_\mathrm{ps}^F\left(z_\tau\right) \nonumber\\
    &- \frac{2}{t}\left[\mathrm{var}\left(z_t\right) - \mathrm{var}\left(z_0\right)\right]\,.
    \label{eq:ComplicatedSecondMomentFinal}
\end{align}
Combining Eqs.~\eqref{eq:tvarDefinition}, \eqref{eq:CrossCorrelationBCFinal}, and \eqref{eq:ComplicatedSecondMomentFinal} yields the transient finite-time correlation bound
\begin{widetext}
    \begin{align}
        \frac{\left[2\mathrm{cov}\left({V}_t, \overline{{z}}_t\right) - \frac{2}{t}\int_0^t\rmd \tau \mathrm{cov}\left({V}_\tau, z_\tau\right)\right]^2}{t\mathrm{var}(\overline{V}_t)} - \mathcal{D}_t + \frac{2}{t}\left[\mathrm{var}\left(z_t\right) - \mathrm{var}\left(z_0\right)\right]\leq \frac{1}{t}\int_0^t\rmd\tau \Sigma^\mathrm{ps}(\tau)\mathrm{pvar}_\mathrm{ps}^F\left(z_\tau\right).
        \label{eq:TransientCB}
    \end{align}
In NESS, Eq.~\eqref{eq:TransientCB} simplifies to [we suppress time indices for stationary single-time (pseudo/co-)variances]
\begin{align}
    &\frac{4\left[\mathrm{cov}({V}_t, \overline{{z}}_t) - \mathrm{cov}_\mathrm{s}({V}, {z})\right]^2}{t\mathrm{var}(\overline{V}_t)} - \mathcal{D}_t 
    \leq  \frac{\langle A_t^2\rangle_\mathrm{s}}{t}\mathrm{pvar}_\mathrm{ps}^F({z})\leq \frac{\Delta S_\mathrm{tot}(t)}{2t}\mathrm{pvar}_\mathrm{ps}^F({z}).
    \label{eq:StationaryCB}
\end{align}
\end{widetext}
The two point probability density $p_{xy}(\tau)$ is not accessible,
since it requires knowing $\f{L}$, hence the pseudo variance in
Eqs.~\eqref{eq:TransientCB} and \eqref{eq:StationaryCB} needs to be
further bounded.
In the former case, it was assumed that $F(\tau) = \langle
z_\tau\rangle$, while  in the
stationary case the choice of $F(\tau) = F$ is free. Since we consider bounded $z_\tau$ on finite state spaces, there
exist bounds $z_{\max/\min}$ such that for all $x\neq y$ \footnote{It
should be clear that $p_{xx}(\tau) = 0$ for all $x$ and $\tau$, see
Eq.~\eqref{eq:PseudoProbability}.} the function entering the pseudo
variance satisfies $z_{\min} \leq z_x + z_y \leq z_{\max}$. In
steady state with $F = \langle
z_\tau\rangle_\mathrm{ps}/2$, $\mathrm{pvar}_\mathrm{ps}^F({z}) = \mathrm{var}_\mathrm{ps}({z})
\leq (z_{\max} - z_{\min})^2/4$, where we used the Popoviciu's inequality. 

It should also be clear from Eqs.~\eqref{eq:TransientCB} and
\eqref{eq:StationaryCB} that the quality factor for the CB,
$\mathcal{Q}^\mathrm{CB}$, can become negative (and hence the bound is
uninformative). To be precise, let
\begin{align}
    G(t)\equiv\frac{\left[2\mathrm{cov}\left({V}_t, \overline{{z}}_t\right) - \frac{2}{t}\int_0^t\rmd \tau \mathrm{cov}\left({V}_\tau, z_\tau\right)\right]^2}{t\mathrm{var}(\overline{V}_t)},
\end{align}
then clearly $G(t), \mathcal{D}_t, \mathrm{var}\left(z_t\right)\geq 0$ for all $t\geq 0$. 
Therefore, if $G(t) - \mathcal{D}_t+ \frac{2}{t}\left[\mathrm{var}\left(z_t\right) - \mathrm{var}\left(z_0\right)\right]<0$, the inferred entropy production is negative. 
The fact that the quality factor may become negative, giving a lower
bound on the total entropy production that is worse than the second
law of thermodynamics, is not particularly appealing. However, the
CB can be applied in situations where many established methods fail to
infer entropy production. In the following we provide a simple four-state example where CB can be applied successfully but other methods fail.

\textit{Example II:~4-State Toy model.---}Consider a four-state model
$\{1, 2, 3, 4\}$ on a ring, see Fig.~\ref{fig:Models}e. The rates clockwise are unity and
the counterclockwise rates are controlled by the parameter
$\epsilon$. The steady-state distribution $p_i^\mathrm{s}=p=1/4$ is
independent of $\epsilon$. Suppose that we can only distinguish
between a pair of mesostates A and B\textcolor{black}{, i.e.,}
it can only be distinguished
whether the \textcolor{black}{system is} in either states 1 and 2 \textcolor{black}{(mesostate A)} or states 3 and
4 \textcolor{black}{(mesostate B)}. These we refer to mesostates A (containing $\{1,2\}$) and B (containing $\{3,4\}$), see
Fig.~\ref{fig:Models}e.

Inferring the total steady-state entropy production $\Delta
S_\mathrm{tot}(t) = t(\epsilon - 1)\log\epsilon$ by only observing
transitions between states A and B is a challenge where surprisingly
few bounds provide non-trivial estimates. For example,  the TUR (only)
gives the second law, as 
\begin{align}
    \langle J_t\rangle = p\left(\underbrace{\epsilon\kappa_\mathrm{AB}}_{2\to 3} + \underbrace{\kappa_\mathrm{BA}}_{3\to 2} + \underbrace{\epsilon\kappa_\mathrm{BA}}_{4\to 1} + \underbrace{\kappa_\mathrm{AB}}_{1\to 4}\right) = 0\,.
    \label{eq:FourStateMeanCurrent}
\end{align}
Including a density, i.e., using the CTUR Eq.~\eqref{eq:TransientCTUR}, yields the same bound, because the density only contributes in transient dynamics. Similarly, the TB is only non-trivial for transient systems. Using the gTB is not profitable either, due to Eq.~\eqref{eq:FourStateMeanCurrent}. 

\begin{figure*}
    \centering
        \includegraphics[width=.8\linewidth]{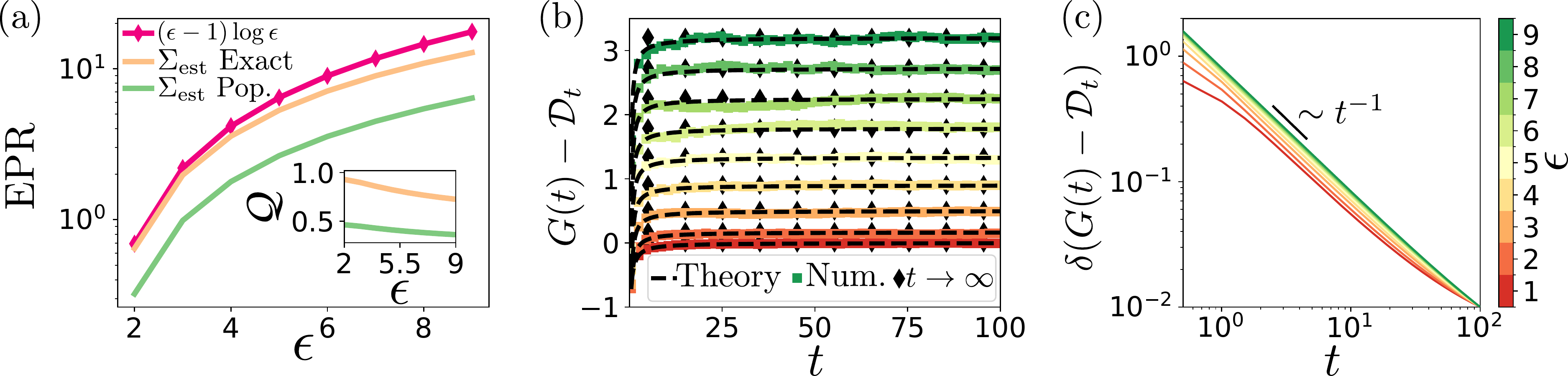}
    \caption{{Correlation bound applied to the driven ring, see Fig.~\ref{fig:Models}e.} In (a), the exact EPR and long-time estimated EPR with and without bounding the (pseudo) variance are shown in dependence of $\epsilon$. Specifically, the choice $F(\tau) = \langle V_x^\mathrm{A}+V_y^\mathrm{A}\rangle_\mathrm{ps}/2$ is used. The inset shows the resulting quality factors. The l.h.s. of the stationary CB, $G(t)-\mathcal{D}_t$, is shown in (b) evaluated analytically and with simulations. Additionally, the long-time limit is included. The deviation from the long-time limit $\delta(G(t)-\mathcal{D}_t)$ is presented in (c) correctly displaying the $\mathcal{O}(t^{-1})$ decay. The simulations are performed with the Gillespie algorithm using $N=50000$ trajectories.}
    \label{fig:FourStateExample}
\end{figure*}

However, the CB can be applied to yield a positive bound on the EP for sufficiently large $t$ and $\epsilon$. The fluctuation scale function is $\mathcal{D}_t = (1+\epsilon)/2$ and for $\mathrm{X}=\mathrm{A},\mathrm{B}$ the steady-state (co)variance is $\mathrm{cov}_\mathrm{s}(V^\mathrm{A}, V^\mathrm{X}) = (\delta_{\mathrm{X},\mathrm{A}}-\delta_{\mathrm{X},\mathrm{B}})/4$. If $\lambda_k$, $k\in\{1, 2, 3, 4\}$, are the eigenvalues of the generator
\begin{align}
    \lambda_1 &= 0\nonumber,\\
    \lambda_2 &= -2(\epsilon + 1)\nonumber,\\
    \lambda_{3/4} &= -1-\epsilon \pm (\epsilon-1)i,\label{CBeigenvalues}
\end{align}
then for times $t\gg \max_{k\in\{2, 3, 4\}}\left(-1/\mathrm{Re}(\lambda_k)\right) = 1/(1+\epsilon)$ we can use
\begin{align}
    t\mathrm{var}(\overline{V}_t^\mathrm{A}) = 2\int_0^t\rmd\tau \mathrm{cov}(V^\mathrm{A}_\tau, V^\mathrm{A}_0) + \mathcal{O}(t^{-1}),
    \label{eq:intermediateCBexample}
\end{align}
and is therefore similar to
\begin{align}
    \mathrm{cov}(V^\mathrm{A}_t, \overline{V}^\mathrm{B}_t) &= -\frac{1}{t}\int_0^t\rmd \tau \mathrm{cov}(V^\mathrm{A}_t, V^\mathrm{A}_\tau) \nonumber\\&= -\frac{1}{t}\int_0^t\rmd \tau \mathrm{cov}(V^\mathrm{A}_\tau, V^\mathrm{A}_0).
\end{align}
Thus, the l.h.s.\ of Eq.~\eqref{eq:StationaryCB} for $t\gg 1/(1+\epsilon)$ reads
\begin{align}
    &\frac{4\left[\mathrm{cov}({V}_t, \overline{{z}}_t) - \mathrm{cov}_\mathrm{s}({V}, {z})\right]^2}{t\mathrm{var}(\overline{V}_t)} - \mathcal{D}_t\nonumber\\
    &=\frac{1}{8\int_0^t \rmd\tau \mathrm{cov}(V^\mathrm{A}_\tau, V^\mathrm{A}_0)} - \frac{1+\epsilon}{2} + \mathcal{O}(t^{-1})
\end{align}
The first term can be simplified further to be $(\epsilon^2+1) /(\epsilon+1)$ (see App.~\ref{AppendixCBdetails}), so that the large $t$ quality factor becomes
\begin{align}
    \mathcal{Q} &\xrightarrow{t\to\infty}\frac{1}{\log\epsilon}\left[\frac{(\epsilon+1)^2 + (\epsilon-1)^2}{(\epsilon^2-1)} - \frac{\epsilon+1}{\epsilon-1}\right]\frac{1}{\mathrm{pvar}_\mathrm{ps}^F(V^\mathrm{A})}\nonumber\\
    &=\frac{(\epsilon-1)}{(\epsilon+1)\log\epsilon}\frac{1}{\mathrm{pvar}_\mathrm{ps}^F(V^\mathrm{A})}\,.
    \label{eq:CBqualityfactorLimit}
\end{align}
Recall that the choice of $F(\tau)$ is free for stationary systems, hence we can choose $F(\tau)=\langle V^\mathrm{A}_x+V^\mathrm{A}_y\rangle_\mathrm{ps}/2$ s.t. $\mathrm{pvar}_\mathrm{ps}^F(V^\mathrm{A}) = \mathrm{var}_\mathrm{ps}(V^\mathrm{A})$. Using Popoviciu's inequality, the pseudo variance (i.e., variance) with the choice $F(\tau)=\langle V^\mathrm{A}_x+V^\mathrm{A}_y\rangle_\mathrm{ps}/2$ (i.e., the two-point variance of $V^\mathrm{A}_x+V^\mathrm{A}_y$) can be bounded by
$\mathrm{var}_\mathrm{ps}^F(V^\mathrm{A})=1/2\leq 1$ (see Appendix~\ref{AppendixCBdetails}). Thus, the
estimated entropy production using Popoviciu's inequality is half of that
estimated with the exact pseudo variance, resulting in
$2\mathcal{Q}^\mathrm{Pop} = \mathcal{Q}$, which is shown in
Fig.~\ref{fig:FourStateExample}a,
where the exact EPR is shown together with the long-time limit of
estimated EPR with and without Popoviciu's inequality. The inset in
Fig.~\ref{fig:FourStateExample}a
shows the corresponding quality factors
Eq.~\eqref{eq:CBqualityfactorLimit} in the limit $t\to\infty$. As a
sanity check, theoretical and simulation results obtained with the celebrated Gillespie algorithm \cite{CelebratedGillespie, CelebratedGillespie2}) of
$G(t)-\mathcal{D}_t$ are compared in
Fig.~\ref{fig:FourStateExample}b,
where a good agreement is found. In addition, the leading order
correction is shown in
Fig.~\ref{fig:FourStateExample}c by
considering  
\begin{align}
    \delta(G(t)-\mathcal{D}_t) = G(t)-\mathcal{D}_t - \frac{(\epsilon-1)^2}{2(\epsilon + 1)}\,,
\end{align}
i.e., subtracting the $t\to\infty$ limit. As expected, the leading
order correction decays as $t^{-1}$.

\section{Response to General Perturbation}\label{sec:Perturbation}

Numerous works addressed the response of Markovian dynamics  to
perturbations by means of various methods in the setting of
equilibrium systems \cite{Muller2020, Basu2018, Baiesi2009},
nonequilibrium steady-states \cite{Speck2016, Owen2020, Aslyamov2024,
  ptaszynski2025nonequilibriumfluctuationresponserelationsstate,
  ptaszynski2025nonequilibriumfluctuationresponserelationsstatecurrent,
  Ptaszynski2024}, as well as general perturbations of single-time
observables \cite{HANGGI1982207} which was recently generalized to path
observables of diffusions~\cite{Klinger2025}. Here we develop the counterpart of~\cite{Klinger2025} for MJP using
the developed stochastic calculus.\\

\subsection{Derivation of Response Formula for MJP}

We are interested in the response of a (generally path-wise) observable
$O(t)=O[(x_\tau)_{0\le\tau\le t}]$ to a perturbation of a control
parameter $\chi\to\chi+\delta$ (e.g., a temperature change $T\to
T+\delta$). We seek for an exact result for
\begin{align}
    \partial_\delta O_t = \lim_{\delta\to0}\frac{1}{\delta}\left(\langle O(t)\rangle_\delta - \langle O(t)\rangle\right),\label{eq:response}
\end{align}
where $\langle O(t)\rangle$ is the expectation w.r.t. the unperturbed
process in Eq.~\eqref{eq:Matrix_eom_MJP}, while $\langle
O(t)\rangle_\delta$ is the average w.r.t. the perturbed process
evolving according to 
\begin{align}
    \rmd \f{n}^\delta(\tau) = \f{R}^\delta(x^\delta_\tau)\rmd \tau + \rmd \f{\varepsilon}^\delta(\tau)\,.
    \label{eq:perteom}
\end{align}
Depending on the observable, the averages $\langle O(t)\rangle$ and
$\langle O(t)\rangle_\delta$ can either be understood as single-time
or path-wise averages. For example, if $O(t)$ measures some
instantaneous quantity at time $t$, e.g., the instantaneous internal
energy of the system, the averages are to be understood as $\langle
O(t)\rangle_{(\delta)} = \sum_{x}O_x(t)p^{(\delta)}_x(t)$, where
$p^{(\delta)}_x(t)$ is the single-time probability evolving according
to the master equation of the (perturbed) system.  
Conversely, if $O(t) = O[(x_\tau)_{0\leq \tau\leq t}]$ is a
path-wise observable depending on the entire trajectory, e.g., a
time-integrated current $J_t $, the average $\langle\cdot\rangle$ is over all paths evolving
according to Eq.~\eqref{eq:Matrix_eom_MJP} [and $\langle\cdot\rangle_{(\delta)}$  over all paths evolving
according Eq.~\eqref{eq:perteom}]
\begin{align}
    \langle O(t)\rangle_{(\delta)} = \sum_{(x_\tau)_{0\leq \tau \leq t}}O[(x_\tau)_{0\leq \tau\leq t}]\textcolor{black}{\mathbb{P}}^{(\delta)}[(x_\tau)_{0\leq \tau\leq t}].
    \label{eq:PathAverage}
\end{align}
The path measures Eq.~\eqref{eq:MJP_PathMeasure} of processes in
Eqs.~\eqref{eq:Matrix_eom_MJP} and \eqref{eq:perteom} can be
conveniently written in terms of the stochastic differentials as
\begin{widetext}
\begin{align}
    \textcolor{black}{\mathbb{P}}^{\delta}[(x_\tau)_{0\leq \tau\leq t}] = p_{x_0}(0)\exp\left[\stoints\sum_{x} r^{\delta}_{xx}\rmd \tau_x(s) + \stoints\sum_{x,y\neq x}\log r_{xy}^{\delta}\rmd n_{xy}(s)\right],\label{eq:PathMeasureDifferentials}
\end{align}
\end{widetext}
where $p_{x_0}(0)$ is the distribution of the initial state
$x_{\tau=0} = x_0$. Note that a given path realization defines a
unique sequence of transitions and hence $\rmd \tau_x(s)$ and $\rmd
n_{xy}(s)$, which in turn 
enter \emph{equally}  $\textcolor{black}{\mathbb{P}}$ in
Eq.~\eqref{eq:MJP_PathMeasure} and $\textcolor{black}{\mathbb{P}}^{\delta}$ in
Eq.~\eqref{eq:PathMeasureDifferentials}. However, because of the different
rates ($r^{\delta}_{xy}$ versus $r_{xy}$), the measure of the same path is generally
different under $\textcolor{black}{\mathbb{P}}$  and $\textcolor{black}{\mathbb{P}}^{\delta}$.  
Moreover, writing
the path measure in terms of stochastic differentials as in
Eq.~\eqref{eq:PathMeasureDifferentials} further has the
advantage of allowing for time-dependent rates. 

Consider the perturbed rates 
\begin{align}
    r_{xy}^\delta = r_{xy}(\chi+\delta) = r_{xy}(\chi) + \delta \tilde{r}_{xy}(\chi)+\mathcal{O}(\delta^2).
\end{align}
Note that we assume that the rates
$r_{xy}=r_{xy}(\chi)$ and $r_{xy}^\delta$ are functions of $\chi$ and
$\chi+\delta$, respectively, ($\chi$ is a placeholder for the
perturbed quantity, e.g., temperature $T$) and that the function is
analytic at $\delta=0$, i.e., that it has a well-defined Taylor
expansion at that point. Therefore, $\textcolor{black}{\mathbb{P}}^\delta$ is absolutely continuous w.r.t.\ $\textcolor{black}{\mathbb{P}}$
for sufficiently small $\delta$
\footnote{For MJP, the perturbed measure $\textcolor{black}{\mathbb{P}}^\delta\ll
\textcolor{black}{\mathbb{P}}$ is absolutely continuous w.r.t.\ the unperturbed measure
$\textcolor{black}{\mathbb{P}}$ as long as for all $x\neq y$ $r_{xy}>0\implies r_{xy}^\delta>0$, i.e., as long as the underlying topology of the MJP remains unchanged.}
and the perturbed average $\langle
O(t)\rangle_\delta$ can be written in terms of the unperturbed process
in Eq.~\eqref{eq:Matrix_eom_MJP} using the Cameron-Martin-Girsanov
theorem with the Radon-Nikodym
derivative~\cite{Liptser_2001} 
\begin{align}
    \langle O(t)\rangle_\delta = \left\langle O(t)\frac{\rmd \textcolor{black}{\mathbb{P}}^\delta}{\rmd \textcolor{black}{\mathbb{P}}}\right\rangle
\end{align}
where we assumed that the processes start from the same initial distribution. The Radon-Nikodym derivative reads
\begin{widetext}
\begin{align}
    \frac{\rmd \textcolor{black}{\mathbb{P}}^\delta}{\rmd \textcolor{black}{\mathbb{P}}}\left[(x_\tau)_{0\leq\tau\leq t}\right] &= \exp\left[\stoints\sum_{x} (r_{xx}^\delta - r_{xx})\rmd \tau_x(s) + \stoints\sum_{x,y\neq x}\log\frac{r_{xy}^\delta}{r_{xy}}\rmd n_{xy}(s)\right]\nonumber\\
    &= \exp\left[-\delta\stoints\sum_{x,y\neq x} \tilde{r}_{xy}\rmd \tau_x(s) + \delta\stoints\sum_{x,y\neq x}\frac{\tilde{r}_{xy}}{r_{xy}}\rmd n_{xy}(s)+\mathcal{O}(\delta^2)\right]\nonumber\\
    &=1 - \delta\stoints\sum_{x,y\neq x}\frac{\tilde{r}_{xy}}{r_{xy}}\left[r_{xy}\rmd \tau_x(s) - \rmd n_{xy}(s)\right] + \mathcal{O}(\delta^2)\nonumber\\
    &= 1 + \delta \stoints\sum_{x,y\neq x}\frac{\tilde{r}_{xy}}{r_{xy}}\rmd \varepsilon_{xy}(s)+ \mathcal{O}(\delta^2).\label{eq:GeneralRadonNikodym}
\end{align}
\end{widetext}
The response to a perturbation in $\chi$ (e.g., the temperature) is therefore evaluated to be
\begin{align}
    \partial_\delta O_t = \left\langle O(t)\stoints\sum_{x,y\neq x}\frac{\tilde{r}_{xy}}{r_{xy}}\rmd \varepsilon_{xy}(s)\right\rangle.\label{eq:GeneralRateResult}
\end{align}
Note that we did not use any particular property of $O(t)$ that would
distinguish between single-time and path-wise
observables. Nevertheless, for the sake of completeness we provide in
Appendix~\ref{sec:pathObs} an
explicit derivation of path-wise observables.

While  elegant, Eq.~\eqref{eq:GeneralRateResult} (similarly to the
continuous-space result \cite{Klinger2025}) cannot generally be
applied to experimental data. This is because of the integration over the noise (which is
never known in experiments) and the effective drift,
including what defines the rates, which renders
Eq.~\eqref{eq:GeneralRateResult} operationally inaccessible. However, Eq.~\eqref{eq:GeneralRateResult}
may be useful in simulations of high-dimensional systems, where the rates and
noise are known but the generator $\f{L}$ cannot be easily
diagonalized. Note that $\rmd \varepsilon_{xy}(s)$ depends on $p_x(s)$ through Eq.~\eqref{depsilonCorr}.

The stochastic-analysis route to Eq.~\eqref{eq:GeneralRateResult} can
be seen as an alternative way of evaluating responses to perturbations
of observables complementing well established existing methods
\cite{HANGGI1982207}. The limit in Eq.~\eqref{eq:response} can in fact
be evaluated directly using a type of Dyson identity
\cite{HANGGI1982207} (see App.~\ref{sec:pathObs} for details). However,
Eq.~\eqref{eq:GeneralRateResult} allows for the same result to be
obtained using correlations of stochastic differentials without
invoking perturbative calculations.

\subsection{Equilibrium Response to Temperature Perturbations}
For MJP in equilibrium we have $p_i^\mathrm{eq}\propto
\mathrm{e}^{-E_i/T}$ in terms of free energies $E_i$. In addition, we
have detailed balance, which, however, does \emph{not} uniquely specify the rates. We choose the rates to be
\begin{align}
    r_{xy} = D\mathrm{e}^{\left(\lambda E_x -
      (1-\lambda)E_y\right)/T}\label{eq:rates}
\end{align}
for any $\lambda\in[0,1]$ (commonly used are $\lambda=1$ and
$\lambda=1/2$) and $D$ may scale with $T$ ($D=\tilde{D} T$ with
$\tilde{D}=\mathrm{const.}$). The rates~\eqref{eq:rates} satisfy
detailed balance.

The aim here is to relate the response in Eq.~\eqref{eq:GeneralRateResult} of a \textcolor{black}{single-time observable $O(t)$} to the equilibrium correlation function,
\begin{align}
    C_{OE}(t) = \left(\langle OE\rangle_\mathrm{eq} - \langle O(t)E(0)\rangle_\mathrm{eq}\right).
\end{align}
We start from Eq.~\eqref{eq:GeneralRateResult} and use the noise-time
lemma in Eq.~\eqref{eq:NoiseTimeCorrelationLemma} to get
\begin{align}
   &\left\langle O(t)\stoints\sum_{x,y\neq x}\frac{\tilde{r}_{xy}}{r_{xy}}\rmd\varepsilon_{xy}(s)\right\rangle_\mathrm{eq}
  =\nonumber\\& \frac{1}{T^2}\!\!\int_0^t\!\!\rmd
  s\!\sum_i\!\!\!\sum_{x,y\neq
    x}\!\!\!O_i\left[P(i,t|y,s)-P(i,t|x,s)\right]\times\nonumber\\&r_{xy}p_x^\mathrm{eq}\left(-\lambda E_x +
      (1-\lambda)E_y
  +{ T}\right)\label{eq:eqres},
\end{align}
since 
\begin{align}
  \frac{\tilde{r}_{xy}}{r_{xy}} &= \frac{\partial_\delta \tilde{D}(T+\delta)\mathrm{e}^{\left(\lambda E_x - (1-\lambda)E_y\right)/(T+\delta)}|_{\delta=0}}{\tilde{D}T\mathrm{e}^{\left(\lambda E_x - (1-\lambda)E_y\right)/T}}\nonumber\\
    &= \frac{ \tilde{D}(1 - \frac{\lambda E_x - (1-\lambda) E_y}{T})\mathrm{e}^{\left(\lambda E_x - (1-\lambda)E_y\right)/T}}{\tilde{D}T\mathrm{e}^{\left(\lambda E_x - (1-\lambda)E_y\right)/T}}\nonumber\\
    &= \frac{T - \lambda E_x + (1-\lambda) E_y}{T^2}.\label{simplifyRate}
\end{align}
The last term in Eq.~\eqref{eq:eqres} including $T$ vanishes because 
\begin{align}
    &\sum_{x,y\neq x}\left[P(i,t|y,s)-P(i,t|x,s)\right]r_{xy}p_x^\mathrm{eq}\nonumber \\=& \sum_{x,y\neq x}P(i,t|x,s)r_{yx}p_y^\mathrm{eq}-\sum_{x,y\neq x}P(i,t|x,s)r_{xy}p_x^\mathrm{eq}\nonumber\\
    =&\sum_{x,y\neq x}P(i,t|x,s)r_{xy}p_x^\mathrm{eq}-\sum_{x,y\neq x}P(i,t|x,s)r_{xy}p_x^\mathrm{eq}\nonumber\\=&0,\label{eq:eqres1}
\end{align}
where we use detailed balance in the third line. Similarly, the {first
  terms} in Eq.~\eqref{eq:eqres} simplifies to
\begin{align}
   & \!\sum_{x,y\neq x}\!\!\left[P(i,t|y,s)-P(i,t|x,s)\right]r_{xy}p_x^\mathrm{eq}\left(-\lambda E_x+(1-\lambda)E_y\right)\nonumber
    \nonumber\\=&\sum_{x,y\neq x}P(i,t|x,s)r_{xy}p_x^\mathrm{eq}\times
    \nonumber\\ &\underbrace{\left(-\lambda E_y+(1-\lambda)E_x + \lambda E_x-(1-\lambda)E_y\right)}_{=E_x-E_y}\nonumber\\
    =&-\sum_x p_x^\mathrm{eq}E_x \underbrace{\sum_{y \neq x}\left[P(i,t|y,s)-P(i,t|x,s)\right]r_{xy}}_{=\sum_{y \neq x}P(i,t|y,s)r_{xy} + P(i,t|x,s)r_{xx}}\nonumber\\&=-\sum_x p_x^\mathrm{eq}E_x \underbrace{\sum_{y}P(i,t|y,s)L_{yx}}_{=-\partial_s P(i,t|x,s)}.\label{eq:eqres2}
\end{align}
Plugging Eqs.~\eqref{eq:eqres1} and \eqref{eq:eqres2} into Eq.~\eqref{eq:eqres}, we finally get
\begin{widetext}
\begin{align}
     \left\langle O(t)\stoints\sum_{x,y\neq x}\frac{\tilde{r}_{xy}}{r_{xy}}\rmd\varepsilon_{xy}(s)\right\rangle_\mathrm{eq} =& \frac{1}{T^2}\int_0^t\rmd s \sum_i\sum_{x}O_iE_xp_x^\mathrm{eq}\partial_s P(i, t|x, s)\nonumber\\
    =& \frac{1}{T^2}\left(\sum_i\sum_{x}O_iE_xp_x^\mathrm{eq}\underbrace{P(i, t|x, t)}_{\delta_{ix}} - \sum_i\sum_{x}O_iE_xp_x^\mathrm{eq}P(i, t|x, 0)\right)\nonumber\\
    =& \frac{1}{T^2}\left(\langle O E\rangle_\mathrm{eq} - \langle O(t) E(0)\rangle \right).\label{eq:discreteFDT}
\end{align}
\end{widetext}
Note that Eq.~\eqref{eq:discreteFDT} does \emph{not} explicitly depend
on $\lambda$, only implicitly through the propagator, i.e., it holds
for every parametrization of the rates in
Eq.~\eqref{eq:EQrates}. Equation~\eqref{eq:discreteFDT} is the
linear response of an equilibrium MJP to a temperature perturbation.

\textcolor{black}{\subsection{Connection to Response Function Formalism}}
\textcolor{black}{
In this section, we connect the results derived in the previous sections to the well established \emph{response function formalism}, which is often used to describe the fluctuation dissipation theorem. While the results of response functions for MJP has been previously derived in, e.g., \cite{HANGGI1982207, Marconi2008, Speck,zheng2025spatialcorrelationunifiesnonequilibrium, SeifertSpeck2010EPL}
, we show how these may be extracted from our results to bridge the ``gap'' to the literature.}

\textcolor{black}{\subsubsection{Equilibrium Response}}
\textcolor{black}{
For a first order perturbation theory of a single-time observable $B(t)$ in an equilibrium system with perturbation $h(s)$, $0\leq s\leq t$, conjugate to some observable $R(s)$, the response reads \cite{HANGGI1982207, Marconi2008}
\begin{align}
    \Delta B(t)\equiv\langle B(t)\rangle_h - \langle B\rangle_\mathrm{eq} = \int_0^t\mathrm{d}s h(s)\chi(t-s)\,.\label{RFF1}
\end{align}
Here, $\chi(s)$ is the response function (recall that $k_\mathrm{B}=1$ throughout the manuscript) \cite{HANGGI1982207}
\begin{align}
    \chi(s) = -\frac{1}{T}\frac{\rmd}{\rmd s}\langle R(0) B(s)\rangle_\mathrm{eq}\,.\label{ResponseFunction}
\end{align}
To connect this to the result we derived in the previous section, we first consider the case a constant perturbation to an equilibrium system, i.e., $h(s)=h$. Comparing Eq.~\eqref{eq:discreteFDT} to Eq.~\eqref{RFF1} (with $\delta\to h$), we see that to first order in $h$
\begin{align}
    \Delta O(t) &= \frac{h}{T^2}\left(\langle O E\rangle_\mathrm{eq} - \langle O(t) E(0)\rangle_\mathrm{eq}\right)\nonumber\\
    &=\frac{1}{T^2}\int_0^t\mathrm{d}s\frac{\rmd}{\rmd s}\langle O(t)E(s) \rangle_\mathrm{eq} h\label{ConstantPert1}\,.
\end{align}
Since the equilibrium correlation is time-translation invariant, we can rewrite the above response Eq.~\eqref{ConstantPert1} as
\begin{align}
    \Delta O(t) =\int_0^t\mathrm{d}s\chi(t-s) h + \mathcal{O}(h^2)\,,
\end{align}
with $\chi(t-s) \equiv \frac{1}{T^2}\frac{\rmd }{\rmd s}\langle {O}(t)E(s) \rangle_\mathrm{eq} = -\frac{1}{T^2}\frac{\rmd}{\rmd t}\langle O(t-s)E(0) \rangle_\mathrm{eq}$ consistent with Eq.~\eqref{ResponseFunction} with $R = T^{-1}E$. \\
\indent For time-dependent $h(s)$ we get to first order
\begin{align}
    \Delta O(t) = \left\langle O(t)\int_{s=0}^{s=t} h(s)\sum_{x\neq y}\frac{\tilde{r}_{xy}}{r_{xy}}\mathrm{d}\varepsilon_{xy}(s)\right\rangle\,.
\end{align}
The simplifications performed in Eqs.~\eqref{simplifyRate},~\eqref{eq:eqres1}, and~\eqref{eq:eqres2} can still be made, as these are independent of the perturbation $h(s)$
\begin{align}
    \Delta O(t) &=  \frac{1}{T^2}\int_{0}^{t}h(s)\sum_{i}\sum_{x} O_i E_x \partial_s P(i, t|x, s)p_x^\mathrm{eq}\mathrm{d}s\nonumber\\
    &=  \frac{1}{T^2}\int_{0}^{t}h(s)\partial_s\langle O(t) E(s)\rangle_\mathrm{eq}\mathrm{d}s\nonumber\\
    &=\int_0^th(s)\chi(t-s)\,,
\end{align}
again with
\begin{align}
    \chi(t-s)&=\frac{1}{T^2}\frac{\rmd }{\rmd s}\langle O(t) E(s)\rangle_\mathrm{eq} \nonumber\\ &= -\frac{1}{T^2}\frac{\rmd}{\rmd t}\langle O(t-s) E(0)\rangle_\mathrm{eq}\,.
\end{align}
The extension of these results, e.g., in case of non-stationary systems, has already been discussed in Ref.~\cite{HANGGI1982207}.  
}
\textcolor{black}{\subsubsection{Response of a Stationary System}}
\textcolor{black}{To further connect our result to responses of (non-equilibrium) steady states, we realize that the perturbation rates $\tilde{r}_{xy}$ are the elements of the perturbation generator $\f L^1$, i.e., $\f L = \f L_0 + h\f L^1$ with $L_{xy}=r_{yx}$ and $L^1_{xy}=\tilde{r}_{yx}$. Using the correlation lemma Eq.~\eqref{eq:NoiseTimeCorrelationLemma}, the response of a single-time observable in steady-state reads to first order in $h$
\begin{align}
 \Delta_\mathrm{s}O(t)\equiv&\langle O(t)\rangle_{h} - \langle O(t)\rangle_\mathrm{s}\nonumber\\
 =& \int_0^t\rmd sh(s)\sum_i O_i\sum_{x,y\neq x}\frac{\tilde{r}_{xy}}{r_{xy}}\nonumber\\&\times\left[P(i,t|y,s)-P(i,t|x,s)\right]r_{xy}p_x^\mathrm{s}\,.\label{RespNESS2}
\end{align}
The last double-sum simplifies as follows
\begin{align}
    &\sum_{y,x\neq y}\tilde{r}_{xy}[P(i,t|y,s)-P(i,t|x,s)]p_x^\mathrm{s}\nonumber\\
    =&\sum_{y,x\neq y}\tilde{r}_{yx}P(i,t|x,s)p_y^\mathrm{s}-\underbrace{\sum_{y,x\neq y}\tilde{r}_{xy}}_{ \tilde{r}_{xx}}P(i,t|x,s)]p_x^\mathrm{s}\nonumber\\
    =& \sum_{y,x}\tilde{r}_{yx}P(i,t|x,s)p_y^\mathrm{s}\,,\label{RespNESS1}
\end{align}
where we split the sum and relabel $x\leftrightarrow y$ in the first line, and then use the definition of the diagonal elements to simplify the expression. As a consequence, we can now write the conjugate observable $R_x^\mathrm{s} \equiv \sum_y L_{xy}^1p_y^\mathrm{s}/p_x^\mathrm{s}$, and identify that inserting Eq.~\eqref{RespNESS1} into Eq.~\eqref{RespNESS2} yields to first order
\begin{align}
    \Delta_\mathrm{s}O(t)=& \int_0^t\rmd sh(s)\sum_i O_i\sum_{y,x}\underbrace{\tilde{r}_{yx}}_{L_{xy}^1}P(i,t|x,s)p_y^\mathrm{s}\nonumber\\
    =& \int_0^t\rmd sh(s)\sum_i O_i\sum_{x}P(i,t|x,s)\underbrace{\sum_yL_{xy}^1\frac{p_y^\mathrm{s}}{p_x^\mathrm{s}}}_{R^\mathrm{s}_x}p_x^\mathrm{s}\nonumber\\
    =& \int_0^t\rmd sh(s)\sum_{i,x}O_iP(i,t|x,s)R^\mathrm{s}_xp_x^\mathrm{s}\,.\label{RespNESS3}
\end{align}
Here, we can identify the stationary response function
\begin{align}
    \chi_\mathrm{s}(t-s)\equiv \textcolor{black}{\mathbb{1}}_{s<t}\langle O(t) R^\mathrm{s}(s)\rangle_\mathrm{s} = \textcolor{black}{\mathbb{1}}_{s<t}\langle O(t-s) R^\mathrm{s}(0)\rangle_\mathrm{s}\,,
\end{align}
so that Eq.~\eqref{RespNESS3} can be written as
\begin{align}
    \Delta_\mathrm{s}O(t) = \int_0^t\rmd s h(s)\chi_\mathrm{s}(t-s)+\mathcal{O}(h^2)\,.
\end{align}
This is consistent with the results from Ref.~\cite{SeifertSpeck2010EPL}.
}
\textcolor{black}{\subsubsection{Response of Path-Wise Observables}\label{subsec:respose pathwise}}
\textcolor{black}{Lastly, we consider the response of, possibly transient, path-wise observables. Specifically, consider the observables
\begin{align}
 O_1(t) &= \int_0^t\sum_{x,y\neq x}b_{xy}(s)\rmd\varepsilon_{xy}(s)\,,\label{PathObserv1}\\
 O_2(t) &= \int_0^tg_s\rmd s\,.
\end{align}
Using Eq.~\eqref{depsilonCorr} the response of Eq.~\eqref{PathObserv1} can be stated directly to first order
\begin{align}
    \Delta O_1(t)\equiv&\langle O_1(t)\rangle_h - \langle O_1(t)\rangle\nonumber\\
    =& \int_0^t\mathrm{d}z \sum_{x,y\neq x} b_{xy}(z)\int_0^t\rmd s h(s)\sum_{i,j\neq i}\frac{\tilde{r}_{ij}}{r_{ij}}\nonumber\\
    &\times \delta_{ix}\delta_{jy}\delta(s-z) r_{ij}p_i(s)\nonumber\\
    =&\int_0^t\mathrm{d}z \sum_{x,y\neq x} b_{xy}(z) h(z)\Tilde{r}_{xy}p_x(z)\,.
\end{align}
Similarly, using Eq.~\eqref{eq:NoiseTimeCorrelationLemma}, the response of $O_2(t)$ becomes to first order (note that this result is compared to the approach used in Ref.~\cite{HANGGI1982207} in App.~\ref{sec:pathObs})
\begin{align}
    \Delta O_2(t)\equiv&\langle O_2(t)\rangle_h - \langle O_2(t)\rangle\nonumber\\
    =&\int_0^t\rmd z\int_0^t\rmd s\textcolor{black}{\mathbb{1}}_{s<z}h(s) \sum_{i, x, y\neq x}\\&\times g_i(z)\tilde{r}_{xy}p_x(s)\left[P(i,z|y, s) - P(i,z|x,s)\right]\nonumber.
\end{align}
The sum over $x,y$ can again be simplified using Eq.~\eqref{RespNESS1}, so that
\begin{align}
     \Delta O_2(t)=&\int_0^t\rmd z\int_0^t\rmd s\textcolor{black}{\mathbb{1}}_{s<z}h(s) \sum_{i, x, y}g_i(z)\tilde{r}_{yx}p_y(s)P(i,z|x,s)\nonumber\\
     &+\mathcal{O}(h^2)\,.
\end{align}
Defining ${R_x(s) = \sum_y \tilde{r}_{yx}p_y(s)/p_x(s) = \sum_y L_{xy}^1p_y(s)/p_x(s) }$, the response functions for $O_1(t)$ and $O_2(t)$ become
\begin{align}
    \chi_{O_1}(t,s) &= \textcolor{black}{\mathbb{1}}_{t>s}\sum_{x,y\neq x}b_{xy}(s)\tilde{r}_{xy}p_x(s), \\
    \chi_{O_2}(t,s) &= \int_0^t\rmd z \textcolor{black}{\mathbb{1}}_{z>s}\langle g_z R(s)\rangle,
\end{align}
so that the response to \emph{any} additive functional $O_\mathrm{p}(t) = O_1(t)+O_2(t)$ becomes
\begin{align}
    \Delta O_\mathrm{p}(t) \equiv& \langle O_\mathrm{p}\rangle_h - \langle O_\mathrm{p}\rangle \nonumber\\=& \int_0^t\rmd s h(s)\chi_{O_\mathrm{p}}(t,s)+\mathcal{O}(h^2)\,,\nonumber\\
    \chi_{O_\mathrm{p}}(t,s) =& \chi_{O_1}(t,s)+\chi_{O_2}(t,s)\,. \label{response pathwise}
\end{align}
Note that observables of integrals over $\rmd \f n (s)$ is a special case where $g_x(s)$ and $b_{xy}(s)$ have been chosen appropriately to satisfy Eq.~\eqref{eq:eom_MJP}.
}\\

\subsection{Examples}

\begin{figure}
    \centering
        \includegraphics[width=\linewidth]{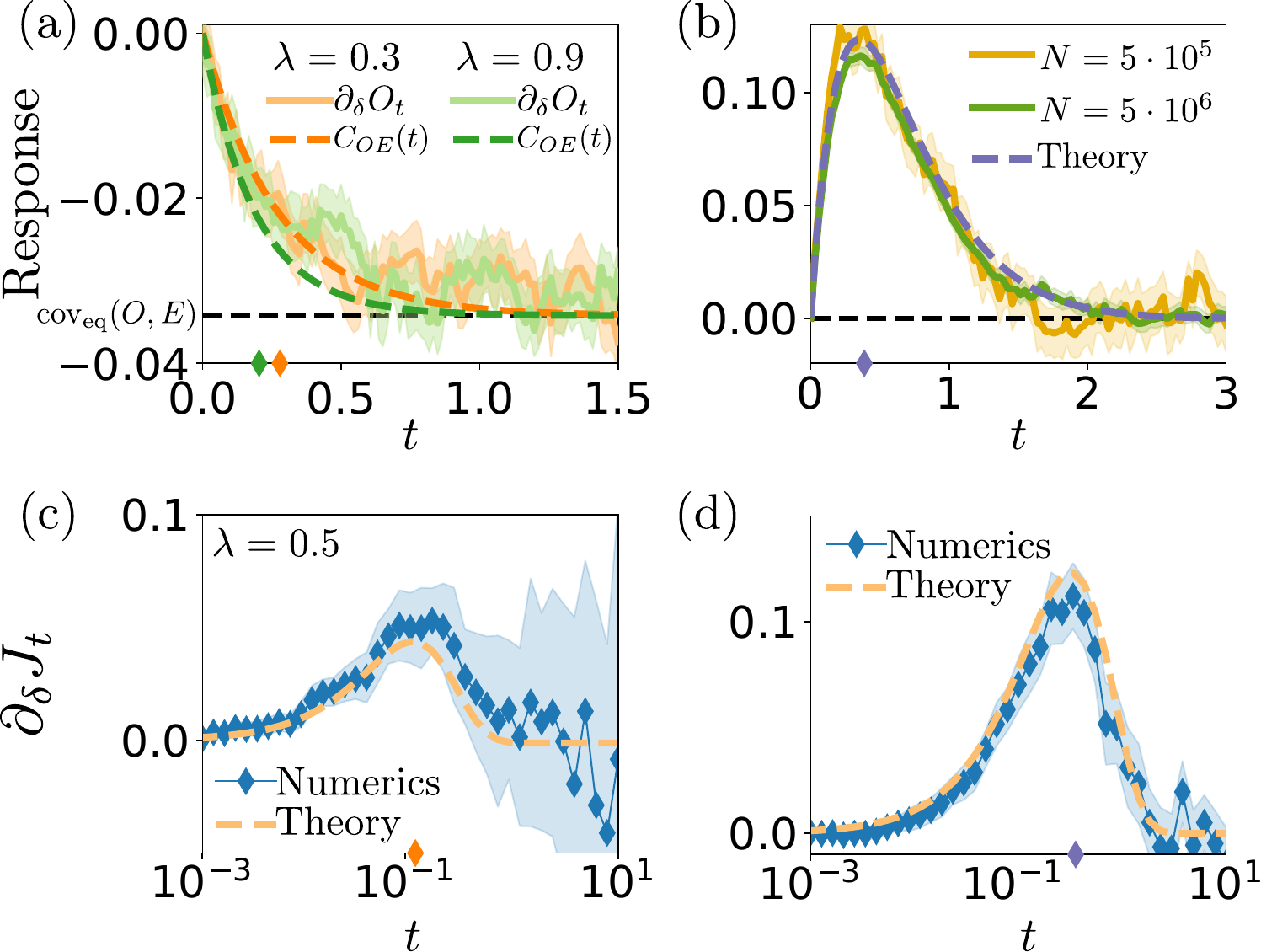}
    \caption{Perturbations of the equilibrium and driven rings. Analytical and numerical response curves of $O=(1,1,0,0)$ for equilibrium (a) and transient (b) dynamics on a four state ring. In (a), we compare the numerical simulations (solid lines) with theoretical prediction Eq.~\eqref{eq:discreteFDT} (dashed lines) for $\lambda=0.3$ (orange) and $\lambda=0.9$ (green). The remaining parameters are $T=1$, $\delta = 0.1$, $N=5\cdot 10^6$, and $\Dt=1$. In (b), the initial distribution is $p^\mathrm{init}=(0, 0, 1, 0)$ and the rates are $r_{12}=r_{23}=r_{34}=r_{41}=T\mathrm{e}^{-\log\epsilon/T}$ and $r_{21}=r_{32}=r_{43}=r_{14}=T\mathrm{e}^{-\log\alpha/T}$ with $\epsilon=3$ and $\alpha=2$, respectively. The theoretical result Eq.~\eqref{eq:result} (dashed purple line) is compared with numerical simulations using $N=5\cdot 10^5$ (solid yellow line) and $N=5\cdot 10^6$ (solid green line) trajectories, respectively. The remaining parameters are $T=2$, $\delta=0.1$, and $\Dt=1$. The markers in (a) and (b) denote respective the largest relaxation time-scales in the systems. The dashed black lines are the expected $t\to\infty$ values $\mathrm{cov}_\mathrm{eq}(A, E)$ in (a) and $0$ in (b). Figure (c) shows the response to a current with weight $\kappa_{ij} = \delta_{i4}\delta_{j1}-\delta_{i1}\delta_{j4}+\delta_{i3}\delta_{j2}-\delta_{i2}\delta_{j3}$ for the aforementioned equilibrium rates with $T=2$, $\delta=0.05$, $N=2\cdot10^6$, $\lambda=0.5$, and initial condition as in (b). In (d), the same parameters as in (c) are used, with the same rates as in (b). The dashed orange lines in (c) and (d) are the theoretically predicted values, while the blue markers are values extracted from numerical simulations. In (a-d) the uncertainty bands are extrapolated from bootstrapping the trajectories into $20$ samples with $30\%$ of the respective $N$s. The free energies, $E_i$, used in (a) and (c) are listed in Tab.~\ref{tab:FreeEnergiesRing}. The simulations are performed with the celebrated Gillespie algorithm.}
    \label{fig:ResponseCurves}
\end{figure}

To visualize the results in Eqs.~\eqref{eq:GeneralRateResult} and
\eqref{eq:discreteFDT}, we consider the response of a single time
observable $O(t)=(1,1,0,0)$ of a four-state ring in an
equilibrium system (see Fig.~\ref{fig:ResponseCurves}a) and in a transient system with constant driving
(see Fig.~\ref{fig:ResponseCurves}b), as well as of a current $J_t$
with weight $\kappa_{ij} =
\delta_{i4}\delta_{j1}-\delta_{i1}\delta_{j4}+\delta_{i3}\delta_{j2}-\delta_{i2}\delta_{j3}$
in transient systems with equilibrium (see Fig.~\ref{fig:ResponseCurves}c) rates and constant
driving (see Fig.~\ref{fig:ResponseCurves}d).

For simplicity we set $D=T$ in Eq.~\eqref{eq:rates}. In
Fig.~\ref{fig:ResponseCurves}a and Fig.~\ref{fig:ResponseCurves}c we
use Eq.~\eqref{eq:EQrates} with the free energies listed in
Tab.~\ref{tab:FreeEnergiesRing}. In Fig.~\ref{fig:ResponseCurves}a, we
compare $\lambda=0.3$ and $\lambda=0.9$, while in
Fig.~\ref{fig:ResponseCurves}c we use $\lambda=0.5$. The rates used in
Fig.~\ref{fig:ResponseCurves}b and Fig.~\ref{fig:ResponseCurves}d are
$r_{12}=r_{23}=r_{34}=r_{41}=T\mathrm{e}^{-\log\epsilon/T}$ in one
direction and
$r_{21}=r_{32}=r_{43}=r_{14}=T\mathrm{e}^{-\log\alpha/T}$ in the
opposite direction with $\epsilon=3$ and $\alpha=2$, respectively. The
temperature is set to $T=1$ and equilibrium initial conditions
$p^\mathrm{init}_i=p^\mathrm{eq}_i\propto e^{-E_i/T}$ are
assumed in Fig.~\ref{fig:ResponseCurves}a, while $T=2$ and initial
condition $p^\mathrm{init}=(0, 0, 1, 0)$ is used in
Fig.~\ref{fig:ResponseCurves}b-Fig.~\ref{fig:ResponseCurves}d. The diamonds on the \textcolor{black}{horizontal axis} denote the largest relaxation time of the respective generators.

As a visual aid, the value of the equilibrium covariance between $O$
and $E$, $\mathrm{cov}_\mathrm{eq}(O, E) = \langle
OE\rangle_\mathrm{eq} - \langle O\rangle_\mathrm{eq}\langle
E\rangle_\mathrm{eq}$, is included in Fig.~\ref{fig:ResponseCurves}a
to show how, independently of $\lambda$, the response $\partial_\delta
O_t\to\mathrm{cov}_\mathrm{eq}(O, E) $ for $t\to\infty$.

\section{Continuum limit}\label{sec:ContinuousLimit}
While we treated continuous-space (Langevin) and discrete-state MJP
dynamics separately so far, it is well known that the two
descriptions should agree in a particular limit. Namely, a
$d$-dimensional continuous description can be approximated by a jump
process on a $d$-dimensional grid (here, we assume a hypercubic lattice) with grid spacing $\Dx$, where the MJP
description reproduces the dynamics as $\Dx\to 0$ or for observations
on scales much larger than $\Dx$ (see, e.g.,
Ref.~\cite{ConsistentNumericalSolver}). For thermodynamic consistency, we here show
that both, the entropy Eq.~\eqref{eq:EP} and pseudo entropy
Eq.~\eqref{eq:PseudoEntopyPoduction} approach the continuous-space
expression Eq.~\eqref{eq:ContinuousEP} as $\Dx\to0$ for any jump
dynamics on a grid approaching the continuous-space dynamics in this
limit. In particular, this implies that the pseudo entropy
Eq.~\eqref{eq:PseudoEntopyPoduction} and entropy Eq.~\eqref{eq:EP}
agree in this limit (as they do for continuous dynamics). This
in turn implies
that the bounds for jump processes can also become saturated in this
continuum limit,
exactly as they can be saturated for continuous-space dynamics.  

We now introduce jump rates for a given $d$-dimensional continuous dynamics
$\rmd\x_\tau=\f F(\x_\tau)\rmd \tau+\sqrt{2D}\rmd\f W_\tau$ and given
$\Dx$. For simplicity, we first consider dynamics settling into an
equilibrium for long times, i.e., $\f F(\x)=-D\nabla U(\x)$ for some
sufficiently confining potential $U(\x)$ (in
  principle the approach equally works for irreversible (i.e., $D^{-1}\f F$ not a
  potential field) and even non-ergodic
  dynamics). Following Ref.~\cite{ConsistentNumericalSolver}, the
rate from $\x$ to a neighboring state $\x' = \x + \fDx$ 
can be written for small $\Dx$ as 
\begin{align}
    r_{\x\x'}
    &=\frac{D}{\Dx^2}\exp\left[-\frac{U(\x')-U(\x)}2\right] \nonumber\\
    &\approx\frac{D}{\Dx^2}\exp\left[-\frac{\nabla U(\x)\cdot \fDx}2\right]\,. \label{rate_discretized}
\end{align}
Note that there are other possible choices that give rise to the
correct continuum limit, which all give the same result in the following calculations, since they have to agree up to second order for $\Dx\to0$ \cite{ConsistentNumericalSolver}. Given a continuous-space probability density $p(\x,\tau)$, the
consistent discretization is $p_\x(\tau)\approx p(\x,\tau)\Dx^d$. To address the thermodynamic entropy production defined in Eq.~\eqref{eq:EP}, note that (in the following,``$\approx$'' denotes equality for $\Dx\to0$)
\begin{align}
    \!\!\!\!\!\!\ln\left(\frac{p_\x(\tau)}{p_{\x'}(\tau)}\right)
    &\approx\ln[p(\x,\tau)]-\ln[p(\x,\tau)+\nabla p(\x,\tau)\cdot\fDx] \nonumber\\
     \!\!\!\!\!\!&\approx-\nabla \ln[p(\x,\tau)]\cdot\fDx\,,\nonumber\\
    \ln\left(\frac{r_{\x\x'}}{r_{\x'\x}}\right)
    & \overset{{\rm Eq.}~\eqref{rate_discretized}} = -\nabla U(\x)\cdot\fDx/2 + \nabla U(\x')\cdot\fDx/2 \nonumber\\
    &= -\nabla U(\x)\cdot\fDx\,,\nonumber\\
    \ln\left(\frac{p_\x(\tau)r_{\x\x'}}{p_{\x'}(\tau)r_{\x'\x}}\right) &=[-\nabla U(\x)-\nabla \ln[p(\x,\tau)]\cdot\fDx\nonumber\\
    &=\frac{\f j(\x,\tau)}{Dp(\x,\tau)}\cdot\fDx\,.
\end{align}
Similarly, we obtain
\begin{align}
    &p_\x(\tau)r_{\x\x'}-p_{\x'}(\tau)r_{\x'\x}\nonumber\\
    &\approx \Dx^dp(\x,\tau)r_{\x\x'}-\Dx^d[p(\x,\tau)+\nabla p(\x,\tau)\cdot\fDx]r_{\x'\x}\nonumber\\
    &\approx \Dx^dp(\x,\tau)\left(r_{\x\x'}-r_{\x'\x}-\nabla\ln[p(\x,\tau)]\cdot\fDx\right)\nonumber\\
    &\approx \Dx^{d-2}Dp(x,\tau)\left(-\nabla U(\x)-\nabla\ln[p(\x,\tau)]\right)\cdot\fDx\nonumber\\
    &=\Dx^{d-2}\f j(\x,\tau)\cdot\fDx\,,
\end{align}
and
\begin{align}
    &p_\x(\tau)r_{\x\x'}+p_{\x'}(\tau)r_{\x'\x} \nonumber\\
    &\approx \Dx^d p(\x,\tau)r_{\x\x'}+[p(\x,\tau)+\nabla p(\x,\tau)\cdot\fDx]r_{\x'\x}\nonumber\\
    &\approx \Dx^{d-2}2Dp(x,\tau)\,.
\end{align}
For the simple hypercubic lattice, the increment vectors $\fDx$ to neighboring states are $\fDx=\pm\Dx\f e_i$. Here, $\f e_i$ with $i\in\{1,2,\dots,d\}$ is a unit vector in the $i$-th coordinate, projecting onto individual components as $\f j\cdot\fDx=\pm j_i\Dx$. The relevant terms from the definitions in
Eqs.~\eqref{eq:PseudoEntopyPoduction} and \eqref{eq:EP} for
the pseudo-entropy and entropy turn out to be
\begin{align}
    &\frac12[p_\x(\tau)r_{\x\x'}-p_{\x'}(\tau)r_{\x'\x}]\ln\left[\frac{p_\x(\tau)r_{\x\x'}}{p_{\x'}(\tau)r_{\x'\x}}\right]\\&\approx\Dx^d\frac{j_i(\x,\tau)^2}{2D p(\x,\tau)}\,,\nonumber\\
    &\frac{[p_\x(\tau)r_{\x\x'}-p_{\x'}(\tau)r_{\x'\x}]^2}{p_\x(\tau)r_{\x\x'}+p_{\x'}(\tau)r_{\x'\x}} \approx \Dx^d\frac{j_i(\x,\tau)^2}{2D p(\x,\tau)}\,.
\end{align}
A summation over $i$ yields $\sum_{i=1}^d j_i(\x,\tau)^2=\f j(\x,\tau)^2$. Upon summation over $\x,\x'$  as done in
Eqs.~\eqref{eq:PseudoEntopyPoduction} and \eqref{eq:EP} for the
pseudo-entropy and entropy production, respectively, we find that both
quantities approach the continuous-space formula for the entropy
production in Eq.~\eqref{eq:ContinuousEP}. Following
Ref.~\cite{ConsistentNumericalSolver} one can directly extend the
argument to non-isotropic diffusion and non-conservative dynamics by
using different ``pseudopotentials'' for each entry of the vector $\f
F(\x)$ and thus the above statements about the limit generalize. 

In particular, this implies that the inequality pseudo-entropy$\le$entropy for jump processes becomes saturated in the continuous limit (even far from equilibrium).

\section{Time-Dependent Driving}\label{t-inhomo}

We now show how most of the results shown so far generalize to systems
with time-dependent driving (i.e., time-inhomogeneous systems), i.e.,
where the generator is explicitly time-dependent $\f{L}(v\tau)$ with
some (constant) protocol velocity $v$. Notably, this means that the
propagator is no longer time-translation invariant, $P(i, \tau'|x,
\tau;v)\neq P(i, \tau'-\tau|x, 0;v)$, as it reads
\cite{HANGGI1982207,kwon2024unifiedframeworkclassicalquantum}
\begin{align}
    P(i, \tau'|x, \tau;v) = \left[\mathcal{T}\exp\left\{\int_\tau^{\tau'}\rmd s \f{L}(vs)\right\}\right]_{ix},
\end{align}
where $\mathcal{T}$ is the time ordering operator. A straightforward
calculation reveals that the expressions for increment correlations \eqref{eq:Xi1}, \eqref{eq:Xi2}, and \eqref{eq:Xi3},
observable covariances \eqref{eq:TransientCovariances}, and transport bound \eqref{eq:TB} remain unchanged apart form additional arguments for $v$. For instance, the observable covariances become
\begin{widetext}
\begin{align}
    \mathrm{cov}({\rho}_t^k, {\rho}_t^l) =& \hat{{I}}^{t}_{V^k\delta, V^l\delta}\left[P(x,\tau;i,\tau';v) - p_x(\tau;v)p_i(\tau';v)\right]\nonumber,
    \\
    \mathrm{cov}({J}_t^k, {\rho}_t^l) =& \hat{{I}}^{t}_{\kappa^k, V^l\delta}\left[\textcolor{black}{\mathbb{1}}_{\tau>\tau'}r_{xy}P(x, \tau| i, \tau';v)p_{i}(\tau';v) + \textcolor{black}{\mathbb{1}}_{\tau<\tau'}r_{yx}(v\tau) P(i, \tau'| y, \tau;v)p_y(\tau;v)- r_{xy}(v\tau)p_x(\tau;v)p_i(\tau';v)\right]\nonumber,
    \\
    \mathrm{cov}(J_t^k, {J}_t^l) =&  \int_0^t\rmd\tau\sum_{x, y}\kappa_{xy}^k(v\tau) \kappa_{xy}^l(v\tau) r_{xy}(v\tau)p_x(\tau;v)\nonumber\\&+ \hat{{I}}^{t}_{\kappa^k, \kappa^l}\left[\textcolor{black}{\mathbb{1}}_{\tau>\tau'}r_{xy}(v\tau)r_{ji}(v\tau') P(x, \tau| j, \tau';v)p_{i}(\tau';v) + \textcolor{black}{\mathbb{1}}_{\tau<\tau'}r_{ij}(v\tau')r_{yx}(v\tau) P(i, \tau'| y, \tau;v)p_y(\tau;v) \right.\nonumber\\&-\left. r_{xy}(v\tau)p_x(\tau;v)r_{ij}(v\tau')p_i(\tau';v)\right]\label{eq:TimeDepTransientCovariances},
\end{align}
\end{widetext}
where the functions $V^{l/k}(v\tau)\delta$ and $\kappa^{l/k}(v\tau)$
inside the integration operator $\hat{{I}}^{t}_{\cdot, \cdot}$ are
time dependent as well. 

However, this does \emph{not} hold for the CTUR as shown in Refs.~\cite{kwon2024unifiedframeworkclassicalquantum, TURTimeDependentDriving}, since the TUR with time-dependent driving reads \cite{kwon2024unifiedframeworkclassicalquantum, TURTimeDependentDriving}
\begin{align}
    \frac{\left[\left(t\partial_t - v\partial_v\right)\langle J_t\rangle\right]^2}{\mathrm{var}(J_t)}\leq \frac{\Delta S_\mathrm{tot}(t)}{2},\label{eq:TimeDepTUR}
\end{align}
and thus requires some special care. 
The observable current and entropy production entering Eq.~\eqref{eq:TimeDepTUR} read
\begin{align}
    \Delta S_\mathrm{tot}(t) &= \int_0^t\rmd \tau \sum_{x,y\neq x}r_{xy}(v\tau)p_x(\tau;v)\log\frac{r_{xy}(v\tau)p_x(\tau;v)}{r_{yx}(v\tau)p_y(\tau;v)},\nonumber\\
    J_t &= \StoInt\mathrm{Tr}\left[\f{\kappa}(v\tau)^T\rmd \f{n}(\tau)\right].
\end{align}
Note that although we write $\rmd \f{n}(\tau)$, the differential is
\emph{not} independent of $v$. This can be seen, e.g., in the
statistical properties, such as the mean $\langle \rmd
n_{xy}(\tau)\rangle_{x_\tau = i} = \delta_{ix}r_{xy}(v\tau)\rmd
\tau$. However, as the differential $\rmd \f{n}(\tau)$ is ``observed''
at time $\tau$, regardless of $v$, we choose to suppress the $v$
argument. Central to the proof of the TUR for driven systems is the identity \cite{kwon2024unifiedframeworkclassicalquantum}
\begin{align}
    &\int_0^{\tau'}\rmd \tau \sum_x \left[\partial_\tau P(i, \tau'|x, \tau;v)\right]p_x(\tau;v)\nonumber \\&= -(\tau'\partial_{\tau'} - v\partial_v)p_i(\tau';v)\,.
    \label{eq:TimeDepDerIdentity}
\end{align}
With this identity, we can extend the time-dependent TUR in
Eq.~\eqref{eq:TimeDepTUR} to also include densities,
\begin{align}
    \rho_t &= \StoInt\textcolor{black}{{V}_{v\tau}\rmd {\tau}},
\end{align}
as well.  To do so, we require $\langle A_t\rho_t\rangle$, which after some simplification using Eq.~\eqref{eq:NoiseTimeCorrelationLemma} reads
\begin{widetext}
   \begin{align}
    \langle A_t \rho_t\rangle 
    =&\int_0^t\rmd\tau^\prime \int_0^t \rmd\tau \sum_{i}V_{i}(v\tau^\prime)\textcolor{black}{\mathbb{1}}_{\tau < \tau^\prime} \label{eq:TimeDepRhoCor}\sum_{x,y}\left[P(i, \tau^\prime|y, \tau;v) - P(i, \tau^\prime|x, \tau;v)\right]p_x(\tau;v) r_{xy}(v\tau)Z_{xy}(v\tau).
\end{align} 
Rearranging the inner sum ($x,y$), we find that Eq.~\eqref{eq:TimeDepRhoCor} simplifies to
\begin{align}
    \langle& A_t \rho_t\rangle \label{eq:TimeDepRhoCor2}
    \int_0^t\rmd\tau^\prime \int_0^{\tau^\prime } \rmd\tau \sum_{i, x}V_{i}(v\tau^\prime)P(i, \tau^\prime|x, \tau;v)\partial_\tau p_x(\tau;v)
    = - \int_0^t\rmd\tau^\prime \int_0^{\tau^\prime } \rmd\tau \sum_{i,x}V_{i}(v\tau^\prime)\left[\partial_\tau P(i, \tau^\prime|x, \tau;v)\right] p_x(\tau;v)\nonumber
    ,
\end{align}
\end{widetext}
where we performed a partial integration over $\tau$ to obtain the third line. The boundary term vanishes because $\sum_x P(i, \tau^\prime|x, \tau';v)p_x(\tau';v) = \sum_x P(i, \tau^\prime|x, 0;v)p_x(0;v)$. Using Eq.~\eqref{eq:TimeDepDerIdentity}, the integral over $\tau$ and sum over $x$ vanish, such that
\begin{align}
    \langle& A_t \rho_t\rangle =\int_0^t\rmd\tau^\prime \sum_{i}V_{i}(v\tau^\prime)\left(\tau'\partial_{\tau'} - v\partial_v\right)p(\tau';v).
\end{align}
Another integration by parts and recognizing $\tau'\partial_{\tau'}V_i(v\tau') = v\partial_v V_i(v\tau')$ finally yields
\begin{align}
    \langle& A_t \rho_t\rangle = \left(t\partial_t - v\partial_v - 1\right)\langle \rho_t\rangle.
\end{align}
Hence, the Cauchy-Schwarz inequality delivers (for the first time) the CTUR for time-dependent driving
\begin{align}
    \frac{\left(\hat{\Lambda} \langle J_t\rangle - c(t)\left\{\left[\hat{\Lambda} - 1\right]\langle \rho_t\rangle \right\}\right)^2}{\mathrm{var}(J_t-c(t)\rho_t)}\leq \frac{\Delta S_\mathrm{tot}}{2},\label{eq:TimeDependentcTUR}
\end{align}
with differential operator $\hat{\Lambda} \equiv t\partial_t -
v\partial_v$. 
Notably, we are still able to saturate the Cauchy-Schwarz inequality,
although the practical inaccessibility of the required transition weights and
state function remains. However, one may again optimize the CTUR
Eq.~\eqref{eq:TimeDependentcTUR} w.r.t. $c(t)$. The results are again
given by Eq.~\eqref{eq:optimal_c} albeit with $a(t) = \hat{\Lambda} \langle J_t\rangle$ and $b(t) = (\hat{\Lambda} -1 )\langle \rho_t\rangle$.

Conversely, the correlation bound  is \emph{not} easily generalizable
to systems with time-dependent driving. Specifically, since
$z_x^{F(\tau)}$ in general explicitly depends on time $\tau$ through
$F(\tau)$, Eq.~\eqref{eq:TimeDepDerIdentity} \emph{cannot} be applied
to shift the time derivative of the propagator. Hence, the expectation
$\langle B_tC_t\rangle$ [see Eq.~\eqref{eq:CrossCorrelationBC}] does
\emph{not} reduce to an accessible quantity.  

\textcolor{black}{\section{Connection to Quantum Unraveling}\label{Quantum}}
\textcolor{black}{Stochastic calculus has also found traction 
in the theory of open quantum systems \cite{Belavkin1990,Gardiner_1992,Caiaffa_2017,Carollo_2019, Bassi_2006}, i.e., in the context of quantum unraveling. In the following, we provide some insight into how the classical stochastic-calculus approach presented here is connected to quantum unraveling. Before doing so, we introduce some basic notation and theory. Subsequently, we highlight the relation to our work.\newline\indent Let $\rho_s^\mathrm{u}$ be a rank-1 matrix denoting a pure state in a discrete-state open quantum system at time $s$. It evolves according to the Belavkin equation \cite{Carollo_2019}
\begin{align}
    \rmd \rho_s^\mathrm{u} = \mathcal{B}(\rho_s^\mathrm{u})\rmd s + \sum_i\left(\frac{\mathcal{J}_i(\rho_s^\mathrm{u})}{\mathrm{Tr}[\mathcal{J}_i(\rho_s^\mathrm{u})]} - \rho_s^\mathrm{u}\right)\rmd \tilde{n}_{is},\label{Belavkin}
\end{align}
where $\mathcal{J}_i(\rho_s^\mathrm{u}) = \hat{J}_i\rho_s^\mathrm{u}\hat{J}_i^\dagger$ with jump operator $\hat{J}_i$ for the $i$th quantum jump. The first term in Eq.~\eqref{Belavkin} has the form
\begin{align}
    \mathcal{B}(\rho_s^\mathrm{u}) = -i\hat{H}_\mathrm{eff}\rho_s^\mathrm{u} + \rho_s^\mathrm{u}\hat{H}_\mathrm{eff}^\dagger - \rho_s^\mathrm{u}\mathrm{Tr}(-i\hat{H}_\mathrm{eff}\rho_s^\mathrm{u} + \rho_s^\mathrm{u}\hat{H}_\mathrm{eff}^\dagger),
\end{align}
where the Hamiltonian $\hat{H}$ of the system enters through the
effective Hamiltonian $\hat{H}_\mathrm{eff} = \hat{H} -
\frac{i}{2}\sum_j\hat{J}_j^\dagger\hat{J}_j$. Lastly, $\rmd
\tilde{n}_{is}$ is a stochastic differential (i.e., a sequence of step
functions) which takes values 1 if
the $i$th transition occurs in $[s,s+\rmd s]$ and 0 otherwise.
Since the probability of more than one transition in the interval
$\rmd s$ vanishes faster than $\rmd s$ \cite{Gardiner_QM}, we have $ \rmd \tilde{n}_{is}\rmd
\tilde{n}_{js} = \delta_{ij}\rmd \tilde{n}_{is}$ and $\rmd
\tilde{n}_{is}\rmd s=0$. Furthermore, we have
$\langle \rmd \tilde{n}_{is}\rangle =
\mathrm{Tr}(\mathcal{J}_i(\rho_s))$. 
Taking the average
over noise histories in Eq.~\eqref{Belavkin}, the resulting density
matrix $\rho_s = \langle \rho_s^\mathrm{u}\rangle$ evolves according
to the Lindblad equation, often referred to as ``quantum master
equation''.
\newline\indent With this short insight into quantum SDE's, we now turn to the question of how these concepts are related to the work we present in this article. Since classical stochastic dynamics should correspond to a special case of quantum open-system evolutions \cite{Breuer_2007}, 
one may at first think that Eq.~\eqref{eq:eom_MJP} describes the classical counterpart to Eq.~\eqref{Belavkin}; they both aim to describe a trajectory (the former a classical sequence of discrete states and the latter the evolution of pure states in a Hilbert space) and both include Poissonian noise $\rmd \f{n}$ and $\rmd\tilde{n}$, respectively.  
\\ 
\indent However, Eq.~\eqref{eq:eom_MJP} is \emph{not} the classical counterpart to Eq.~\eqref{Belavkin}, which can be seen from two observations: (i) Eq.~\eqref{eq:eom_MJP} is an SDE describing the evolution of transitions and (ii) Eq.~\eqref{Belavkin} is a \emph{functional} of the transitions describing the evolution of the states. Moreover, Eq.~\eqref{eq:eom_MJP} allows us to consider the (stochastic) evolution of transitions occuring in a discrete systems and further enables us to form and study additive functionals of the dynamics. Hence, using (i) enables us to study functionals such as (ii). To be precise on this last point, we can construct a classical functional of the transitions which is the classical counterpart to Eq.~\eqref{Belavkin}. It turns out that this functional simply is $\rmd \textcolor{black}{\delta_{x_s x}}\equiv\frac{\rmd}{\rmd \tau}\textcolor{black}{\delta_{x_s x}}\rmd\tau$, see Eq.~\eqref{id_functional}. Explicitly, using Eq.~\eqref{eq:eom_MJP} in Eq.~\eqref{id_functional} yields
\begin{align}
    \rmd \textcolor{black}{\delta_{x_s x}} &= \sum_{y\neq x} (\rmd n_{yx} - \rmd n_{x y})\label{ClassicalBelavkin}\\
    &= \sum_{y\neq x} (r_{yx}\delta_{x_s y}\rmd s - r_{x y}\delta_{x_s x}\rmd s) +  \sum_{y\neq x} (\rmd \varepsilon_{yx} - \rmd \varepsilon_{x y}).\nonumber
\end{align}
The first sum is the deterministic evolution of the state, i.e., the master operator acting on the ``population vector'' $\delta_{x_s x}$ and the second sum is the zero-mean stochastic noise in the system. Shifting the mean of the noise increments by adding and subtracting $\sum_{y\neq x} (\langle\rmd n_{yx}\rangle - \langle\rmd n_{x y}\rangle)$ [this corresponds to the trace terms in Eq.~\eqref{Belavkin}], we recover the classical analogue of Eq.~\eqref{Belavkin}
\begin{align}
    \rmd \delta_{x_s x} =& \sum_{y\neq x}\left(r_{yx}\delta_{x_s y}\rmd s - \langle \rmd n_{yx}\rangle -r_{xy}\delta_{x_s x}\rmd s + \langle \rmd n_{xy}\rangle \right)\nonumber\\
    &+\sum_{y\neq x}(\rmd \varepsilon_{yx} + \langle \rmd n_{yx}\rangle - \rmd \varepsilon_{x y}-\langle \rmd n_{xy}\rangle).
\end{align}
Indeed, taking the average of Eq.~\eqref{ClassicalBelavkin} w.r.t.\ the noise history recovers the master equation, i.e., the classical analogue of the Lindblad equation, further showing the correspondence between Eqs.~\eqref{Belavkin} and~\eqref{ClassicalBelavkin}.\newline\indent
For completeness it should be mentioned that the quantum analogue of Eq.~\eqref{eq:eom_MJP} has already been identified in Ref.~\cite{kwon2024unifiedframeworkclassicalquantum} as
\begin{align}
    \rmd \tilde{n}_{it} =  \mathrm{Tr}(\mathcal{J}_i(\rho_t))\rmd t + \rmd \tilde{\varepsilon}_{it},
\end{align}
where the noise $\rmd \tilde{\varepsilon}_{it}$ corresponds to Eq.~\eqref{eq:DiscreteNoise} and the statistical properties can be evaluated in a similar fashion; $\langle \rmd \tilde{\varepsilon}_{it}\rangle = 0$ and $\langle \rmd \tilde{\varepsilon}_{it}\rmd \tilde{\varepsilon}_{jt'}\rangle = \delta_{ij}\delta(t-t')\mathrm{Tr}(\mathcal{J}_i(\rho_t))\rmd t \rmd t'$. While the connection to the unraveling of the Lindblad equation and, thus, the Belavkin equation is therefore established, the question of how this in turn relates to the classical counterpart remains elusive, further motivating our discussion here.}\\

\section{Outlook}\label{Outlook}
We developed a stochastic-calculus for path-wise observables (i.e., functionals) of
Markov-jump processes that unifies the different approaches to Markov-jump
dynamics, as well as the descriptions of diffusion and
jump dynamics. We presented the approach as an exact
parallelism with the continuous-space diffusion counterpart, 
starting from a ``Langevin equation for Markov-jump processes''
together with a central noise-time-correlation Lemma, defined general
path-wise observables and determined their complete (co)variation
structure, and identified (generalized) Green-Kubo relations.  
The approach includes general transients and time-inhomogeneous dynamics
(e.g., time-dependent driving). 
We used the stochastic
calculus to prove \emph{directly} the known thermodynamic inequalities (TUR,
correlation TUR, transport bounds, bounds on correlations etc.) in their most
general form and, afforded by the directness of the approach,
discussed their saturation conditions. We showed that these inequalities
follow directly from the equations of motion, which establishes them
as an inherent property of stochastic (Markov-jump) equations of motion.  
 
We further derived the response
of any (generally path-wise) observable to a general (incl.\ thermal)
perturbation 
\textcolor{black}{and constructed a corresponding response-function formalism for single-time as well as path-wise observables. Subsequently}, we established the continuum limit to achieve the complete unification of diffusion and jump dynamics. \textcolor{black}{Lastly, we derived the classical analogue of the Belavkin equation and commented on the analogy with open quantum systems.} Our results 
ultimately place functionals of diffusion and jump processes on an equal
footing on the level of individual stochastic realizations, and hence achieve a
``contraction'' of two until now disjoint frameworks in time-average 
statistical mechanics. \textcolor{black}{Our newly developed framework has recently already been applied in \cite{zheng2026}.}

The results enable, and are expected to inspire, new
directions of research. In particular, one may extend the
jump-dynamics results to
account for dynamics in phase space (i.e., momentum
exchange) in the spirit of \cite{Nicolis_Prigogine_1977} (see Chapter
11 therein) on the discrete-state
side. Conversely, one may also deepen the results on underdamped dynamics (i.e., unify the
transport bound and the TUR derived in
\cite{CrutchfieldUnderdampedTUR}). Moreover, we expect our results to
inspire research in the direction of discrete-state analogs of generative
diffusion models as in Ref.~\cite{Rotskoff2025,x5vj-8jq9} and the learning of
stochastic thermodynamics from fluctuating discrete-state trajectories
as recently carried out for diffusions in Ref.~\cite{lyu}. There the
power of stochastic calculus unfolds in a very prominent manner.  
\vspace{0.2cm}\\
\textcolor{white}{Uffbassa! Mir hen da wos g'scheid's gmochdd!}\vspace{0.2cm}\\

\section{Data availability}
The data that support the findings of this article are openly
available at \cite{stutzer_2025_16572781}.

\section*{Acknowledgments}
\textcolor{black}{We thank Jorge "Jefe" Tabanera-Bravo for discussions on quantum systems. }The financial support from the European Research Council (ERC) under
the European Union’s 
Horizon Europe research and innovation program (Grant Agreement
No.\ 101086182 to AG) and
the German Research Foundation (DFG) through the Heisenberg Program
(grant GO 2762/4-1 
to AG) is gratefully acknowledged.

\appendix

\section{Proofs and Derivations}\label{Proofs}

In the Appendix we provide some technical results that are required to
understand the detailed derivations but are not critical for
a conceptual understanding of the results. 

\subsection{Proof of Noise-Time Correlation Lemma}\label{sec:NoiseTimeProof}
\begin{figure}
    \centering
     \includegraphics[width=.6\linewidth]{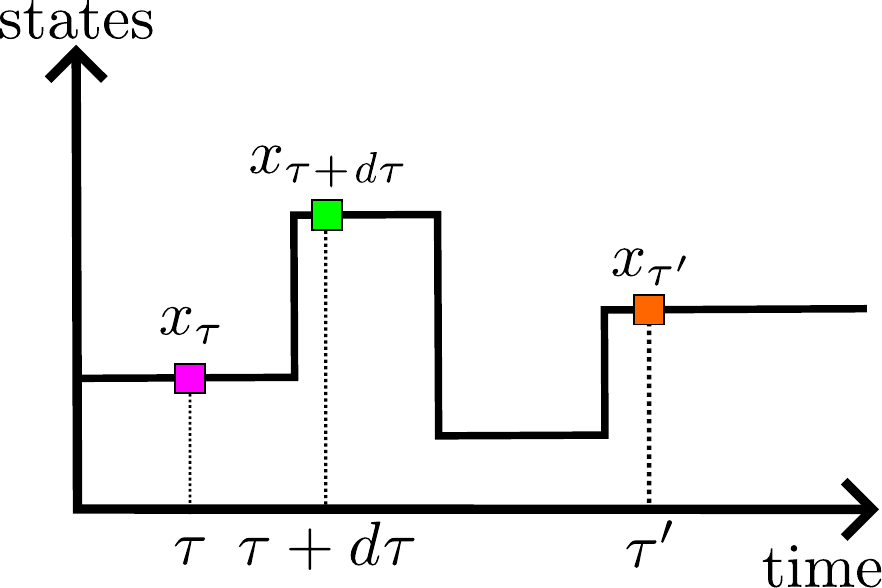}
    \caption{Visualization of the intermediate point considered in the proof of Noise-Time correlation lemma. The general idea to evaluate the noise in $x_\tau$ at $\tau$ and being in state $x_{\tau'}$ at time $\tau'>\tau$ is to include the intermediate point $x_{\tau+\rmd \tau}$.}
    \label{fig:Proof_sketch}
\end{figure}

To prove how the noise Eq.~\eqref{eq:DiscreteNoise} and time spent in a
state in Eq.~\eqref{eq:dtau} correlate, i.e., $\langle \rmd {\varepsilon}_{xy}(\tau)\rmd{\tau}_i(\tau^\prime)\rangle$, we consider three cases: the transition from $x$ to $y$ happens (i) before $\tau^\prime$, (ii) at $\tau^\prime$, or (iii) after $\tau^\prime$, corresponding to $\tau < \tau^\prime$, $\tau = \tau^\prime$, and $\tau > \tau^\prime$, respectively. Specifically, we can write 
\begin{align}
    \langle \rmd {\varepsilon}_{xy}(\tau)\rmd{\tau}_i(\tau^\prime)\rangle =& \textcolor{black}{\mathbb{1}}_{\tau \geq \tau'}\langle \rmd {\varepsilon}_{xy}(\tau)\rmd{\tau}_i(\tau^\prime)\rangle\nonumber\\ &+\textcolor{black}{\mathbb{1}}_{\tau < \tau'}\langle \rmd {\varepsilon}_{xy}(\tau)\rmd{\tau}_i(\tau^\prime)\rangle,
\end{align}
where the first term can be written as
\begin{align}
    &\textcolor{black}{\mathbb{1}}_{\tau \geq \tau'}\langle \rmd {\varepsilon}_{xy}(\tau)\rmd{\tau}_i(\tau^\prime)\rangle \nonumber\\&= \underbrace{\sum_{\rmd \varepsilon_{xy}}\rmd \varepsilon_{xy} p(\rmd \varepsilon_{xy})}_{=0} P(x, \tau|i, \tau')p_i(\tau').
\end{align}
Since the system is Markovian, (ii) and (iii) do not contribute. Hence, without loss of generality we can write
\begin{align}
    \langle \rmd {\varepsilon}_{xy}(\tau)\rmd{\tau}_i(\tau^\prime)\rangle = \textcolor{black}{\mathbb{1}}_{\tau < \tau'}\langle \rmd {\varepsilon}_{xy}(\tau)\rmd{\tau}_i(\tau^\prime)\rangle.
    \label{eq:MarkovianNoiseTimeCorrelation}
\end{align}
For small $\rmd \tau\to 0$, the noise increment can only take two values
\begin{align}
    \rmd \varepsilon_{xy}(\tau) = \textcolor{black}{\delta_{x_\tau x}}(\tau)
    \begin{cases}
        1 - r_{xy}\rmd \tau & \text{transition},\\
        -r_{xy}\rmd \tau & \text{no transition}.
    \end{cases}
\end{align}
These values occur with probability $r_{xy}\rmd \tau$ and $1-r_{xy}\rmd\tau$, respectively. Defining the state function
\begin{align}
    \gamma(\rmd \varepsilon_{xy}(\tau)) =
    \begin{cases}
       y & \text{transition},\\
        x& \text{no transition},
    \end{cases}
\end{align}
allows us to explicitly evaluate Eq.~\eqref{eq:MarkovianNoiseTimeCorrelation} by introducing an intermediate point (see Fig.~\ref{fig:Proof_sketch}), such that to leading order
\begin{widetext}
    \begin{align}
        \langle \rmd {\varepsilon}_{xy}(\tau)\rmd{\tau}_i(\tau^\prime)\rangle 
        &= \textcolor{black}{\mathbb{1}}_{\tau < \tau^\prime}\rmd\tau^\prime\sum_{\rmd{\varepsilon}_{xy}(\tau)} p(\rmd{\varepsilon}_{xy}(\tau)) \rmd{\varepsilon}_{xy}(\tau)P(i, \tau^\prime|\gamma(\rmd{\varepsilon}_{xy}(\tau)), \tau+\rmd\tau) p_x(\tau)\nonumber\\
        &=\textcolor{black}{\mathbb{1}}_{\tau < \tau^\prime}\rmd\tau^\prime p_x(\tau) r_{xy}\rmd\tau\left(1-r_{xy}\rmd\tau\right)\left[P(i, \tau^\prime|y, \tau + \rmd\tau) - P(i, \tau^\prime|x, \tau + \rmd\tau)\right]\nonumber\\
        &= \textcolor{black}{\mathbb{1}}_{\tau < \tau^\prime}\rmd\tau^\prime \left[P(i, \tau^\prime|y, \tau) - P(i, \tau^\prime|x, \tau)\right]p_x(\tau) r_{xy}\rmd\tau\,.
    \end{align}    
\end{widetext}

\subsection{Derivation of Increment Correlations}
\subsubsection{Jump-Time Correlations\label{sec:JumpTimeProof}}
We see for an explicit expression for $\langle \rmd n_{xy}(\tau)\rmd
\tau_i(\tau')\rangle$. We can decompose this expectation using
Eq.~\eqref{eq:eom_MJP} as
\begin{align}
    \langle \rmd n_{xy}(\tau)\rmd \tau_i(\tau')\rangle = \langle \rmd \varepsilon_{xy}(\tau)\rmd \tau_i(\tau')\rangle + r_{xy} \langle \rmd \tau_x(\tau)\tau_i(\tau')\rangle.
\end{align}
Using the last two equations in Eqs.~\eqref{eq:ListExpectations}, we get
\begin{align}
    \frac{\langle \rmd n_{xy}(\tau)\rmd \tau_i(\tau')\rangle}{r_{xy}\rmd \tau \rmd \tau'} =&\textcolor{black}{\mathbb{1}}_{\tau<\tau'}P(i, \tau'|y, \tau)p_x(\tau) \nonumber\\&+ \textcolor{black}{\mathbb{1}}_{\tau\geq\tau'}P(x, \tau|i, \tau')p_i(\tau').
\end{align}

\subsubsection{Jump-Jump Correlations\label{sec:JumpJumpProof}}
Similar to the jump-time correlation, we again decompose the jump-jump
correlation as
\begin{align}
    \langle \rmd n_{xy}(\tau)\rmd n_{ij}(\tau')\rangle =& \langle \rmd \varepsilon_{xy}(\tau)\rmd \varepsilon_{ij}(\tau')\rangle \nonumber\\
    &+ r_{xy} \langle\rmd \tau_x(\tau)\rmd\varepsilon_{ij}(\tau')\rangle\nonumber\\
    &+r_{ij} \langle\rmd \varepsilon_{xy}(\tau)\rmd\tau_{i}(\tau')\rangle \nonumber\\
    &+ r_{xy}r_{ij} \langle \rmd \tau_x(\tau)\tau_i(\tau')\rangle.
\end{align}
Using Eqs.~\eqref{eq:ListExpectations} therefore yields
\begin{align}
    \frac{\langle \rmd n_{xy}(\tau)\rmd n_{ij}(\tau')\rangle}{r_{xy}r_{ij}\rmd \tau \rmd \tau'}=&\textcolor{black}{r_{ij}}\delta_{ix}\delta_{jy}\delta(\tau-\tau')p_x(\tau)\nonumber\\
    &+ \textcolor{black}{\mathbb{1}}_{\tau<\tau'}P(i, \tau'|y, \tau)p_x(\tau) \nonumber\\
    &+ \textcolor{black}{\mathbb{1}}_{\tau\geq\tau'}P(x, \tau|j, \tau')p_i(\tau').
\end{align}

\subsection{From Equations of Motion to Ensemble Description \label{sec:EoM_to_Ensemble}}

\subsubsection{Continuous Space: Langevin to Fokker-Planck}
To highlight the analogy between continuous- and discrete space
dynamics,  we first recapitulate how the Fokker-Planck equation follows directly from the  Langevin
Equation in Eq.~\eqref{eq:LangevinEq}. The calculation is well \textcolor{black}{k}nown
(see e.g., Ref.~\cite{gardiner2004handbook}). Let $K(x)$
be an arbitrary test function that is sufficiently smooth and obeys the boundary
conditions. Considering the expectation $\langle K(\f{x}(t))\rangle =
\int \rmd\f{x} K(\f{x}) P(\f{x},t)$ for the dummy function at some
time $t$, the temporal derivative can be evaluated using It\^{o}'s
lemma \cite{gardiner2004handbook}
\begin{align} 
    &\frac{\rmd}{\rmd t}\langle K(\f{x}(t))\rangle = \langle \f{F}(\f{x})\cdot\nabla K(\f{x}) + \nabla\cdot\f{D}\nabla K(\f{x})\rangle\label{eq:FokkerPlanckDer1}\\
    &=\int \rmd \f{x}\textcolor{black}{ P(\f{x},t)}\left[\f{F}(\f{x})\cdot\nabla K(\f{x}) + \nabla\cdot\f{D}\nabla K(\f{x})\right].\nonumber
\end{align}
Equivalently, the derivative can be written as
\begin{align}
    \frac{\rmd}{\rmd t}\langle K(\f{x}(t))\rangle = \int \rmd \f{x} K(\f{x})\frac{\partial}{\partial t} P(\f{x}, t).
    \label{eq:FokkerPlanckDer2}
\end{align}
By equating Eqs.~\eqref{eq:FokkerPlanckDer1} and
\eqref{eq:FokkerPlanckDer2} we find 
\begin{align}
    \int& \rmd \f{x} K(\f{x})\frac{\partial}{\partial t} P(\f{x}, t) \nonumber\\=& \int \rmd \f{x}\textcolor{black}{ P(\f{x},t)}\left[\f{F}(\f{x})\cdot\nabla K(\f{x}) + \nabla\cdot\f{D}\nabla K(\f{x})\right] \\
    =& \int \rmd \f{x} K(\f{x})\left[- \nabla\left\{\f{F}(\f{x})P(\f{x}, t)\right\} \right.+\left. \nabla \cdot \f{D} \nabla P(\f{x}, t)\right]\,,\nonumber
\end{align} 
where we performed a partial integration to obtain the last line (the
boundary terms vanish by the properties assumed on $K(\x)$). Since the
above equation holds for any test function $K(\x)$ we derived the
Fokker-Planck equation
\begin{align}
\partial_t P(\f{x}, t)=[\nabla \cdot \f{D} \nabla -\nabla \f{F}(\f{x})]P(\f{x}, t).
  \label{FPE}
  \end{align}

\subsubsection{Discrete Space: Jumps to Master Equation}

To show how the master equation emerges from the stochastic equations
of motion in Eq.~\eqref{eq:eom_MJP}, we first need to consider the indicator function $\textcolor{black}{\delta_{x_\tau x}}$ of being in state $x$ at time $\tau$. Consider $\textcolor{black}{\delta_{x_{\tau+\rmd\tau} x}} = \textcolor{black}{\delta_{x_\tau x}} + \frac{\rmd}{\rmd \tau}\textcolor{black}{\delta_{x_\tau x}}\rmd \tau + \mathcal{O}(\rmd\tau^2)$, where the derivative should be understood as
\begin{align}
    \frac{\rmd}{\rmd \tau}\textcolor{black}{\delta_{x_\tau x}} = \lim_{\rmd \tau\to0^+}\frac{\textcolor{black}{\delta_{x_{\tau+\rmd\tau} x}} - \textcolor{black}{\delta_{x_\tau x}}}{\rmd \tau}.
\end{align}
By the non-explosive property of MJP \cite{Bremaud1999} a finite number of jumps
occur in any finite time interval with probability 1. Therefore, we do
not need to consider any higher order correction, as the probability
of a second transition occurring in an infinitesimal increment
vanishes. The term $\frac{\rmd}{\rmd \tau}\textcolor{black}{\delta_{x_\tau x}}\rmd \tau$
can therefore take on values
\begin{align}
    \frac{\rmd}{\rmd \tau}\textcolor{black}{\delta_{x_\tau x}}\rmd \tau = 
    \begin{cases}
    0 & \text{no transition in }[\tau, \tau+\rmd \tau],\\
    1 & x_\tau=y \neq x \text{, $y\to x$ in }[\tau, \tau+\rmd \tau],\\
    -1 & x_\tau = x \text{,  $x\to y\neq x$ in }[\tau, \tau+\rmd \tau].\\
    \end{cases}
\end{align}
As a consequence, we can write
\begin{align}
    \frac{\rmd}{\rmd \tau}\textcolor{black}{\delta_{x_\tau x}}\rmd \tau = \sum_{y\neq x}\textcolor{black}{[}\rmd n_{yx}(\tau) - \rmd n_{xy}(\tau)\textcolor{black}{]},\label{id_functional}
\end{align}
because at most one jump differential will be non-zero as $\rmd \tau\to 0$. The probability of being in a state $x$ at time $\tau$ is $p_x(\tau) = \langle \textcolor{black}{\delta_{x_\tau x}}\rangle$. Therefore, 
\begin{align}
    p_x(\tau+\rmd\tau) - p_x(\tau) &= \left\langle \textcolor{black}{\delta_{x_{\tau+\rmd\tau} x}} -\textcolor{black}{\delta_{x_\tau x}} \right\rangle\nonumber\\
    &= \left\langle \frac{\rmd}{\rmd \tau} \textcolor{black}{\delta_{x_\tau x}} \rmd \tau \right\rangle\label{eq:EoMtoMaster}\\
    &= \left\langle \sum_{y\neq x}\textcolor{black}{[}\rmd n_{yx}(\tau) - \rmd n_{xy}(\tau)\textcolor{black}{]} \right\rangle\nonumber\\
    &= \rmd \tau \sum_{y\neq x}\textcolor{black}{[} r_{yx}p_y(\tau) - r_{xy}p_x(\tau)\textcolor{black}{]}\nonumber
\end{align}
Rearranging Eq.~\eqref{eq:EoMtoMaster} and taking the limit
$\rmd\tau\to0$ yields the master equation~\eqref{eq:MasterEquation}  in a way equivalent to the
Fokker-Planck equation~\eqref{FPE}.

\subsection{Derivation of \texorpdfstring{Eq.~\eqref{eq:AJ2correlatorFinal}}{}\label{sec:Current2AcorrelationExplicit}}
Starting from Eq.~\eqref{eq:AJ2correlator}, we can simplify the sum over $x,y$ by using the symmetry of $Z_{xy}(\tau)[P(i, \tau'|y, \tau) - P(i, \tau'|x, \tau)]$ to get
\begin{align}
    \langle& A_t J_t^\mathrm{II}\rangle \nonumber\\
    =&\frac{1}{2}\int_0^t\rmd\tau^\prime \int_0^t \rmd\tau \sum_{i,j}\kappa_{ij}(\tau^\prime)r_{ij}\textcolor{black}{\mathbb{1}}_{\tau < \tau^\prime} \nonumber\\&\times\sum_{x,y}\left[P(i, \tau^\prime|y, \tau) - P(i, \tau^\prime|x, \tau)\right]\left[p_x(\tau) r_{xy} - p_y(\tau) r_{yx}\right]\nonumber\\
    =&-\int_0^t\rmd\tau^\prime \int_0^t \rmd\tau \sum_{i,j}\kappa_{ij}(\tau^\prime)r_{ij}\textcolor{black}{\mathbb{1}}_{\tau < \tau^\prime}\label{eq:intermediateCalc1} \\&\times\sum_{x}P(i, \tau^\prime|x, \tau)\sum_{y, y\neq x}\left[p_x(\tau) r_{xy} - p_y(\tau) r_{yx}\right]\nonumber.
\end{align}
With Eq.~\eqref{eq:MasterEquation} and an integration by parts, Eq.~\eqref{eq:intermediateCalc1} reduces to
\begin{align}
    \langle& A_t J_t^\mathrm{II}\rangle \nonumber\\
    =&\int_0^t\rmd\tau^\prime  \sum_{i,j}\kappa_{ij}(\tau^\prime)r_{ij}\left[-\sum_x P(i, \tau'|x, 0)p_x(0)\right.\nonumber\\
    &-\left.\int_0^t\rmd \tau \sum_xp_x(\tau)\partial_\tau\left\{P(i, \tau'|x, \tau)\textcolor{black}{\mathbb{1}}_{\tau<\tau'}\right\}\right]\nonumber\\
    =& - \langle J_t\rangle + \int_0^t\rmd\tau'\sum_{i,j}\kappa_{ij}(\tau')r_{ij}\partial_{\tau'}\left[\tau'p_i(\tau')\right],
\end{align}
where we use that $\partial_\tau P(i, \tau'|x, \tau) = - \partial_{\tau'} P(i, \tau'|x, \tau)$, $\partial_\tau \textcolor{black}{\mathbb{1}}_{\tau<\tau'} = - \partial_{\tau'}\textcolor{black}{\mathbb{1}}_{\tau<\tau'}$, and $\int_0^{\tau'}\rmd \tau \textcolor{black}{\mathbb{1}}_{\tau<\tau'} = \tau'$. With another integration by parts, the correlator finally becomes
\begin{align}
    \langle A_t J_t^\mathrm{II}\rangle  = &- \langle J_t\rangle + \underbrace{t\sum_{i,j}\kappa_{ij}r_{ij}p_i(t)}_{=t\partial_t \langle J_t\rangle} \\
    &- \int_0^t\rmd\tau'\sum_{i,j}\partial_{\tau'}\left[\kappa_{ij}(\tau')\right]r_{ij}\partial_{\tau'}\tau'p_i(\tau').\nonumber
\end{align}

\subsection{Indispensability of the Modified Current}

\begin{figure}[ht]
    \centering
        \includegraphics[width=.9\linewidth]{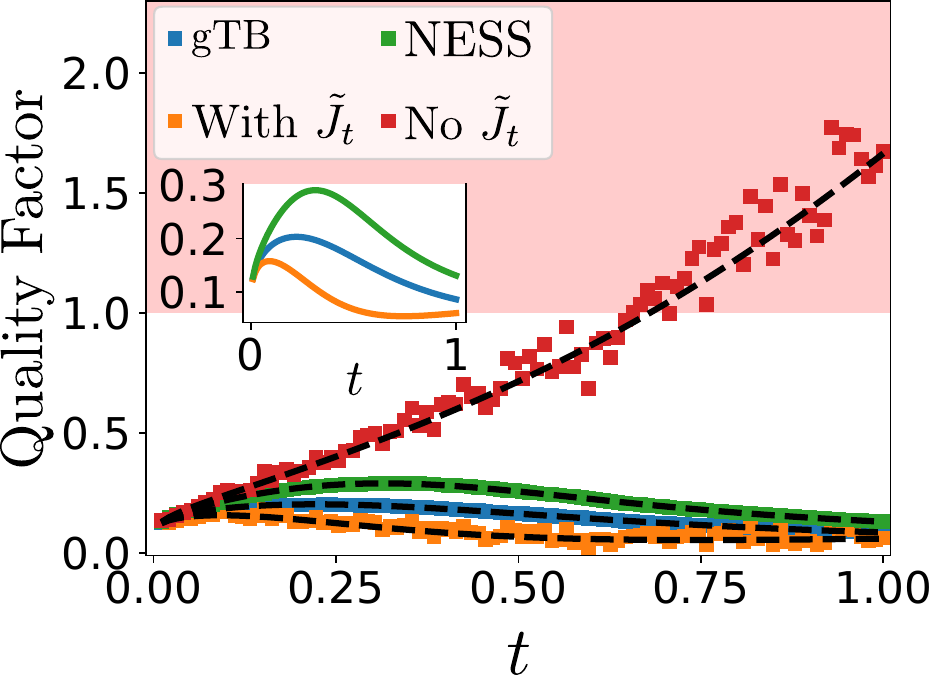}
    \caption[Necessity for modified current]{Validity and comparison of generalized transport bound and various TUR quality factors. A three-state system with generator Eq.~\eqref{eq:MoCurGenerator} and initial condition $p_i(0) = (\delta_{i1} + 2\delta_{i2})/3$. The transition weights is $\kappa_{ij}(\tau) = \mathrm{e}^{4\tau}\left(\delta_{i1}\delta_{j2}-\delta_{i2}\delta_{j1}\right)$. The black dashed lines are the theoretical predictions to the corresponding numerical quality factors of generalized transport bound Eq.~\eqref{eq:gTB} (blue), nonequilibrium steady-state TUR (green), transient TUR with $ {\tilde{J}}_t$ (orange), and transient TUR without $ {\tilde{J}}_t$ (red). A clear violation (shaded red region) of $\mathcal{Q}\leq 1$ is seen for the transient TUR without $ {\tilde{J}}_t$. The inset shows the theory of the former three quality factors to better visualize their values. The numerical values are extracted from $N=5\cdot 10^5$ trajectories using the Gillespie algorithm.}
    \label{fig:QualityFactorBreaking}
\end{figure}

Consider a three-state system with generator
\begin{equation}
    \begin{aligned}
        \f{L} = 
        \begin{pmatrix}
        -4 & 2 & 1\\
        1 & -3 & 1\\
        3 & 1 & -2
        \end{pmatrix}\,,
    \end{aligned}
    \label{eq:MoCurGenerator}
\end{equation}
with eigenvalues $0, -4, -5$. Let the current defining transition weights be
$\kappa_{ij}(\tau) =
\mathrm{e}^{4\tau}\left(\delta_{i1}\delta_{j2}-\delta_{i2}\delta_{j1}\right)$. Figure~\ref{fig:QualityFactorBreaking}
shows various quality factors of the transient system with initial
distribution $p_i(0) = (\delta_{i1} + 2\delta_{i2})/3$. Explicitly,
the generalized transport bound in Eq.~\eqref{eq:gTB}, nonequilibrium steady state TUR, and the transient TUR in
Eq.~\eqref{eq:TransientTUR} with and without modified current
${\tilde{J}}_t$, see Eq.~\eqref{eq:ModifiedCurrent}. As can be seen in
Fig.~\ref{fig:QualityFactorBreaking}, one needs ${\tilde{J}}_t$ in
order to ensure that the bound $\mathcal{Q}\leq 1$ is not violated. To leading order $t\partial_t\langle {J}_t\rangle \sim t\mathrm{e}^{4t}$ for large $t$, while $t\partial_t\langle {J}_t\rangle - \langle {\tilde{J}}_t\rangle \sim \mathrm{e}^{4t}$. Additionally, $\mathrm{var}({J}_t)\Delta S_\mathrm{tot}(t)\sim t\mathrm{e}^{8t}$ for large $t$, so that
\begin{equation}
    \begin{aligned}[b]
        \frac{2\left(t\partial_t\langle {J}_t\rangle\right)^2}{\mathrm{var}({J}_t)\Delta S_\mathrm{tot}(t)}&\sim t\,,\\
        \frac{2\left(t\partial_t\langle {J}_t\rangle - \langle {\tilde{J}}_t\rangle\right)^2}{\mathrm{var}({J}_t)\Delta S_\mathrm{tot}(t)}&\sim \frac{1}{t}\,.
    \end{aligned}
\end{equation}
Hence, the inclusion of the modified current is indeed necessary for
the quality factor to remain bounded, as it otherwise diverges for
$t\to\infty$, which obviously is a violation of the upper bound of the
quality factor. Note that $t$ in Fig.~\ref{fig:QualityFactorBreaking}
is not large enough to observe the $t^{-1}$ decay. 

\subsection{Long-time limit of driven
    ring}\label{AppendixCBdetails}
\subsubsection*{Integrated Covariance}
Here we present details on the large-$t$ limit of the stationary correlation
bound in Eq.~\eqref{eq:StationaryCB} applied to the driven ring
Sec.~\ref{models} with the assumption that $\epsilon>1$.
We start by realizing that \textcolor{black}{$V^\Delta_\tau = V_\tau - \langle V_\tau\rangle = \f \delta(\tau)^T \Delta \f V$ with $\f\delta(\tau)^T=(\delta_{x_\tau 1}, \delta_{x_\tau 2}, \delta_{x_\tau 3}, \delta_{x_\tau 4})$ and} $\Delta \f{V}\equiv \f{V}^\mathrm{A} - \langle {V}^\mathrm{A}\rangle \f{1} =\frac{1}{2} (1, 1, -1, -1)^T$. Hence, we can write
\begin{align}
    \int_0^t\rmd \tau\mathrm{cov}_\mathrm{s}[V^\mathrm{A}_\tau, V^\mathrm{A}_0] =  \frac{1}{4}\int_0^t\rmd \tau \Delta \f{V}^T \f{P}(\tau)\Delta\f{V},  \label{AppendixCB1}
\end{align}
because $p_i^\mathrm{s}=1/4$. Writing the propagator as
\begin{align}
    \f P(\tau) = 
    \begin{pmatrix}
    \f{P}_1(\tau) & \f{P}_2(\tau)\\
    \f{P}_3(\tau) & \f{P}_4(\tau)\\
    \end{pmatrix},
\end{align}
with $\f{P}_i(\tau)$ are $2\times 2$ matrices. Equation~\eqref{AppendixCB1} therefore reads
\begin{align}
    &\int_0^t\rmd \tau\mathrm{cov}_\mathrm{s}[V^\mathrm{A}_\tau, V^\mathrm{A}_0]\label{AppendixCB2}\\&=  \frac{1}{16}\int_0^t\rmd \tau \f{1}^T\left[\f{P}_1(\tau)+\f{P}_4(\tau)-\f{P}_2(\tau)-\f{P}_3(\tau)\right]\f{1}. \nonumber
\end{align}
All that remains is to evaluate the matrices $\f P_i(\tau)$ and solve the integral. Consider the generator
\begin{align}
    \f{L} = 
    \begin{pmatrix}
        -1-\epsilon & 1 & 0 & \epsilon\\
        \epsilon &-1-\epsilon &1 & 0\\
        0 &\epsilon &-1-\epsilon  & 1\\
        1 & 0 & \epsilon & -1-\epsilon 
    \end{pmatrix},
\end{align}
which can be written as $\f{L} = \f{U} \f{E}\f{U}^{-1}$ with matrices
\begin{align}
    \f U &= 
    \begin{pmatrix}
        1 & i & -i & -1\\
        1 & -1 & -1 & 1\\
        1 & -i & i & -1\\
        1 & 1 & 1 & 1
    \end{pmatrix},\\
    \f E &= \mathrm{diag}(0, \lambda_4, \lambda_3, \lambda_2),\\
    \f U^{-1} &= \frac{1}{4}
    \begin{pmatrix}
        1 & 1 & 1 & 1\\
        -i & -1& i& 1\\
        i & -1& -i& 1\\
        1 & 1& -1& 1
    \end{pmatrix}.
\end{align}
The eigenvalues of $\f{L}$ are listed in Eq.~\eqref{CBeigenvalues}. Hence, the propagator is simply $\f{P}(\tau) = \f U \exp{\{\f{E}\tau\}}\f U^{-1}$ and Eq.~\eqref{AppendixCB2} can explicitly be evaluated to be
\begin{widetext}
\begin{align}
    \int_0^t\rmd \tau\mathrm{cov}_\mathrm{s}[V^\mathrm{A}_\tau, V^\mathrm{A}_0] \nonumber=&  \frac{4}{16}\int_0^t\rmd \tau \rme^{-\tau(1+\epsilon)}\cos[\tau(\epsilon-1)]\nonumber\\=&  \frac{1}{4}\frac{\rme^{-\tau(1+\epsilon)}\left((1+\epsilon)\rme^{\tau(1+\epsilon)}-(1+\epsilon)\cos[(\epsilon-1)t]+(\epsilon-1)\sin[(\epsilon-1)t]\right)}{2(1+\epsilon^2)} \nonumber\\
    =&\frac{1}{8}\frac{1+\epsilon}{1+\epsilon^2}+\mathcal{O}(\rme^{-(1+\epsilon)t}).  \label{AppendixCB3}
\end{align}
\end{widetext}

\subsubsection*{(Pseudo-)Variance}
Consider $\mathrm{pvar}_\mathrm{ps}^F(V^\mathrm{A})$ with the choice $F(\tau)=\langle V^\mathrm{A}_x+V^\mathrm{A}_y\rangle_\mathrm{ps}/2$, i.e., the variance $\mathrm{var}_\mathrm{ps}(V^\mathrm{A})$, for the stationary driven four-state ring. The two-point probability is
\begin{align}
    \f p^\mathrm{ps} = \frac{1}{4(\epsilon + 1)}
    \begin{pmatrix}
        0 & \epsilon & 0& 1\\
        1& 0 & \epsilon& 0\\
        0 & 1& 0 & \epsilon\\
        \epsilon & 0 & 1& 0
    \end{pmatrix},
\end{align}
since $\Sigma_\mathrm{ps}=(\epsilon-1)^2/(\epsilon+1)$, $p_x^\mathrm{s}=1/4$, and, for $x\neq y$, $Z_{xy}^2=(\epsilon-1)^2/(\epsilon+1)^2$
Hence,
\begin{align}
    2F =& \frac{1}{4(1+\epsilon)}\left(\underbrace{2\epsilon}_{x=1, y=2}+\underbrace{1}_{x=1, y=4}+\underbrace{2}_{x=2, y=1}\right.\nonumber\\&+\left.\underbrace{\epsilon}_{x=2, y=3}+\underbrace{1}_{x=3, y=2}+\underbrace{\epsilon}_{x=4, y=1}\right)\nonumber\\
    =& 1,
\end{align}
and similarly
\begin{align}
    \left\langle \left(V^\mathrm{A}_x+V^\mathrm{A}_y\right)^2\right\rangle_\mathrm{ps}  =& \frac{1}{4(1+\epsilon)}\left(\underbrace{4\epsilon}_{x=1, y=2}+\underbrace{1}_{x=1, y=4}+\underbrace{4}_{x=2, y=1}\right.\nonumber\\&+\left.\underbrace{\epsilon}_{x=2, y=3}+\underbrace{1}_{x=3, y=2}+\underbrace{\epsilon}_{x=4, y=1}\right)\nonumber\\
    =& 1.5,
\end{align}
so that
\begin{align}
    \mathrm{var}_\mathrm{ps}(V^\mathrm{A})&= \left\langle \left(V^\mathrm{A}_x+V^\mathrm{A}_y\right)^2\right\rangle_\mathrm{ps} - 4F^2= 0.5.
\end{align}
\indent We may immediately recognize that $0\leq V_x^\mathrm{A}+V_y^\mathrm{A}\leq 2$, the former inequality saturates when $x,y\in\{3, 4\}$ and the latter inequality saturates when $x,y\in\{1,2\}$. Thus, Popoviciu's inequality yields
\begin{align}
    \mathrm{var}_\mathrm{ps}(V^\mathrm{A})\leq \frac{(2-0)^2}{4}=1.
\end{align}

\subsection{Physically Motivated Rates - Pseudo Potential Form}
We specifically assume the transition rates to be of the form \cite{ConsistentNumericalSolver}
\begin{align}
    r_{xy} = D_{xy}\rexp=\Dt T\rexp,\label{eq:rateform}
\end{align}
where $D_{xy}$ is an edge specific ``diffusion coefficient" and
$V_{xy}$ is a pseudo-potential on the edge. This pseudo-potential
allows for general nonequilibrium systems to be implemented and therefore allows,
under given conditions, for taking the continuum limit. Temperature
perturbations enter the rates as 
\begin{align}
    r_{xy}^\delta = \Dt (T+\delta)\mathrm{e}^{-V_{xy}/(T+\delta)}\label{eq:pertrate}
\end{align}
By expanding Eq.~\eqref{eq:pertrate} around $\delta=0$ we get
\begin{align}
    r_{xy}^\delta = r_{xy} + \Dt^\mathrm{eff}\rexp \delta + \mathcal{O}(\delta^2),
\end{align}
where $\Dt^\mathrm{eff} \equiv \Dt(1+V_{xy}/T)$, so that
$r_{xy}^\delta - r_{xy} =\Dt^\mathrm{eff}\rexp \delta +
\mathcal{O}(\delta^2)$. We can plug everything into
Eq.~\eqref{eq:GeneralRateResult} to obtain
\begin{align}
    \partial_\delta O_t =& \lim_{\delta\to0}\frac{1}{\delta}\left(\left\langle O(t)\frac{\rmd \mathbb{P^\delta}}{\rmd \textcolor{black}{\mathbb{P}}}\right\rangle - \langle O(t)\rangle\right)\nonumber\\
    =& \frac{1}{T^2}\left\langle O(t)\stoints\sum_{x,y\neq x}\left(V_{xy} + T\right)\rmd\varepsilon_{xy}(s)\right\rangle\label{eq:result}
\end{align}
Note that we can remove the $T$ in the bracket in the last line if we
assume the $D_{xy}$ \emph{not} to scale with temperature, i.e., only a perturbation in the exponential in Eq.~\eqref{eq:pertrate}. By identifying the effective drift $-(V_{xy}+T)$, the result Eq.~\eqref{eq:result} corresponds to the result from Ref.~\cite{Klinger2025}.

\subsection{Perturbations of Path Observables}\label{sec:pathObs}
We already stated that Eq.~\eqref{eq:result} is valid for both,
single-time and path observables. The latter, however, perhaps need
some clarification in the context of evaluating expectations over path
ensembles. To clarify this, consider some hollow matrix $\f{b}$,
\begin{align}
    O(t) = \stoints \mathrm{Tr}[\f{b}^T\rmd\f\varepsilon(s)]
    \label{eq:ANoiseIntegral}.
\end{align}
In this case, Eq.~\eqref{eq:GeneralRateResult} simplifies to
\begin{align}
    \partial_\delta O_t = \int_0^t\rmd s \sum_{x,y\neq x} b_{xy}\tilde{r}_{xy}p_x(s),\label{eq:AExpectedNoisePerturbation}
\end{align}
which in general is non-zero. However, the definition
of the derivative in Eq.~\eqref{eq:response} contains averages of
Eq.~\eqref{eq:ANoiseIntegral}. One may, falsely, assume both averages
to vanish (recall that $\langle \rmd \varepsilon_{xy}(t)\rangle = 0$ for all $x\neq y$ and $t\geq 0$). Only $\langle O(t)\rangle=0$, while $\langle O(t)\rangle_\delta\neq 0$ in general. To further understand this, we can decompose Eq.~\eqref{eq:ANoiseIntegral} into jump and dwell-time integrals using Eq.~\eqref{eq:Matrix_eom_MJP}
\begin{align}
    O(t) = &\stoints \sum_{x,y\neq x}b_{xy}\rmd n_{xy}(s)
    \nonumber\\
    - &\stoints \sum_{x,y\neq x}b_{xy}r_{xy}\rmd \tau_{x}(s).\label{eq:ASplitIntegral}
\end{align}
Taking the average w.r.t. the perturbed system, it can be recognized that
\begin{align}
    \langle O(t)\rangle_\delta =& \left\langle\stoints \sum_{x,y\neq x}b_{xy}\rmd n_{xy}(s)\right\rangle_\delta\nonumber\\
    &- \left\langle\stoints \sum_{x,y\neq x}b_{xy}r_{xy}\rmd \tau_{x}(s)\right\rangle_\delta\nonumber\\
    =&\int_0^t\rmd s \sum_{x,y\neq x}b_{xy}r_{xy}^\delta p^\delta_x(s) \nonumber\\&- \int_0^t\rmd s \sum_{x,y\neq x}b_{xy}r_{xy} p^\delta_x(s)\nonumber\\
    =& \int_0^t\rmd s \sum_{x,y\neq x}b_{xy}\underbrace{(r_{xy}^\delta-r_{xy})}_{\delta \tilde{r}_{xy}} p^\delta_x(s)\nonumber\\
    =& \delta \int_0^t\rmd s \sum_{x,y\neq x}b_{xy}\tilde{r}_{xy} p_x(s) + \mathcal{O}(\delta^2),
    \label{eq:APerturbedAverage}
\end{align}
where $p^\delta_x(s)$ is the probability of $x^\delta_s=x$ in the perturbed system. The discrepancy of probabilities $\Delta p_x(s) = p_x^\delta(s) - p_x(s)$ in any state $x$ at time $s$ is $\mathcal{O}(\delta)$ \cite{Klinger2025}, so that $p^\delta_x(s) = p_x(s) + \Delta p_x(s) = p_x(s)+\mathcal{O}(\delta)$.

Dividing Eq.~\eqref{eq:APerturbedAverage} by $\delta$ and taking the
limit $\delta\to 0$ delivers
Eq.~\eqref{eq:AExpectedNoisePerturbation}. In other words, it is
important to account for the perturbation in the stochastic
differentials only, since the observable is nominally independent of the perturbation. i.e., $b_{xy}$ and $b_{xy}r_{xy}$ in the first and second term of Eq.~\eqref{eq:ASplitIntegral}, respectively, are invariant under perturbations (even though the unperturbed rates enter the latter term).

Now let the observable $O(t)$ be a dwell-time integral
\begin{align}
    O(t) = \stoints \textcolor{black}{ g_s\rmd s}
    \label{eq:ATimeIntegral},
\end{align}
with some state function \textcolor{black}{$g_\tau\equiv\sum_k \delta_{x_\tau k}g_k$}. The response is 
\begin{align}
    \partial_\delta O_t 
    &=\lim_{\delta\to 0}\frac{1}{\delta}\int_0^t\rmd s\sum_x g_x\left[ p^\delta_x(s) - p_x(s) \right]\label{eq:IntermediatePathResult1}
\end{align}
Using the Dyson identity for the perturbed generator $\f{L}^\delta =
\f{L} + \delta\tilde{\f{L}}+\mathcal{O}(\delta^2)$ with
$(\tilde{\f{L}})_{ij} = \tilde{r}_{ji}$ \cite{HANGGI1982207} yields
\begin{align}
    \rme^{\f{L}^\delta t} = \rme^{\f{L} t}  + \delta \int\rmd s \rme^{\f{L}(t-s)} \tilde{\f{L}}\rme^{\f{L}^\delta s}+\mathcal{O}(\delta^2).
\end{align}
To linear order in $\delta$, the probability is \cite{HANGGI1982207}
\begin{align}
    p^\delta_x(t) &= p_x(t) + \delta \sum_{i,j,n}\int_0^t\rmd s P(x, t|i, s)\tilde{r}_{ji}P(j, s|n, 0)p_n(0)\nonumber\\
    &= p_x(t) + \delta \sum_{i,j}\int_0^t\rmd s P(x, t|i, s)\tilde{r}_{ji}p_j(s).
\end{align}
Hence, the limit in Eq.~\eqref{eq:IntermediatePathResult1} reduces to
\begin{align}
    \lim_{\delta\to0}\frac{1}{\delta}\left[p_x^\delta(s) - p_x(s)\right] = \sum_{i,j}\int_0^s\rmd z P(x, s|i, z)\tilde{r}_{ji}p_j(z),
\end{align}
so that Eq.~\eqref{eq:IntermediatePathResult1} becomes
\begin{widetext}
\begin{align}
    \partial_\delta O_t&=\int_0^t\rmd s\sum_x g_x\sum_{i,j}\int_0^s\rmd z P(x, s|i, z)\tilde{r}_{ji}p_j(z)\nonumber\\
    &=\int_0^t\rmd s\sum_x g_x\int_0^s\rmd z \left[\sum_{i,j\neq i}P(x, s|i, z)\tilde{r}_{ji}p_j(z) + \sum_iP(x, s|i, z)\tilde{r}_{ii}p_i(z)\right]\nonumber\\
    &=\int_0^t\rmd s\sum_x g_x\int_0^s\rmd z \sum_{i,j\neq i}\left[P(x, s|i, z)\tilde{r}_{ji}p_j(z) - P(x, s|i, z)\tilde{r}_{ij}p_i(z)\right]\nonumber\\
    &=\int_0^t\rmd s\sum_x g_x\int_0^s\rmd z \sum_{i,j\neq i}\left[P(x, s|j, z) - P(x, s|i, z)\right]\tilde{r}_{ij}p_i(z).\label{eq:DysonResult}
\end{align}
\end{widetext}
Conversely, employing Eqs.~\eqref{eq:GeneralRateResult} and \eqref{eq:NoiseTimeCorrelationLemma} yields
\begin{align}
    \partial_\delta O_t = \sum_{i, x, y\neq x}\int_0^t\rmd \tau&\int_0^t\rmd s \textcolor{black}{\mathbb{1}}_{s<\tau}g_i \tilde{r}_{xy}p_x(s)\\
    &\times \left[P(i, \tau|y, s) - P(i, \tau|x, s)\right],\nonumber
\end{align}
which agrees with the Eq.~\eqref{eq:DysonResult}. 

Since the jump increments $\rmd \f{n}$ are a linear combination of the
dwell-time and noise increments, it follows immediately that
Eq.~\eqref{eq:GeneralRateResult} applies for jump-integrated
observables, e.g., currents,  as well.

Lastly, although not explicitly written, the matrix $\f{b}$ and \textcolor{black}{functions $g_k$} may also depend on
time. Equations~\eqref{eq:AExpectedNoisePerturbation} and
\eqref{eq:DysonResult} are written to readily accommodate for this.

\subsection{Time-Dependent Rates}
The result in Eq.~\eqref{eq:GeneralRateResult} easily allows for
time-dependent rates to be included. In fact, the Radon-Nikodym
derivative in Eq.~\eqref{eq:GeneralRadonNikodym} is already written in
a form that allows to accommodate time-inhomogeneous dynamics. If $\tilde{r}_{xy}(t)$ depends on $t$, then it may even be seen as a time-dependent perturbation. 

\subsection{Optimization of CTUR}\label{sec:optimal_c_cTUR}
\begin{figure*}[ht]
    \centering
        \includegraphics[width=.7\linewidth]{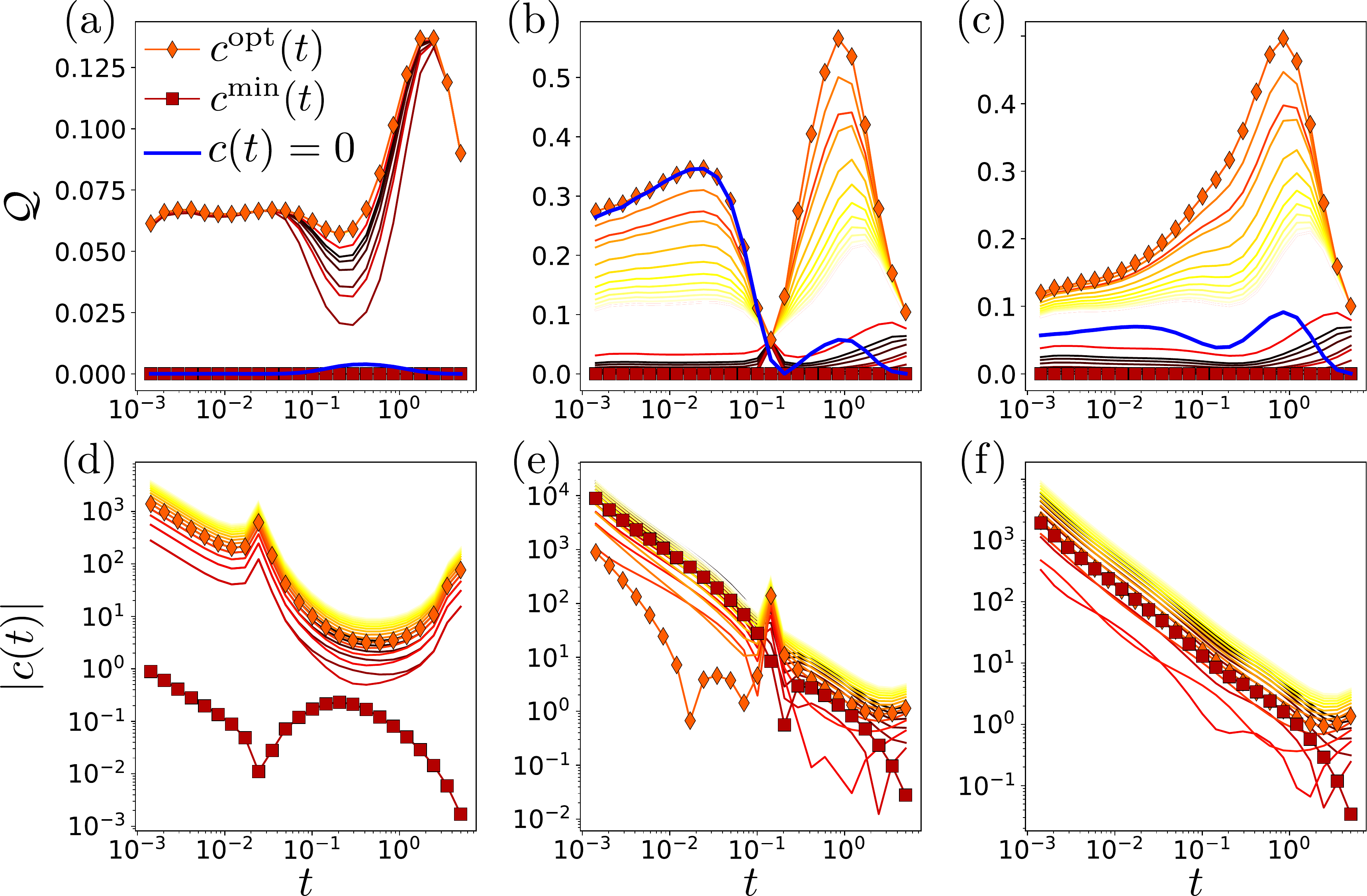}
    \caption{Quality factor of the CTUR in (a-c) using values of $c(t)$ shown in (d-f) in the calmodulin model. The values of $c(t)=c^{\textcolor{black}{\text{opt}}}(t) + \frac{\gamma}{5}(c^{\textcolor{black}{\text{opt}}}(t) - c^{\textcolor{black}{\min}}(t))$ are shown for values $-10 \leq \gamma \leq 10$, where $\gamma=-10$ is black and $\gamma=10$ is yellow. The initial condition is $p_i(0) = \delta_{i6}$ and the density uses $V_i(\tau)=\delta_{i3}$ in each panel. The transition weights are $\f{\kappa}^{(12)}$ in (a) and (d), $\f{\kappa}^{(56)}$ in (b) and (e), and $\f{\kappa}^{(26)}$ in (c) and (f). The most optimal $c(t)=c^{\textcolor{black}{\mathrm{opt}}}(t)$ is marked by diamonds and the least optimal $c(t)=c^{\textcolor{black}{\min}}(t)$ is marked by squares. In (a-c) the $\mathcal{Q}$ for $c=0$ is included as a blue line.}
    \label{fig:c_opt_TUR_comp}
\end{figure*}
In Fig.~\ref{fig:c_opt_TUR_comp} we further elaborate on the influence
of $c(t)$ in the CTUR. The quality factor $\mathcal{Q}$ for the
calmodulin system (see Fig.~\ref{fig:Models}) for currents with
$\f{\kappa}^{(12)}$, $\f{\kappa}^{(56)}$, and $\f{\kappa}^{(26)}$
(recall the notation
$\kappa_{ij}^{(kl)}=\delta_{ik}\delta_{jl}-\delta_{il}\delta_{jk}$)
and density with $V_i(\tau)=\delta_{i3}$ is shown in Fig.~\ref{fig:c_opt_TUR_comp}a-Fig.~\ref{fig:c_opt_TUR_comp}c.
Various values of $c(t)$ are used, including $c(t)=0$, $c^{\textcolor{black}{\mathrm{opt}}}(t)$,
and $c^{\textcolor{black}{\mathrm{min}}}(t)$. The absolute values of the considered $c(t)$ are shown in
Fig.~\ref{fig:c_opt_TUR_comp}d-Fig.~\ref{fig:c_opt_TUR_comp}f. As can be easily seen, the behavior of $c^{\textcolor{black}{\mathrm{opt}}}(t)$
and $c^{\textcolor{black}{\mathrm{min}}}(t)$ is highly nontrivial and changes greatly in magnitude.

\subsection{Simulation Parameters}\label{AppSimPara}

\begin{table}[ht]
    \centering
    \caption{\textcolor{black}{Transition rates used in the calmodulin example from Ref.~\cite{FirstPassageRick} which are adapted from Ref.~\cite{CalmodulinStigler}.}}
    \begin{tabular}{cc}
         \hline 
         Transition $i\to j$ & Rate $r_{ij}$\\
         \hline
         $1\to2$ &  5.997\\
         $2\to1$ &  0.774\\
         
         $1\to4$ &  13.439\\
         $4\to1$ &  127.968\\
         
         $1\to5$ &  15.330\\
         $5\to1$ &  0.121\\
         
         $5\to6$ &  3.749\\
         $6\to5$ &  13.326\\
         
         $2\to3$ &  1514.820\\
         $3\to2$ &  53.0661\\
         
         $2\to6$ &  13.441\\
         $6\to2$ &  2.922\\
         \hline
    \end{tabular}
    \label{tab:Calmodulin_Rates}
\end{table}

\begin{table}[ht]
    \centering
    \caption{\textcolor{black}{Values of fixed parameters in SAT simulations which are adapted from Ref. \cite{Berlaga2022}.}}
    \begin{tabular}{cc}
         \hline 
        Symbol &  Value\\
         \hline
         $\gamma$ & 5\\
         $x$ & 2\\
         $l_A$ & 20\\
         $l_B$ & 1\\
         $e^\mathrm{out}_A$ & 30\\
         $e^\mathrm{in}_A$ & 10\\
         $e^\mathrm{in}_B$ & 2\\
         \hline
    \end{tabular}
    \label{tab:ParameterSAT}
\end{table}

\begin{table}[ht]
    \centering
    \caption{\textcolor{black}{Overview of free energies for the four state ring with equilibrium rates.}}
    \begin{tabular}{c c}
        \hline
        State & $E_i$\\
        \hline
        $1$ & $0.1$ \\
        $2$ & $0.3$ \\
        $3$ & $0.5$ \\
        $4$ & $0.2$ \\
        \hline
    \end{tabular}
    \label{tab:FreeEnergiesRing}
\end{table}

\bibliography{sources}

\end{document}